\documentclass[aps,prb,twocolumn]{revtex4} 

\usepackage{graphicx}
\usepackage{dcolumn}
\usepackage{bm}
\usepackage{amsmath} 

\newcommand{\comment}[1]{}

\begin{document}

\title{Introduction to the Modern Theory of Bose-Einstein Condensation,
Superfluidity, and Superconductivity}


\author{Phil Attard}
\affiliation{ {\tt phil.attard1@gmail.com}  8 Sept 2025, \today }



\begin{abstract}
The modern theory of Bose-Einstein condensation,
superfluidity, and superconductivity is reviewed.
The thermodynamic principle for superfluid flow
and the equation of motion for condensed bosons are given.
Computer simulations of Lennard-Jones $^4$He
give the $\lambda$-transition and the superfluid viscosity.
The statistical mechanical theory of
high-temperature superconductivity is presented.
Critical comparison is made with older approaches,
such as ground energy state condensation,
irrotational superfluid flow, and the macroscopic wavefunction.
\end{abstract}



\maketitle

%
\section{Introduction}
\setcounter{equation}{0} \setcounter{subsubsection}{0}
\renewcommand{\theequation}{\arabic{section}.\arabic{equation}}
%

In recent years the author has advocated a new approach
to Bose-Einstein condensation, superfluidity, and superconductivity
that provides a detailed, molecular-level understanding
of these phenomena and their physical basis.
Aspects of the new theory contradict conventional views,
particularly in regard to the nature of condensation
and the origin of superfluidity.
The new perspective has in part come from a formally exact formulation
of quantum statistical mechanics in classical phase space.
This has enabled discussion in terms of particles,
which are more intuitively appealing than quantum wavefunctions.
It has also facilitated the development of efficient computer algorithms
for the simulation of the $\lambda$-transition for interacting  $^4$He,
and for the simulation of the shear viscosity in the superfluid regime.

The modern theory goes beyond
Einstein's (1924, 1925) idea of Bose-Einstein condensation
into the ground energy state.
F. London's (1938) ideal boson model
for the $\lambda$-transition has likewise been re-interpreted,
and complemented with computer simulations for interacting $^4$He.
Most significantly,
Landau's (1941) theory of superfluidity
has been replaced by
the new picture of Bose-Einstein condensation
and the general thermodynamic principle
that drives superflows and supercurrents.
The equation of motion for condensed bosons has been obtained,
which explains physically why superfluid flow is flow without viscosity,
and why supercurrents have no resistance.
It has also given rise to a quantum molecular dynamics algorithm,
which, for the first time,
shows quantitatively the reduction in viscosity in helium~II.
In the case of superconductivity,
an explanation for the Meissner-Ochsenfeld (1933) effect has been obtained,
and a candidate for the pairing mechanism in high-temperature
superconductivity has been identified.

A recent book (Attard 2025a)
documents developments prior to mid 2024.
However the mathematical detail in that book
may not be required by those who would prefer to begin with
an overview of the subject.
Also, the modern theory is developing rapidly,
and recent progress has taken some topics out of book.
The purpose of this review is to consolidate the newer aspects
of the theory,
with comprehensive coverage beginning at an elementary level.
The focus is on the broad concepts and the main arguments
while avoiding much of the mathematical and experimental detail.

%
\section{The $\lambda$-Transition}
\setcounter{equation}{0} \setcounter{subsubsection}{0}
\renewcommand{\theequation}{\arabic{section}.\arabic{equation}}
%

\subsection{Bose-Einstein Condensation} \label{Sec:BEC}

\subsubsection{Boson Configurations}

\begin{figure}[t]
\centerline{ \resizebox{8cm}{!}{ \includegraphics*{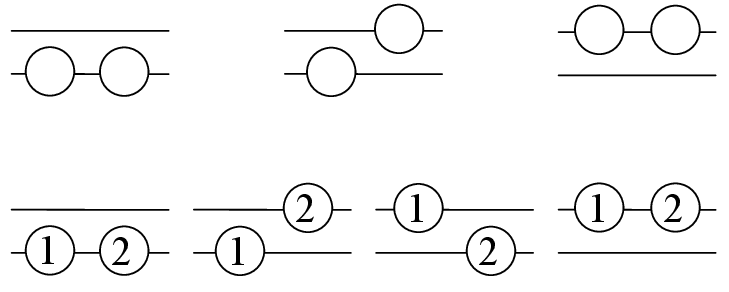} } }
\caption{\label{Fig:occ}
The three possible occupancies (upper),
and the four possible configurations (lower)
of two bosons in two single-particle states.
}
\end{figure}

The simplest way to understand Bose-Einstein condensation
is in terms of states and their occupancy.
For identical bosons,
it is the occupancy of the quantum states that matters
when it comes to counting the number of possible arrangements of the system
(Fig.~\ref{Fig:occ}, upper).
But it is more natural to think
in terms of the number of configurations of labeled particles
(Fig.~\ref{Fig:occ}, lower).
This is the way that we perceive the macroscopic world,
it allows individual particles to be followed over time,
and it yields the most straightforward mathematical formulation
of the physical situation.
The issue is that since the two viewpoints
---occupancy and configuration---
reflect the same physical situation,
they have to be consistent in the way that they count number.
Specifically, in the lower example of Fig.~\ref{Fig:occ},
the configuration with the two bosons in different states is counted twice,
and so each must have half the weight of a configuration with the two bosons
in the same state.
In other words,  configurations with  bosons in the same state
have more weight than configurations with bosons in different states.

More generally,
a given configuration of $N$ bosons
with $N_{\bf a}$ occupying the single-particle state ${\bf a}$
has weight proportional to $ \prod_{\bf a} N_{\bf a}!$.
Because of the factorial,
a configuration with a few highly-occupied states
has much more weight
than a configuration with many few-occupied states.
This is the origin of Bose-Einstein condensation.

The greater weight of multiply-occupied states
shows that Bose-Einstein condensation is driven by entropy.
According to Boltzmann,
the entropy of a state is the logarithm of the weight of molecular
configurations in that state.
Boltzmann dealt with the simplest case where weight
equals the number of equally weighted configurations,
but more generally entropy is also defined
for non-uniformly weighted configurations (Attard 2002, 2012).

It is important to note in the above explanation
that it is the occupancies (unlabeled particles)
that are counted equally,
not the configurations (labeled particles).
This reflects the fundamental quantum property
that the wavefunction  for identical bosons must be fully symmetric
with respect to particle interchange.
The two configurations with bosons in different states
(Fig.~\ref{Fig:occ}, lower)
are not symmetric with respect to interchange,
but the single corresponding occupancy (Fig.~\ref{Fig:occ}, upper) is.
The unsymmetrized wavefunction,
$\Phi_{\bf p}({\bf q}) = \prod_{j=1}^N \phi_{{\bf p}_j}({\bf q}_j)$,
where particle $j$ at ${\bf q}_j$
is in the single-particle state ${\bf p}_j$,
is in configuration form.
It changes if particles $j$ and $k$ 
are transposed,
$ \phi_{{\bf p}_j}({\bf q}_j)\phi_{{\bf p}_k}({\bf q}_k)
\ne \phi_{{\bf p}_k}({\bf q}_j)\phi_{{\bf p}_j}({\bf q}_k)$,
$ {\bf p}_j \ne {\bf p}_k$.
To satisfy the quantum requirement,
the wavefunction is symmetrized by summing over all possible permutations,
\begin{equation}
\Phi^+_{\bf p}({\bf q}) =
\frac{1}{\sqrt{N! \chi_{\bf p}^+}}
\sum_{\hat{\rm P}} \Phi_{\hat{\rm P}{\bf p}}({\bf q}) .
\end{equation}
Normalization is ensured by the symmetrization factor,
$\chi_{\bf p}^+ =   \prod_{\bf a} N_{\bf a}({\bf p})!  $,
with the occupancy being
$N_{\bf a}({\bf p}) = \sum_{j=1}^N \delta_{{\bf p}_j,{\bf a}}$.
The logarithm of the symmetrization factor
gives the occupation entropy of the configuration.

\begin{figure}[t]
\centerline{ \resizebox{8cm}{!}{ \includegraphics*{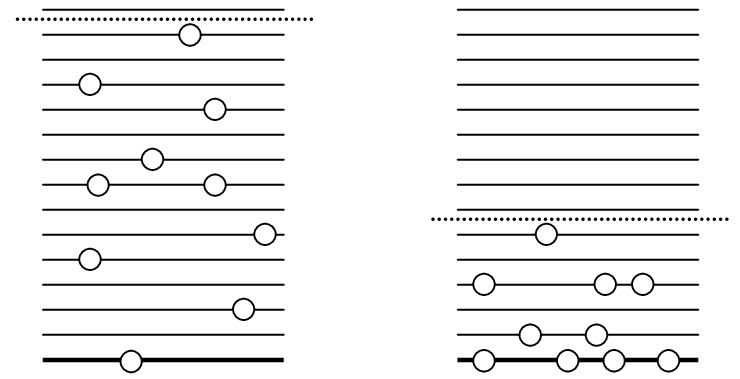} } }
\caption{\label{Fig:access}
Quantum single-particle states occupied by 10 bosons
at high temperatures (left) and at low temperatures (right),
with the dotted line delimiting the accessible states.
}
\end{figure}

Whether or not Bose-Einstein condensation occurs
is determined by  two competing entropic effects.
The number of accessible momentum states increases with temperature,
and at high temperatures there are many more accessible states
than there are particles (Fig.~\ref{Fig:access}, left).
(The reason for focussing upon momentum states will be made clear shortly.)
In this regime entropy dictates that almost all accessible states
are empty, with a minority being singly occupied,
and with almost none being multiply occupied.
This is the uncondensed or classical regime.
As the temperature is lowered,
the number of accessible momentum states is reduced.
When their number is comparable to the number of bosons
then many will be multiply occupied,
although singly occupied and empty states are still present
(Fig.~\ref{Fig:access}, right).
The reason for empty states
is that the permutation weight discussed above
makes it favorable for bosons in nearby states to condense
into the same state, thereby emptying the neighbors.
The reason that there is not a unique, macroscopically occupied state
(eg.\ the ground state) is that entropy drives
the spreading out of the occupancies
into any and all of the accessible states,
(admittedly with a bias toward low-lying states).
Also the relative fluctuations in occupancy are on the order of unity,
so that a state that is highly occupied at one instant
may be empty at another.

\begin{figure}[t]
\centerline{ \resizebox{8cm}{!}{ \includegraphics*{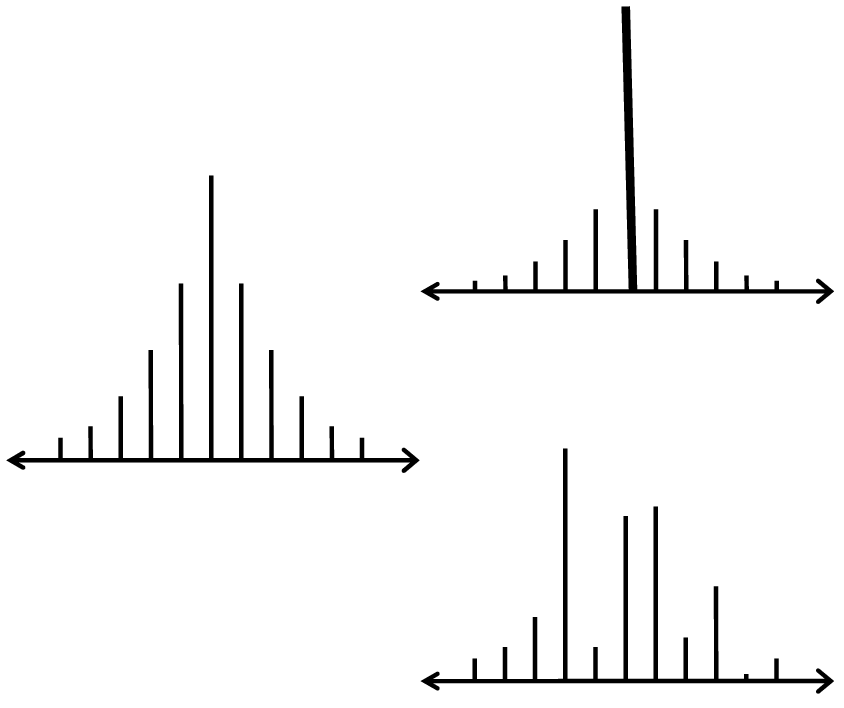} } }
\caption{\label{Fig:Con}
Occupancy of states prior to condensation (left),
and after condensation according to Einstein (right, upper),
and according to the present author (right, lower).
}
\end{figure}

The spacing between momentum states
is inversely proportional to the size of the system,
$\Delta_p = 2\pi \hbar /L$,
where the volume is $V=L^3$.
This is infinitesimal for a macroscopic system,
which means that momentum can usually be regarded
as belonging to the continuum,
the exception being when the occupancy of the momentum states
has to be accounted for.
This infinitesimal spacing between momentum states
explains why empty states can be intercalated with highly occupied states
at little or no energy cost,
and why the ground momentum state is not materially different
to other low-lying momentum states (Fig.~\ref{Fig:Con}).

This also explains why there is no latent heat
at the $\lambda$-transition (Donnelly and Barenghi 1998).
In the present model a macroscopic number of bosons can condense
without a macroscopic energy change.
(A macroscopic number is required to account for
the behavior of the heat capacity, which is an extensive variable,
and for the discontinuous appearance of superfluidity,
which can be observed with the naked eye.)
If the macroscopic number of bosons
condensed solely into the ground energy state,
then there would be a discontinuous change in energy
and a latent heat at the transition.

This  idea differs from Einstein's (1924, 1925)
conception of Bose-Einstein condensation
in two, dare I say, ground-breaking ways.
Einstein asserted that condensation was into energy states,
specifically into the ground energy state.
However, both aspects of this idea are wrong.
First, condensation cannot be into energy states
because for interacting particles the occupancy of energy states
cannot be defined mathematically  (Attard 2025b Appendix~A).
Penrose and Onsager (1956 p.~577) state
`the average number of particles in the lowest single-particle level
\ldots
has meaning for noninteracting particles only,
because single-particle energy
levels are not defined for interacting particles.'
And second, condensation cannot be solely into the ground state,
whether energy or other,
because for a macroscopic system
the spacing between states is negligible compared
to the thermal energy.
Also, the occupancy of a single-particle state
is an intensive thermodynamic variable
(Attard 2025a \S2.5).
A sketch of condensation as conceived by Einstein
and by the present author is given in Fig.~\ref{Fig:Con}.

It is difficult to overstate the damage caused
by Enstein's (1924, 1925) assertion that Bose-Einstein condensation
was solely into the ground energy  state.
The conventional understanding
of superfluidity and of superconductivity
in its entirety is based upon it.
The many limitations of conventional theory
can be blamed on this misconception.
I stress that Einstein is to be admired for the original idea
of boson condensation,
and also for making a first approximation to describe it.
But those who came after Einstein
failed to critically examine his assumptions:
their respect for him as an authority
should have made them take his work seriously;
instead they took it literally.
Once a sufficient number of scientists had taken up his idea,
peer group pressure proved overwhelming
and no one ever questioned it.
The many deleterious consequences of this error
will be discussed in detail in several places below.

\subsubsection{Momentum States, Decoherence,
and the Quantum-Classical Transition}

It is worth pausing to take a closer look at the foundations
that underly the above description of Bose-Einstein condensation.
Most notable, perhaps, is the formulation in terms of momentum states
rather than the energy states that are ubiquitous
in quantum mechanical analysis.
The emphasis on configurations rather than wave functions
also stands out.
There are both practical and conceptual reasons for these choices.

As mentioned,
occupancy can only be defined for single-particle states
such as momentum states.
For interacting particles the occupancy of an energy state is undefined
(Attard 2025b Appendix~A, Penrose and Onsager 1956 p.~577).
Since Bose-Einstein condensation concerns the multiple occupancy of states,
it would be pointless to attempt to describe it in terms of energy states.

A related point is that momentum states and eigenfunctions
are universal,
since they depend only upon the size, and perhaps the geometry,
of the system.
They can be given exactly and explicitly.
In contrast, for interacting particles the energy eigenfunctions
and energy eigenvalues always involve approximations of one sort or another,
and they change depending upon the nature of the interactions.
It is obviously a very great advantage to formulate a theory
in terms that are universally applicable,
and this no doubt reflects the universal nature
of Bose-Einstein condensation itself.

The momentum eigenfunction,
$\phi_{\bf p}({\bf q}) = V^{-N/2} e^{-{\bf p}\cdot{\bf q}/{\rm i}\hbar}$,
associates a complex number $\phi$ with each configuration of
the system $\{{\bf q},{\bf p}\}$,
which is a  point in the $6N$-dimensional classical phase space.
This is the natural link between
the wave functions of quantum mechanics
and the configurations of classical mechanics.
In the opinion of the present author,
the great advantage of describing a system in terms of
the configurations of its particles
is that these align with natural thought processes:
human brains have evolved to make sense of the world
as perceived through the fives senses,
and these respond only to macroscopic classical stimuli.
A second advantage,
as is demonstrated by the numerical results below,
is that it is easier and more efficient
to formulate computer simulation algorithms for quantum systems
in classical phase space than it is in wave space.

Of course quantum and classical phenomena can be quite different,
and not all quantum phenomena can be usefully described
in classical terms.
In the present context there are four quantum effects
beyond classical experience
that are directly relevant to the modern understanding
of Bose-Einstein condensation, superfluidity and superconductivity.
These are:
({\bf 1}) the indistinguishability of particles,
which must be reconciled with configurations;
({\bf 2})  the superposition of quantum states due to the linearity
of wavefunctions and operators;
({\bf 3}) Heisenberg's uncertainty principle
for position and momentum,
which casts doubt on the reality of phase space configurations
and of the trajectories of particles in time;
and ({\bf 4}) quantum non-locality.
The mathematical treatments of these are
of varying complexity and sophistication.
The real issue is whether
these quantum attributes directly influence the measured phenomena,
and, if so, whether a classical or quasi-classical formulation contributes
to the understanding of them.
What modifications to classical notions are required,
and is such a quasi-classical formulation more useful
than \emph{ab initio} quantum analysis?

{\bf 1.}\
The treatment of indistinguishability and occupancy
in the configuration picture has been dealt with above.
It is mathematically exact,
and the extra effort required to weight the multiple occupancy
of momentum states with the symmetrization factor
is more than justified by the transparency afforded by configurations.
In addition it gives a very direct measure of condensation
and it allows an interpretation in particulate terms
of many of the relevant physical phenomena.

{\bf 2.}\
Schr\"odinger's cat is dead.
The suppression of superposition
in the transition from the quantum world to the classical world
is ultimately due to the macroscopic nature
of most classical systems of interest.
Quantum mechanics is restricted to closed quantum systems
of few particles, and in these the superposition of states
due to the coherent nature of the wave function is evident.
But in reality most systems of interest
are open macroscopic quantum systems that interact with their environment
(or with other parts of themselves).
The conservation laws due to exchange with the environment
create an entangled, decoherent wavefunction
such that  superposition collapses
(Joos and Zeh 1985, Schlosshauer 2005, Zurek 1991).
This is the basis for the trace formulation
of quantum statistical mechanics (Attard 2018, 2021).
For this reason the quantum mechanics of closed quantum systems
is of limited use in applications of Bose-Einstein condensation
to the $\lambda$-transition, superfluidity,
and high-temperature superconductivity.
One exception is the temporary superposition states
in the molecular dynamics of superfluidity (\S\ref{Sec:SMSF}).
A system composed of Avogadro's number of particles
should be described statistically rather than mechanically.
The differences between quantum mechanics for few-body closed quantum systems,
and quantum statistical mechanics for macroscopic open quantum systems,
go beyond the absence of superposition states,
and they can be significant.

{\bf 3.}\
The lack of simultaneity for position and momentum
technically refers to the non-commutativity
of the position and momentum operators,
and the consequent Heisenberg uncertainty principle
for the product of the variances of their expectation values.
This is not directly relevant for classical phase space;
evaluating the momentum eigenfunction
$\phi_{\bf p}({\bf q})$ at a precise position configuration ${\bf q}$
for a precise momentum eigenvalue ${\bf p}$
is a well-defined mathematical operation
whose physical interpretation must be judged by its  consequences.
The mathematical derivation of quantum statistical mechanics
from quantum mechanics
accounts for the non-commutativity of the position and momentum operators
with what I call the commutation function
(Attard 2017, 2018, 2021),
but which might be better called the Wigner-Kirkwood function
(Wigner 1932, Kirkwood 1933).
This  short-ranged function
is neglected in the following
because Bose-Einstein condensation
is dominated by long-range, non-local effects.
This function is identically zero
for non-interacting particles,
and the fact that the ideal boson model gives a passingly good description
of the $\lambda$-transition in $^4$He (see next)
tends to confirm that neglecting it in general is reasonable.
An exception is the computational values
for the saturated liquid density for Lennard-Jones $^4$He,
which are overestimated because of the neglect
of this repulsive function
(\S \ref{Sec:WKfn}).

{\bf 4.}\
Quantum non-locality is manifest in phenomena such as
({\bf a}) the dependence
of the wave function and eigenvalues on the boundaries of the subsystem,
({\bf b}) the entanglement of the subsystem with the environment,
and ({\bf c}) the effects of multiple occupancy of momentum states irrespective
of the separation of the particles involved.
This last phenomenon is obviously directly
relevant to Bose-Einstein condensation,
but in fact all three instances of non-locality
have important consequences.
The reason that the classical world appears to be localized
is that the effects of non-locality are absent.
({\bf a}.)
The spacing between momentum states is inversely proportional
to the distance between the boundaries,
and since classical systems are almost all macroscopic,
momentum appears to be continuous.
There are some important exceptions to this in the following,
as in the treatment of the superfluid critical velocity,
the calculations of the occupancy of the momentum states,
and the simulations of the superfluid viscosity.
({\bf b}.)
As already discussed, non-local entanglement
collapses the subsystem into a decoherent mixture of pure quantum states,
suppresses superposition states,
and gives rise to quantum statistical mechanics.
({\bf c}.)
Bose-Einstein condensation is fundamentally non-local.
At high temperatures the momentum states are empty or singly occupied
and this is the classical regime
where the  non-local occupancy effects  are non-existent.
At low temperatures multiple occupancy of momentum states occurs,
which makes this the quantum regime where non-locality is apparent.

We make one further observation about
the classical phase space formulation of quantum statistical mechanics
that is used for the modern theory
of Bose-Einstein condensation, superfluidity,
and high-temperature superconductivity.
It can be argued that the $\lambda$-transition in liquid $^4$He,
which represents the onset of Bose-Einstein condensation and of superfluidity,
marks the boundary between the quantum and the classical domains.
The uncondensed regime is the classical regime,
and the condensed regime is the quantum regime.
This provides some motivation for formulating the problem
in classical, or quasi-classical, terms:
momentum eigenfunctions are the portal between
the quantum and the classical worlds.

\subsection{Ideal Boson Model} \label{Sec:Ideal}

It was F. London (1938)
who made the link between Bose-Einstein condensation
and the $\lambda$-transition in $^4$He.
He used the ideal (non-interacting, free) boson model,
which can be solved analytically.
Using the measured saturated liquid density,
he predicted a transition temperature, $T_{\rm min}^{\rm id} = 3.13$\,K,
that was close to the measured one, $T_\lambda = 2.17$\,K.
This convinced most workers
that Bose-Einstein condensation was real,
and that it was the cause of superfluidity
and, by analogy, of superconductivity.

Unfortunately, it also convinced workers that condensation
was into the ground energy state,
as assumed in the ideal boson calculations
(Attard 2025a Ch.~2, F London 1938, Pathria 1972 \S7.1).
This notion was used as the basis for other results,
such as the equation for the superfluid fountain pressure
(H. London 1939) (see \S\ref{Sec:Fountain}),
and the two-fluid model for hydrodynamic superfluid flow
(Tisza 1938) (see \S\ref{Sec:TwoFluid}).
The quantitative success of these in describing experimental data
strongly reinforced the idea of ground energy state condensation.

Landau,
who was awarded the Nobel Prize in physics in 1962 for `pioneer
investigations in the theory of condensed matter and especially of
liquid helium',
never accepted Bose-Einstein condensation
as the basis of superfluidity,
presumably because he thought an ideal gas model of liquid $^4$He unrealistic.
Nevertheless Landau
was apparently familiar with the work of the London brothers
and of Tisza (Balibar 2014, 2017)
and all of Landau's work on superfluidity
was based on the assumption that it was carried by $^4$He atoms
in the ground energy state
(Landau  1941, Landau and Lifshitz  1955).
I shall criticize the theories of Landau in detail below.

There is a real question as to the applicability
of the ideal boson model to the dense liquid.
Does the agreement with measurement for the  transition temperature
reflect the underlying physics captured by the model,
or is it simply a coincidence due to some sort of cancelation of errors,
or some form of parameter fitting?

In the ideal boson model there is only kinetic energy,
and it can be equally formulated
in terms of energy or momentum states.
When the continuum approximation is made,
the average density of bosons is predicted to be
$\overline N^{\rm id}/V
= \Lambda^{-3}  g_{3/2}(z)
\le \Lambda^{-3} \zeta(3/2)$,
where $\Lambda = [2\pi\hbar^2\beta/m]^{1/2}$ is the thermal wavelength,
$z^{\beta\mu} <1$ is the fugacity,
and $\beta = 1/k_\mathrm{B}T$ is the inverse temperature
(Attard 2025a \S2.3.2, Pathria 1972 \S7.1).
For the atomic mass of  $^4$He,
the maximum value at $z=1$
becomes less than the  measured saturated
liquid density of $^4$He below $T_{\rm min}^{\rm id} = 3.13$\,K.
This is the lowest temperature
that still gives the liquid density.
F. London (1938) identified this
as the $\lambda$-transition temperature.

Because the volume element,
$4\pi p^2 {\rm d}p$, vanishes at the origin,
Pathria (1972 \S7.1) and others
assert that the continuum approximation
neglects the occupancy of the ground energy state.
They say that the calculated value
reflects only bosons in excited states,
$\overline N_*^{\rm id}(z,T)=V\Lambda^{-3} g_{3/2}(z)$.
It is further asserted that bosons only begin to condense
into the ground energy state for $T \le T_{\rm min}^{\rm id}$,
and then their number is
$N_0 =  V[ \rho_{\rm l}^{\rm sat} -  \Lambda^{-3} \zeta(3/2)]$,
where $\rho_{\rm l}^{\rm sat}$
is the measured saturation liquid density.

In terms of mathematical rigor,
it is true that the conversion
of the sum over discrete momenta to the continuum integral
breaks down in the joint limit $p\to 0$ and $z\to 1$.
The proposed fix of explicitly adding the ground momentum state
contribution to the continuum integral
works well in a practical sense,
as comparison with exact enumeration of the momentum states shows
(Attard 2025a \S2.4).
However, the physical interpretation is another matter,
since the fact that the density from the continuum integral
is less than the measured liquid density
is a consequence of the ideal boson model
neglecting the attractive interactions between $^4$He atoms.
Setting the fugacity to unity at and below this point
is required by the ideal boson model to get the measured density,
but it conflicts with the measured chemical potential
(Attard 2025a \S4.4.2, Donnelly and  Barenghi  1998),
which gives a fugacity substantially less than unity.

Ascribing the `excess' measured bosons below $ T_{\rm min}^{\rm id}$
to the ground state of the ideal boson model
has the appearance of condensation by Einstein's (1924, 1925) criterion.
But in reality there is limited mathematical justification for this;
indeed it would be more consistent with thermodynamics
and with the continuum approximation
to assign them to a range of low-lying momentum states (Attard 2025a \S2.5).
London's (1938) expression for the ground state occupancy
below the transition temperature,
$N_0 =  V[ \rho_{\rm l}^{\rm sat} -  \Lambda^{-3} \zeta(3/2)]$,
is a serious violation of the fundamental thermodynamic principle
that the occupancy of a single-particle state
is an intensive thermodynamic variable,
which means that it cannot be macroscopic
(Attard 2025a \S2.5).
The simplest way to see this is that since the volume of each momentum state
is inversely proportional to the volume of the system,
doubling the volume and number of bosons in the system
(ie.\ constant density)
halves the volume of the momentum states,
which means that twice as many bosons go into twice as many states,
leaving the occupancy of each comparable state unchanged.
For this reason the fundamental definition
of Bose-Einstein condensation given by Penrose and Onsager (1956 p.~583),
`B.E. condensation is present
whenever a finite fraction of the particles
occupies one single-particle quantum state',
is quite wrong as it contradicts this basic thermodynamic requirement.
In fact, thermodynamics demands that
the fraction of bosons condensed in any one state, ground or otherwise,
goes to zero in the thermodynamic limit,
even deep in the Bose-Einstein condensation regime.
It is only a macroscopic range of states
that can be occupied by a macroscopic (ie.\ non-infinitesimal fraction)
number of condensed bosons.

Because of this requirement that the occupancy
of the ground momentum state cannot be macroscopic,
one should not take F. London's ideal boson calculations too literally.
The exact enumeration of states for the ideal boson model
(Attard 2025a \S2.4) shows that the occupancy
of the ground momentum state is non-zero above $T_{\rm min}^{\rm id}$,
and that the so-called excited state bosons, $N_*$,
undergo permutations with non-zero weight
both above and below $T_{\rm min}^{\rm id}$.

\begin{figure}[t]
\centerline{ \resizebox{8cm}{!}{ \includegraphics*{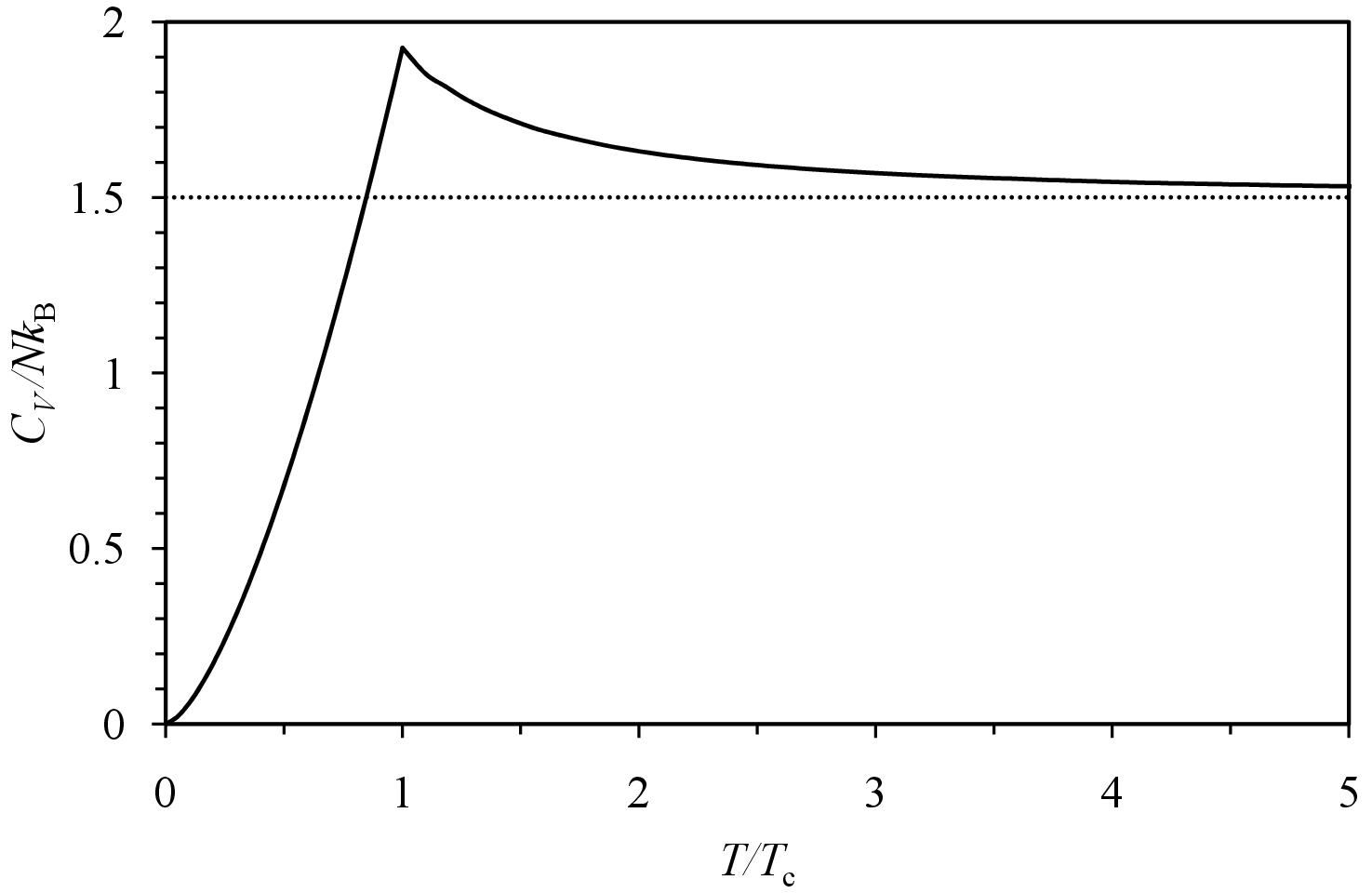} } }
\caption{\label{Fig:IdealCv}
Specific heat capacity for ideal bosons.
The dotted line is the classical ideal gas result.
}
\end{figure}

Figure~\ref{Fig:IdealCv}
shows the heat capacity predicted by the ideal boson model
(Attard 2025a \S2.3, F. London 1938, Pathria 1972 \S7.1).
Unlike the sharp divergence that signifies the $\lambda$-transition
in laboratory measurements,
the model gives a finite peak of rather broad width.
The existence of the peak and of the first order discontinuity
are due solely to two different models being joined at
$ T_{\rm min}^{\rm id}$:
ground state occupancy is taken to be $N_0=0$ above this temperature,
and $N_0 =  V[ \rho_{\rm l}^{\rm sat} -  \Lambda^{-3} \zeta(3/2)]$
below it.

On the far side of the $\lambda$-transition,
the ideal boson heat capacity goes like $(T/T_{\rm min}^{\rm id})^{3/2}$,
which is consistent with the measured data.
Hence in the condensed regime
the ideal boson model largely works,
and,
apart from the macroscopic occupancy of the ground  state,
its description of Bose-Einstein condensation
reflects some elements of reality.
One can conclude that
in the condensed regime
interactions play only a secondary r\^ole.
This supports  the idea  that condensation is into momentum states,
which are independent of the interaction potential.

It is notable that Bose-Einstein condensation is non-local:
the permutation of bosons in the same momentum state
pays no regard to where those bosons are located in the system.
Since the number of pairs of bosons separated by $r$ grows as
$ V \rho_{\rm l}^2 4\pi r^2 {\rm d}r$,
condensation is dominated by bosons at macroscopic separations
beyond the range of the pair potential.
This explains why the ideal boson model works so well
for Bose-Einstein condensation below the $\lambda$-transition,
whereas it would not work at all for gas-liquid condensation.

So how is one to judge the ideal boson model
and F. London's (1938) work on the $\lambda$-transition in $^4$He?
There is no doubt that Bose-Einstein condensation
is responsible for the  $\lambda$-transition and superfluidity,
if for no other reason then that these don't occur in $^3$He,
a fermion, at a comparable temperature.
F. London deserves full credit for making this connection.
The argument that the ideal gas model cannot possibly account
for liquid $^4$He might be answered by suggesting
that the interactions have been subsumed into a one-body mean-field potential
that gives an effective ideal fugacity close to unity
and greater than the measured fugacity.
That the ideal boson model gives qualitatively correct behavior
for the heat capacity below the $\lambda$-transition
is consistent with the non-local nature of Bose-Einstein condensation,
which makes interactions irrelevant.
The approximation that condensation is solely into the ground state
can be remedied by invoking a macroscopic range
of low-lying momentum states.
The conclusions that excited state bosons
are not in multiply occupied states,
or that the ground state is not occupied above the transition,
are contradicted by the exact enumeration of the model.

The measured divergence in the heat capacity at the $\lambda$-transition
is due to interactions between the $^4$He atoms,
as is shown quantitatively next.
Since the ideal boson model is incapable of exhibiting this divergence,
one must conclude that the peak in the ideal heat capacity
is qualitatively different to the $\lambda$-transition in reality,
and that the physical origins of the two are different.
This is a strong argument
that the coincidence of
the lowest temperature that still gives the liquid density,
$T_{\rm min}^{\rm id} = 3.13$\,K,
and the measured $\lambda$-transition temperature,
$T_\lambda = 2.17$\,K,
is a lucky accident.

\subsection{Interacting Bosons and the $\lambda$-Transition}
\label{Sec:IntBos}

\subsubsection{Nature of the $\lambda$-Transition}

The $\lambda$-transition is signified by a spike in the heat capacity
of saturated liquid $^4$He at 2.2\,K. The experimental
evidence is that on the liquid saturation curve the energy,
the density, and the shear viscosity are continuous
functions of temperature at the  $\lambda$-transition; the density
and the shear viscosity have a discontinuity in their first
temperature derivative (Donnelly and Barenghi 1998).
Superfluid flow occurs in thin films and capillaries immediately
below the  $\lambda$-transition, but not above it.

Bose-Einstein condensation
was introduced above as being driven by permutation entropy,
with permutations between bosons in the same momentum state
having unit weight.
These dominate below the condensation transition
and are responsible for superfluidity (and for superconductivity).
However in the immediate vicinity of the transition
permutations between bosons in different momentum states
contribute and in fact they can outweigh or even suppress
same-state permutations.

That these are important can be gleaned from the nature
of the $\lambda$-transition in $^4$He,
which shows an integrable divergence in the heat capacity
(Lipa \emph{et al.}\ 1996).
For the ideal boson model
the specific heat capacity is finite at the $\lambda$-transition
(Fig.~\ref{Fig:IdealCv}).
Since for ideal bosons the only non-zero permutations
are those between bosons in the same momentum or energy state
(Attard 2025a Eq.~(2.4)), 
it is clear that the measured divergence in the heat capacity
must be due to permutations between bosons in different momentum states.
Of course the other difference with the ideal boson model
is the interaction potential between the $^4$He atoms,
and this plays a quantitative r\^ole in the location
of the $\lambda$-transition
and in the behavior of the system in its vicinity.

\subsubsection{Position Permutation Loops}

In general the spacing between quantum states
decreases with increasing system size.
In particular, for momentum states
it is inversely proportional to the size of the system,
$\Delta_p = 2\pi \hbar /L$, where the volume is $V=L^3$.
Hence for macroscopic systems the continuum limit holds,
as is the classical experience.
In the discrete case the symmetrization factor gives the number of
non-zero permutations,
and in the continuum case its analogue,
the symmetrization function,
accounts for the fact that not all permutations are equal.
In this case it is the ratio
of the permuted and the unpermuted wavefunctions that gives the weight.
When bosons in the same momentum state are swapped,
the wavefunction is unchanged,
which is to say that their permutation has unit weight.
But because the momentum eigenfunction
is a Fourier factor, $\phi_{\bf p}({\bf q})
= V^{-N/2} e^{-{\bf p}\cdot{\bf q} /{\rm i}\hbar}$,
permuting bosons in different states gives an oscillatory factor
that averages to zero over small changes in momentum or in position.
These concepts hold as well in the continuum.

As mentioned,
the weight of a permutation is the ratio of the permuted to the unpermuted
wave function.
All permutations may be factored into loops.
A position loop is a cyclic permutation around a ring of
bosons with successive neighbors in close spatial proximity.
The weight of an $l$-loop
after averaging over the momenta
with Maxwellian weight (ie.\ $e^{-\beta p^2/2m}$) is
(Attard 2025a \S3.1)
\begin{eqnarray}
\eta^{(l)}_*({\bf p}^l,{\bf q}^l)
& = &
e^{-{\bf p}_{j_l}\cdot{\bf q}_{j_l,j_{1}} /{\rm i}\hbar}
\prod_{k=1}^{l-1}
e^{-{\bf p}_{j_k}\cdot{\bf q}_{j_k,j_{k+1}} /{\rm i}\hbar}
\nonumber \\
\Rightarrow
\eta^{(l)}_*({\bf q}^l)
& = &
e^{-\pi q_{j_l,j_{1}}^2 /\Lambda^2}
\prod_{k=1}^{l-1} e^{-\pi q_{j_k,j_{k+1}}^2 /\Lambda^2} .
\end{eqnarray}
The independent momentum integrals used for the averages
mean that the permutation
is for bosons not in the same momentum state.
The result shows that if consecutive bosons around the loop
are separated by less than about the thermal wavelength,
$\Lambda \equiv \sqrt{2\pi\hbar^2/mk_{\rm B}T}$,
then the weight is close to unity.
This is called a position permutation loop,
as opposed to the momentum permutation loops (\S\ref{SEC:MtmLoops})
in which all the bosons are in the same momentum state,
as in the ideal boson model.

Keeping only the identity permutation and the pair transposition,
which is the dimer loop,
the symmetrization function
behaves as an effective pair potential,
$v(q_{ij}) = -k_{\rm B}T \ln[ 1 + e^{-\pi q_{ij}^2 /\Lambda^2}] $.
This is attractive and increases the density
above what it would be in the absence of wave function symmetrization.
This gives a leading order correction to classical statistical mechanics
at high temperatures and low densities.

\begin{figure}[t!]
\centerline{ \resizebox{8cm}{!}{ \includegraphics*{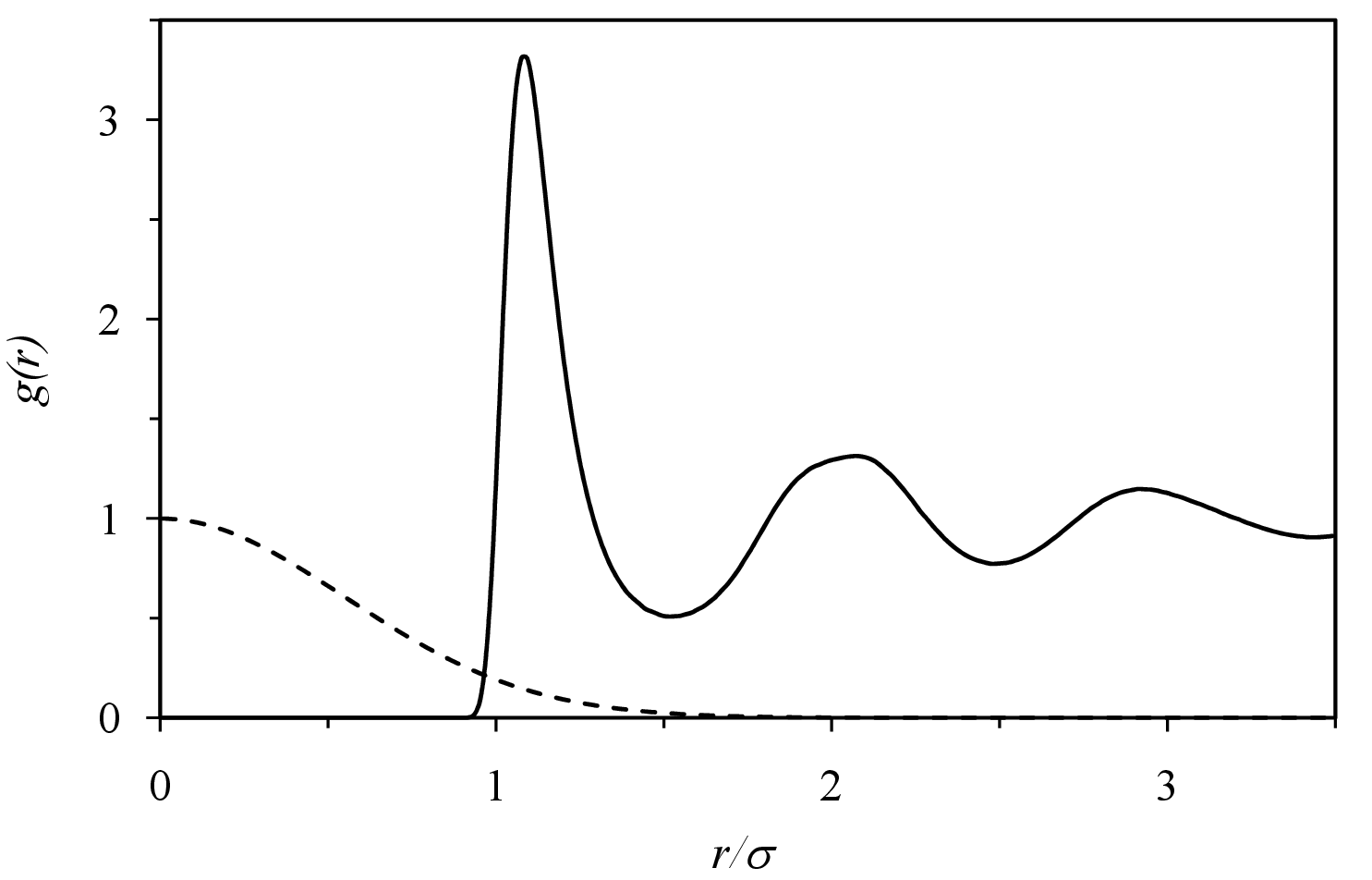} } }
\caption{
Radial distribution function (solid curve)
in saturated Lennard-Jones liquid
($k_{\rm B}T/\varepsilon=0.6$,
$\rho \sigma^3 = 0.8872$, $\Lambda/\sigma = 1.3787$).
The dashed curve is the Gaussian
$e^{-\pi r^2/\Lambda^2}$.
The Lennard-Jones pair potential is
$u(r) = 4 \varepsilon [ (\sigma/r)^{12}  - (\sigma/r)^{6} ]$.
}\label{Fig:g(r)}
\end{figure}

The number and size of position permutation loops
grow with decreasing temperature
as the thermal wavelength increases
and encompasses the first peak in the pair distribution function,
which is at about the diameter of the $^4$He atom
(Fig.~\ref{Fig:g(r)}).
For interacting particles the peak also grows with decreasing temperature.
This suggests that the $\lambda$-transition
is a sort of percolation transition
in which individual loops have grown to span the entire system:
any two bosons in the system belong to
at least one and the same permutation loop with weight close to unity.
Feynman (1953) 
suggested that the superfluid transition
is associated with macroscopic permutation loops
that span the entire system.

It turns out that the sum of the products of permutation loop factors
leads to a grand potential for each size loop (Attard 2018, 2025a \S3.1).
Thermodynamic derivatives of the loop grand potential
leads to a loop series for the energy, heat capacity, etc.
The series diverges approaching the $\lambda$-transition,
which explains the divergence in the heat capacity.

This result for position permutation loops
explains the link between Bose-Einstein condensation
and the structure of the $^4$He liquid.
The reason that the density of the saturated liquid peaks
at the $\lambda$-transition  (Donnelly and Barenghi 1998)
is that the position permutation loops provide an effective attraction
between the $^4$He atoms;
the closer together neighbors around a loop are,
the greater is the weight of the loop, and the more loops can form.
This obviously goes beyond the ideal boson model
in which the  momentum loops
are divorced from the structure of the ideal fluid.

\subsubsection{Momentum Permutation Loops} \label{SEC:MtmLoops}

Momentum loops involve permutations of bosons in the same momentum state,
$ {\bf p}_{j_k}  = {\bf p}_{j_{k+1}} $, $k=1,2,\ldots,l-1$.
This means that the loops are non-local
and have weight unity,
$\eta({\bf p}^l,{\bf q}^l)
=
e^{-{\bf q}_{j_l}\cdot{\bf p}_{j_l,j_{1}} /{\rm i}\hbar}
\prod_{k=1}^{l-1}
e^{-{\bf q}_{j_k}\cdot{\bf p}_{j_k,j_{k+1}} /{\rm i}\hbar}
= 1.$
The cyclic order in the loop is unimportant,
and the sum of all loops for a momentum state is just
the factorial of the occupancy of that state,
$\eta_{\bf a} = N_{\bf a}!$,
where $N_{\bf a}$ is the number of bosons
currently in the single-particle momentum state ${\bf a}$.
This is the only contribution from Bose-Einstein condensation
that is taken into account in the ideal boson model.

In principle one could take into account permutation loops
where the momentum difference between adjacent bosons was small but non-zero.
This is the analogue of the position permutation loops.
In practice there is little need for this
as by the time time momentum loops become dominant,
the occupancy of individual momentum states is sufficiently high
to consider only permutations between bosons in the same state.
In this case, one can write down directly
the total number of permutations for each state,
namely $N_{\bf a}!$,
without actually formulating and summing over the individual loops.

In general it is not a reasonable approximation
to consider permutations between all of the condensed bosons
as if they were in a single state,
which is what the ideal boson model with ground state condensation does.
However this does depend on the application and the accuracy desired.
If there are $N_0 = \sum_{\bf a}\!^{(a < a_0)} N_{\bf a}$ bosons
in low lying momentum states, $a < a_0$,
then the total occupation entropy due to permutations
only within each state is
$S^{\rm occ} /k_{\rm B}=  \sum_{\bf a}\!^{(a < a_0)} \ln N_{\bf a}!
\approx \sum_{\bf a}\!^{(a < a_0)} N_{\bf a} \ln N_{\bf a} - N_0
\approx N_0 \ln \overline N_{\bf a} - N_0$.
The difference between this
and the entropy due to the total number of permutations
as if they were all condensed in a single state is
$[S^{\rm occ}  - \tilde S^{{\rm occ}}] /k_{\rm B} \approx
\sum_{\bf a}\!^{(a < a_0)} N_{\bf a} \ln N_{\bf a} -  N_0 \ln N_0
=
\sum_{\bf a}\!^{(a < a_0)} N_{\bf a} \ln [ N_{\bf a}/N_0 ]
\approx N_0 \ln [\overline N_{\bf a}/N_0] $.
This is large and negative since in the condensed regime
there is a macroscopic number of condensed bosons
($\overline N_0 = {\cal O}(10^{26})$)
in  highly occupied low-lying momentum states
($\overline N_{\bf a} = {\cal O}(10^2)$).

On the one hand permutations between bosons in the same momentum state
are non-local since they do not depend upon where each boson is.
On the other hand, permutations between bosons in nearby momentum states
are only approximately non-local,
since consecutive bosons around each permutation loop
are restricted to smaller separations
as their momentum difference increases.
This means that the above expression for $\tilde S^{{\rm occ}}$
overestimates the total permutation entropy of the condensed bosons.

\subsubsection{The $\lambda$-Transition} \label{Sec:lambda}

The experimental fact that the heat capacity
and the density begins to decline
below the $\lambda$-transition  (Donnelly and Barenghi 1998)
suggests that the position permutation loops themselves
must begin to decline.
This implies that there is a competition
between condensation and position permutation loops,
with condensation and the associated  momentum loops
growing and dominating below the $\lambda$-transition.
This competition can be understood as follows.

Since each permutation is the product of loops,
an individual boson in an individual  permutation belongs to a single loop,
which is to say that the loops in a permutation are disjoint.
The dominant loops (ie.\ those with weight close to unity
after averaging over small changes in position, or momentum, or time)
are either position or momentum loops.
Of course for a given configuration
the weight of all the permutations must be summed.
But it is these individual loops that persist across multiple permutations
that dominate the sum.
A boson in a highly occupied momentum state tends to remain in that state
because of the occupation entropy of the state
(see the mechanism for superfluidity, \S\ref{Sec:SMSF}, below).
Conversely, a boson in a position loop tends to remain
in that structural arrangement
and to sample individually multiple momentum states over time
because of the favorable position loop permutation weight.
Because of these competing requirements on its momentum,
an individual boson in a configuration tends to belong mainly
to one type of permutation loop or the other, but not to both.

In these circumstances
one can understand why position permutation loops
are dominant above the $\lambda$-transition
and momentum  permutation loops below it.
Prior to the first sign of condensation as the temperature is lowered,
the thermal wavelength overlaps
the first peak in the pair distribution function
and position loops begin to form with close to unit weight.
At lower temperatures the number of accessible momentum states becomes
comparable to the number of bosons and momentum loops begin to form
as individual momentum states become highly occupied.
There is active competition between the two types of loops,
because a condensed boson (ie.\ one in a multiply occupied momentum state)
interferes with the formation of position permutation loops
in its neighborhood.
For this reason  position permutation loops
formed on the high-temperature side of the $\lambda$-transition
suppress the formation of momentum loops
and  Bose-Einstein condensation.
Conversely, when momentum loops finally emerge
on the low-temperature side of the $\lambda$-transition
they degrade the number and size of position permutation loops.

This rationalized picture is consistent
with the experimental and numerical evidence.
The measured growth and divergence of the heat capacity
on the high-temperature side of the $\lambda$-transition
is predicted by quantum Monte Carlo computer simulations
for the position loop series in Lennard-Jones $^4$He
(see next, and Attard (2025d)).
Those simulations show the necessity of suppressing condensation
in order to reproduce the sharp peak of the type observed experimentally.
This suppression explains why superfluidity is measured
to be discontinuous at the $\lambda$-transition.
The experimental evidence suggest that the number of condensed bosons
is macroscopic (it affects the heat capacity and causes superfluidity)
and it begins continuously from zero at the $\lambda$-transition
(the heat capacity begins to decline,
the viscosity is continuous,
and the effects of superfluidity are first observed).
The specific heat capacity of position permutation loops
is larger than that of momentum permutation loops
apparently because the bosons in them sample a greater range of momenta
and because neighbors in a loop are close to the pair potential minimum.

\subsubsection{Computer Simulation Results for the $\lambda$-Transition}
\label{Sec:QMCCPS}

Realistic computer simulations of liquid $^4$He
require an interaction potential,
of which the most common is the Lennard-Jones 6--12 pair potential
(Allen and Tildesley 1987).
A different pair potential has been used
in path integral Monte Carlo simulations of $^4$He (Ceperley 1995).
The present author's algorithm,
quantum Monte Carlo in classical phase space,
can be implemented in several related ways (Attard 2025a, 2025d).
The simplest is to carry out a canonical simulation with $N$ identical
$^4$He atoms,
and in the analysis phase to subdivide these into two `species',
with $N_0$ condensed and $N_*$ uncondensed bosons,
the total number being $N=N_0+N_*$ (Attard 2025d).
This binary division is akin to Einstein's (1924, 1925) approximation
of condensation into a single state.
All atoms interact identically with the difference
between the two species in the analysis
being that the condensed bosons participate
in momentum but not position permutation loops,
and the uncondensed bosons participate
in position but not momentum permutation loops.

It is a bit of a misnomer to call all atoms of the species 0 `condensed'
as such a boson can be the sole occupant of its momentum state.
But since at low temperatures the majority of such bosons
will be in multiply occupied momentum states,
the nomenclature is arguably justified.

At high temperatures there is no multiple occupancy of momentum states,
and there are no position permutation loops,
and so there is no distinction between the two species,
$\overline N_0 = \overline N_* = N/2$.
As the $\lambda$-transition is approached,
one or other is favored,
and the optimum number is determined by minimizing the free energy.

The constrained Helmholtz free energy is  (Attard 2025d Eq.~(A.10))
\begin{eqnarray} \label{Eq:Free1}
\lefteqn{
F(N_0|N,V,T)
}  \\
& = &
F_0^\mathrm{id}(N_0,V,T)
+ k_\mathrm{B}T
\ln [ N_*! \Lambda^{3N_*}V^{-N_*}]
\nonumber \\ && \mbox{ }
- k_\mathrm{B}T \ln [V^{-N} Q(N,V,T)]
- N k_\mathrm{B}T \sum_{l=2}^{l_\mathrm{max}}
f_{*}^l g^{(l)} .\nonumber
\end{eqnarray}
The momentum and the position contributions factorize for both species,
with the classical position configuration integral being
$Q(N,V,T) = \int {\rm d}{\bf q}^N e^{-\beta U({\bf q}^N)}$.
For the condensed bosons,
the momentum contribution is the quantum ideal expression,
$F_0^\mathrm{id}(N_0,V,T)
= \Omega_0^\mathrm{id}(z_0,V,T) + N_0 \ln z_0$,
with the ideal quantum grand potential being
$\Omega_0^{\rm id}(z_0,V,T)
= -k_{\rm B}T V \Lambda^{-3} g_{5/2}( z_0)$.
The corresponding ideal average number is
$\overline N_0^{\rm id}( z_0)
=  V \Lambda^{-3} g_{3/2}( z_0)$,
which is not equal to the constrained number, $N_0$.
The  fugacity is taken to be $z_0 = f_0 \rho \Lambda^3$,
with the fraction of condensed bosons being $f_0=N_0/N$,
and the number density being $\rho=N/V$.

The intensive loop Gaussian is
\begin{equation} \label{Eq:gl}
g^{(l)}
=
\frac{1}{N} \left\langle
\sum_{j_1,\ldots,j_l}^{N}\!\!'\;
\eta^{(l)}_*({\bf q}^l)
\right\rangle_{N,V,T}.
\end{equation}
In the free energy expression this is multiplied by $f_{*}^l$,
which is the uncorrelated probability that
all $l$-bosons in the loop are uncondensed.

The derivative at constant $N$ is
\begin{eqnarray}
\frac{\beta}{N} \frac{\partial F(N_0|N,V,T) }{\partial f_0}
& = &
\ln z_0
+ \left[ 1 - z_0^{-1} g_{3/2}(z_0)\right]
 \\ && \mbox{ }\nonumber
- \ln [f_* \rho \Lambda^{3} ]
+  \sum_{l=2}^{l_{\rm max}} l f_*^{l-1} g^{(l)} .
\end{eqnarray}
Setting the derivative to zero gives the optimum fraction of condensed bosons
for fixed density $N/V$.

\begin{figure}[t]
\centerline{ \resizebox{8cm}{!}{ \includegraphics*{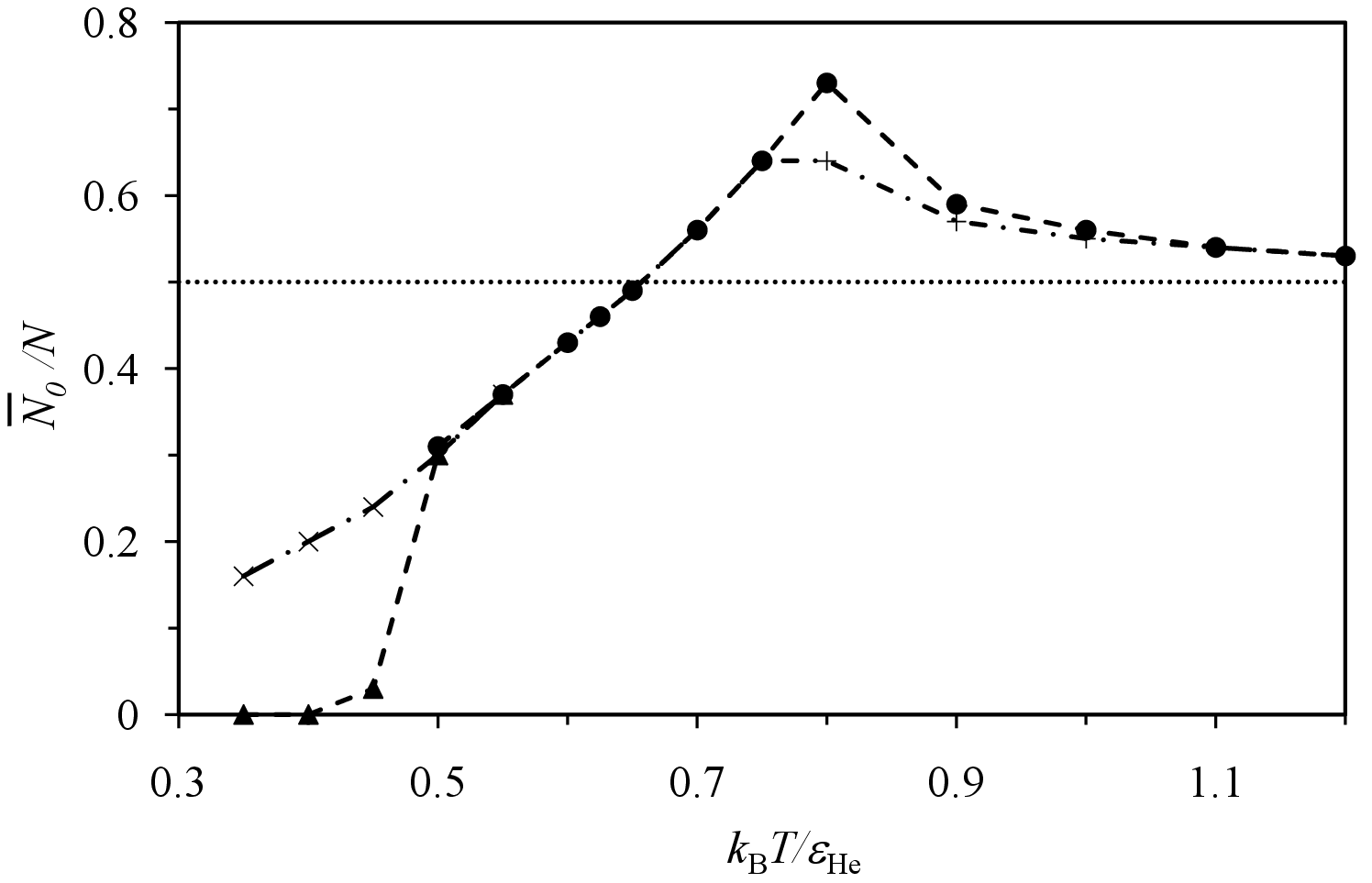} } }
\caption{\label{Fig:N0barA}
Most likely fraction of condensed Lennard-Jones $^4$He atoms.
The filled symbols include position loops only, Eq.~(\ref{Eq:Free1}),
and the algebraic symbols include position loops and chains,
Eq.~(\ref{Eq:Free3}).
The dotted line and the lines connecting the symbols are eye guides.
Note that $\varepsilon_{\rm He}/k_{\rm B} = 10.22$\,K.
}
\end{figure}

Quantum Monte Carlo simulations in classical phase space
were performed for a Lennard-Jones liquid.
The well-depth was $\varepsilon_{\rm He}/k_{\rm B} = 10.22$\,J
and the diameter was $\sigma_{\rm He} = 0.2556$\,nm
(van Sciver 2012).
The density was the simulated
classical saturated Lennard-Jones liquid density at each temperature.
This is a factor of 2--3 times the measured saturated liquid density
for $^4$He.
The number of atoms in the simulations was $N=5,000$.

There are a number of approximations
in the  simulations
such as the use of the Lennard-Jones pair potential,
and only the pair potential,
the use of the classical Lennard-Jones saturated liquid density,
the use of the constant volume heat capacity rather
than that on the line of saturation,
the neglect of the commutation function
(this is the primary reason for the too large density),
the limited number of terms in the loop series,
and the somewhat artificial definition of condensation.

Figure~\ref{Fig:N0barA} shows the simulation results
for the optimum fraction of condensed atoms.
It can be see that with decreasing temperature
the fraction rises slowly from 50\%.
At $T=$ 5.1\,K,
the pure position loops alone give
a sudden drop in the number of condensed bosons to zero.
(Mixed chains are discussed \S\ref{Sec:Chains}.)
This may be called the suppression transition
because below this temperature it is favorable to eliminate condensation
so that all atoms can participate in the position permutation loops.

The current algorithm does not perform reliably at temperatures
lower than those shown,
mainly because the loop series diverges.
For this reason
the (presumed) re-emergence of condensation
at the peak of the $\lambda$-transition
does not appear in the figure.

As mentioned, the nomenclature `condensed' bosons for the $N_0$ bosons
that do not participate in position permutation loops is a little misleading
as it includes bosons that are the sole occupant of their momentum state.
These are unaffected by Bose-Einstein condensation
or occupation entropy.
In fact, for the data in the figure the fraction of condensed bosons
in states occupied on average by more than one boson
does not rise beyond 20\% in the temperature range shown.
This fraction is zero at high temperatures,
and it drops suddenly to zero at the suppression transition.
The word `condensed' would be more accurate for this subset
as they are affected by the occupation entropy.
Obviously one could use a  higher threshold than
$\overline N_{\bf a} > 1$ to define these;
in the discussion of superfluidity
it is argued that bosons in more highly occupied states
are more highly superfluid.

\begin{figure}[t]
\centerline{ \resizebox{8cm}{!}{ \includegraphics*{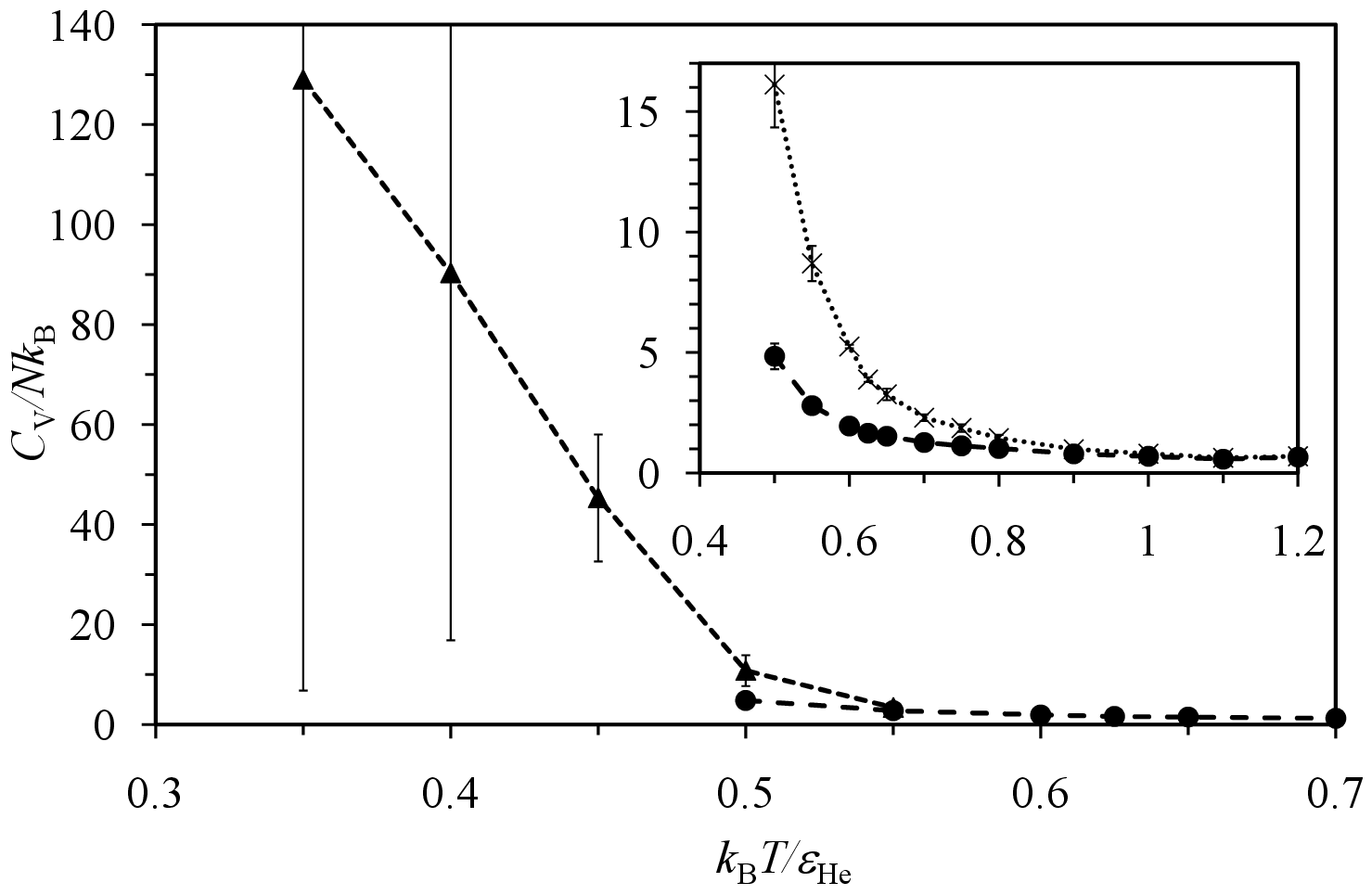} } }
\caption{\label{Fig:CvA}
The specific heat capacity of Lennard-Jones $^4$He
due to the classical contribution plus position loops only,
(cf.\ Eq.~(\ref{Eq:Free1})).
The circles are from a homogeneous liquid
and the triangles are from simulations of a droplet.
The filled symbols have the $\overline N_0$ condensed atoms
excluded from the position permutation loops,
whereas the crosses in the inset have all atoms
included in the position loops.
The lines connecting symbols are eye guides.
The error bars gives the 95\% confidence level.
}
\end{figure}

Figure~\ref{Fig:CvA}
shows the specific heat capacity from the simulations.
Results are shown for when all the bosons are allowed
to participate in the position permutation loops
($N_0=0$, crosses),
and for when the condensed bosons, if present, are excluded
($\overline N_0  \ne 0$, filled symbols).
In the former case the specific heat capacity begins to rise
for $T \alt 9$\,K,
whereas in the latter case there is no increase until the suppression
transition at $T=$ 5.1\,K.
The heat capacity appears to be diverging,
rising to more than 25 times its pre-suppression value
over an interval of 1.5\,K.
(Pre-suppression means that the fraction of condensed bosons is non-zero
and they are excluded from the position permutation loops.
Suppression means that there are no condensed bosons,
and that all bosons in the system can participate
in position permutation loops.)

The results qualitatively agree
with the measured $\lambda$-transition in $^4$He.
The divergence in the heat capacity
is due to the divergence of the position permutation loop series
and is  a definite improvement upon the ideal boson model.
The present simulations are problematic
at low temperatures where the intensive loop Gaussian $g^{(l)}$
increase with $l$.
The series was terminated at $l^{\rm max} = 5$.
This is one reason why the simulations  have not been pursued to
temperatures lower than those shown.
Presumably, it is also the reason why there is no peak
and subsequent decline in the heat capacity.

The Lennard-Jones $^4$He classical saturation liquid density
is about 2.7 times the measured density in actual $^4$He.
(This is primarily due to the neglect of the Wigner-Kirkwood (commutation)
function that is discussed next.)
This causes the parameter $\rho \Lambda^3$
to be overestimated, as well as the peak of the pair distribution function.
(The measured density in the 5,000 atom Lennard-Jones simulations
results in spinodal decomposition into separate liquid and vapor phases.)
This is why the heat capacity diverges
at a higher temperature 
than the measured $\lambda$-transition temperature. 

Path integral Monte Carlo simulations
give a $\lambda$-transition temperature in close agreement
with the measured value (Ceperley 1995),
in part due to using the measured saturation density,
made possible by the small system (64 atoms)
and the implicit Heisenberg uncertainty repulsion.
The path integral estimates of the fraction of condensed bosons
are much lower than the measured values,
because only ground momentum state bosons are counted,
as is discussed along with quantum molecular dynamics results
in \S\ref{Sec:QMD}.

\subsubsection{Wigner-Kirkwood Function} \label{Sec:WKfn}

The Wigner-Kirkwood function (Wigner 1932, Kirkwood 1933)
is also called the commutation function by the present author
(Attard 2017, 2018, 2021).
Since its neglect in the present results
appears to be the main reason for the overestimate
for the Lennard-Jones $^4$He saturation liquid density,
it is worth briefly discussing it
and the prospects for including it in the future.

The Wigner-Kirkwood function
$\omega = e^W$
is defined via
\begin{equation}
e^{-\beta{\cal H}({\bf p},{\bf q})} e^{W({\bf p},{\bf q})}
e^{-{\bf p}\cdot{\bf q}/{\rm i}\hbar}
=
e^{-\beta \hat{\cal H}({\bf q})}
e^{-{\bf p}\cdot{\bf q}/{\rm i}\hbar} .
\end{equation}
If the position and momentum operators commute
(or if there is no potential energy, $U({\bf q}) = 0$),
then $W=0$.
Thus this provides the extra phase space weight
due to their non-commutativity,
and it is the way that the Heisenberg uncertainty principle
is accounted for in the classical phase space formulation
of quantum statistical mechanics.
The effects of the  uncertainty principle are short-range,
and the zero point energy that results
can be attributed to an effective repulsion.

The most useful computational approach to date
takes the inverse temperature derivative
of both sides and develops a recursion relation
for the coefficients in a series expansion in powers of $\hbar$,
$W = \sum_{n=1}^\infty W_n \hbar^n$.
The recursion relation for $n \ge 2$ is (Attard 2017)
\begin{eqnarray}
\frac{\partial W_n}{\partial \beta }
& = &
\frac{\mathrm{i}}{m} {\bf p} \cdot \nabla W_{n-1}
 + \frac{1}{2m}
 \sum_{j=0}^{n-2}  \nabla W_{n-2-j}  \cdot \nabla W_j
\nonumber \\ && \mbox{ }
- \frac{\beta}{m} \nabla W_{n-2}  \cdot \nabla U
 + \frac{1}{2m} \nabla^2 W_{n-2} .
\end{eqnarray}
The first few coefficient functions are
\begin{equation} \label{Eq:w1}
W_1
=
\frac{-\mathrm{i}\beta^2}{2m} {\bf p} \cdot \nabla  U ,
\end{equation}
\begin{equation} \label{Eq:w2}
W_2
 =
\frac{\beta^3}{6m^2}
{\bf p} {\bf p} : \nabla \nabla  U
 + \frac{1}{2m}
\left\{ \rule{0cm}{0.4cm}
\frac{ \beta^3}{3}  \nabla  U \cdot \nabla  U
-\frac{ \beta^2}{2}  \nabla^2 U
\rule{0cm}{0.4cm}\right\} .
\end{equation}
The coefficients grow quickly in complexity;
the third and fourth are given in Attard (2025a \S 7.3.2).

\begin{figure}[t]
\centerline{ \resizebox{8cm}{!}{ \includegraphics*{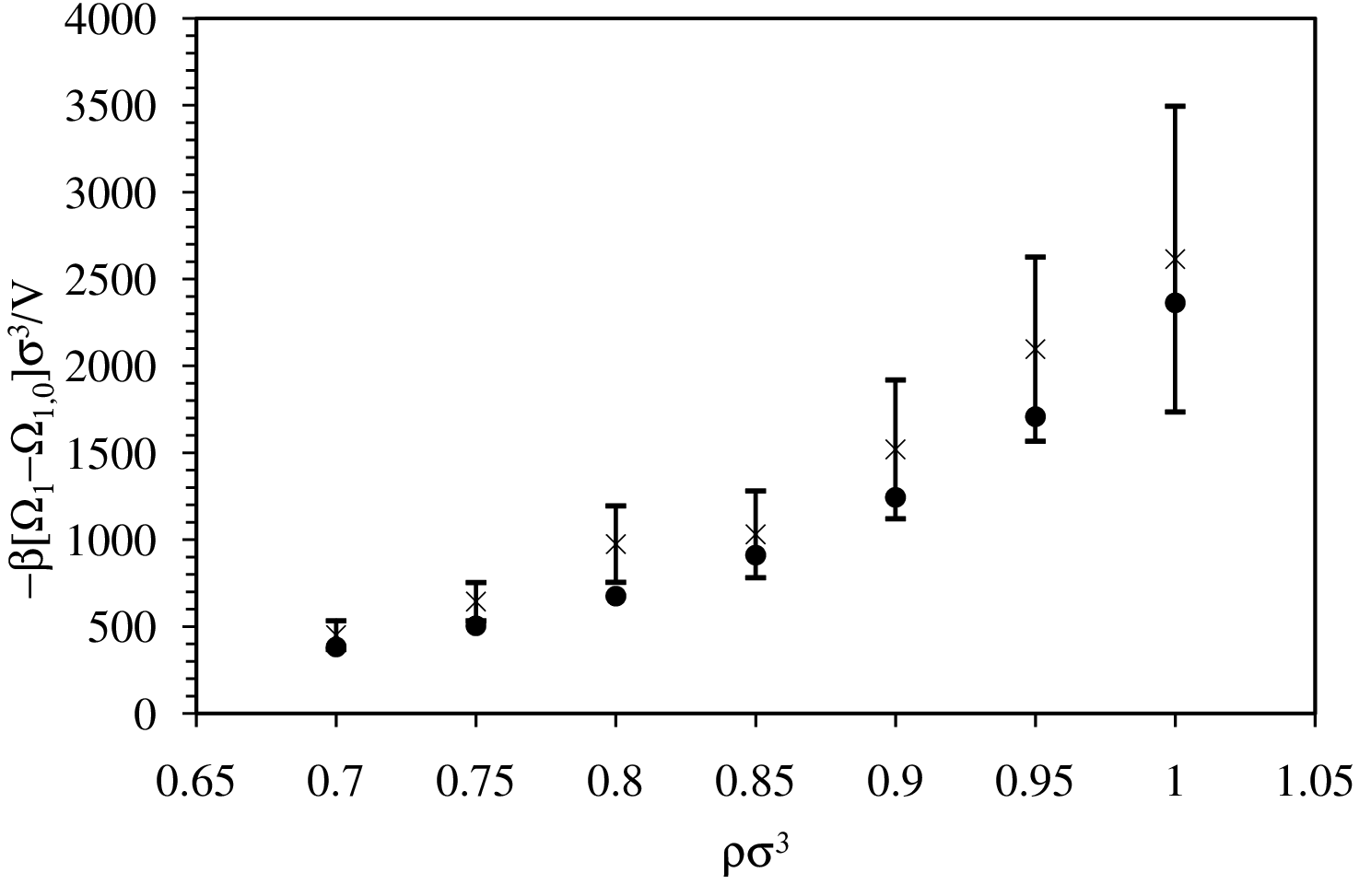} } }
\caption{\label{Fig:He100}
Wigner-Kirkwood quantum correction to the classical pressure
for Lennard-Jones $^4$He at $k_{\rm B}T/\varepsilon = 1.0$,
($\Lambda = 1.0679\sigma$)
using two different fourth order expansions.
The classical pressure at $\rho \sigma^3 =1 $ is $\beta p \sigma^3 = 7.5$.
The error bars gives the 68\% confidence level.
From Attard (2017 Fig.~6).
 }
\end{figure}

Monte Carlo simulation results for Lennard-Jones $^4$He
have been carried out (Attard 2017).
Typical results for the quantum correction
to the classical pressure are shown in Fig.~\ref{Fig:He100}.
These results do not include symmetrization effects,
which are relatively negligible at this temperature.
The agreement between the different  fourth order approximations
give confidence in the results.
It can be seen that the quantum effects due to non-commutativity
are substantial,
being several hundred times the classical pressure  for the same density.
This means that as  an effective potential,
the Wigner-Kirkwood exponent is strongly repulsive,
which accords with an intuitive understanding of
 the effects of the Heisenberg uncertainty principle
on the zero point energy.

Presumably,
with the Wigner-Kirkwood function included at the phase space level,
the liquid in equilibrium with its own vapor will have density
much closer to the experimentally measured value
than in the current simulations.
Of course the Lennard-Jones liquid is relatively incompressible,
as can be seen  in Fig.~\ref{Fig:He100},
and the required reduction in density of a factor of 2--3
will reduce the pressure by a much larger factor.

There are some issues involved in combining
the  Wigner-Kirkwood function and the symmetrization function
in a classical phase space simulation.
That both are complex functions means that the imaginary terms,
which are odd in momentum,
have to be handled in the Monte Carlo algorithm.
The earlier simulations (Attard 2017)
obtained the pressure using various classical phase space averages
of the Wigner-Kirkwood function in which the imaginary parts vanished.

\subsubsection{Position Permutation Chains} \label{Sec:Chains}

The incompatibility between momentum and position permutation loops
was discussed in \S~\ref{Sec:lambda}.
It was argued that this would lead to a suppression
of condensation on the high temperature side
of the $\lambda$-transition,
which appears to be consistent with the experimental data
and with the simulation data in Fig.~\ref{Fig:N0barA}.
Here we explore the consequences
of mixing condensed and uncondensed bosons
in the same permutation loop.

If, instead of being integrated over,
the momentum of a boson in a position permutation loop,
say the last, is close to zero,
${\bf p}_{j_l} \approx {\bf 0}$,
then the associated Fourier factor can be set to unity,
$e^{-{\bf p}_{j_l}\cdot{\bf q}_{j_l,j_{1}} /{\rm i}\hbar }\approx 1$,
and the loop becomes an open-ended chain
\begin{eqnarray}
\eta_{*0}({\bf p}^l,{\bf q}^l)
& = &
\prod_{k=1}^{l-1}
e^{-{\bf p}_{j_k}\cdot{\bf q}_{j_k,j_{k+1}} /{\rm i}\hbar}
\nonumber \\
\Rightarrow
\eta_{*0}({\bf q}^l)
& = &
\prod_{k=1}^{l-1} e^{-\pi q_{j_k,j_{k+1}}^2 /\Lambda^2} .
\end{eqnarray}
The momentum state of the final boson
has been fixed and not averaged over for such chains.
Because the distance between the first and last boson does not enter,
such chains are open-ended.
Therefore,
given the existence of the boson in the low-lying momentum state,
they are more readily formed than loops.

What opposes chains at higher temperatures
is the need for the final boson to be in the ground,
or close to the ground, momentum state.
Given the many accessible momentum states at higher temperatures,
the probability of this is low,
which reduces the weight attached to chains.
Since bosons in low-lying momentum states
are more likely at lower temperatures,
one might expect chains to come into existence
and to dominate at and below the $\lambda$-transition.

Chains are in a sense intermediate between position permutation loops
and momentum loops.
Each of the latter consists of bosons
in the same multiply occupied momentum state.
Low-lying states are more likely to be  multiply occupied
than higher momentum states.
One can therefore say that there is a correlation
between the formation of permutation chains
and the condensation into low-lying momentum states.

This qualifies the discussion in \S\ref{Sec:lambda}
that position permutation loops suppress
condensation because they are disrupted
by bosons in multiply occupied low-lying momentum states.
It would be more correct to say that position loops suppress condensation
until they can no longer do so.
At such a time position chains begin to form
each with a condensed boson at the head.

The number of possible loops grows exponentially with their size.
Chains tend to be shorter than loops as condensation proceeds
because there can be only one `head' boson with low momentum in each chain.
And so as the number of bosons
in multiply occupied  low-lying momentum states increases,
the average length of the chains decreases.
Also, as the number of these condensed bosons increases,
the probability of forming a long loop without them decreases.
This more or less
accounts for the suppression of condensation above the $\lambda$-transition,
and the nucleation of condensation below the transition.
We know that loops and chains exist below the transition
because the measured heat capacity is large
(and decreasing with decreasing temperature),
whereas the specific heat capacity of condensed bosons alone is small.

As for loops,
one can define intensive chain Gaussian
\begin{equation}
\tilde g^{(l)}
=
\frac{1}{N} \left\langle
\sum_{j_1,\ldots,j_l}^{N}\!\!'\;
\eta_{*0}({\bf q}^l)
\right\rangle_{N,V,T}.
\end{equation}
With this the contribution from mixed chains
to be added to the free energy Eq.~(\ref{Eq:Free1}) is
(Attard 2025d Eq.~(A.10))
\begin{equation} \label{Eq:Free3}
F_{0*}(N_0,N_*,V,T) =
- N k_{\rm B}T
\sum_{l=2}^{l_{\rm max}} f_0 f_*^{l-1} \tilde g^{(l)} .
\end{equation}

The results in Fig.~\ref{Fig:N0barA}
show that including mixed chains does not significantly change the results
from those obtained using  pure position loops alone.
Perhaps the most noticeable difference
is that condensation is partially but not entirely suppressed
at the lowest temperatures shown.
One should be cautious about the conclusions drawn
from these free energy expressions because of their approximate nature,
the simplicity of the Lennard-Jones model,
and the limited number of terms that are used
in the loop and chain series.

It seems that the main point to be drawn from
this analysis of mixed chains
is not so much quantitative as conceptual.
As an intermediary between pure position loops composed
of uncondensed bosons,
and pure momentum loops of condensed bosons
in multiply occupied low-lying momentum states,
mixed chains provide a mechanism for the rise of the latter
and the decline of the former
on the far side of the $\lambda$-transition.

%
\section{Superfluidity}
\setcounter{equation}{0} \setcounter{subsubsection}{0}
\renewcommand{\theequation}{\arabic{section}.\arabic{equation}}
%

The macroscopic and microscopic treatments of superfluidity
complement each other.
The macroscopic approach,
based on thermodynamics and hydrodynamics,
is empirical as it derives from the quantitative description
of fountain pressure measurements.
This gives the fundamental thermodynamic principle
that determines superfluid flow,
and it leads to the so-called two-fluid model
that successfully predicts second sound, amongst other things.
The microscopic approach uses statistical mechanics
and computer simulations.
It explains the physical and mathematical basis
for the thermodynamic results,
it yields molecular equations of motion
that explain how superfluid flow occurs without viscosity,
and it provides the basis for a computer simulation algorithm
for obtaining the viscosity of $^4$He below the $\lambda$-transition.


\subsection{Thermodynamics of Superfluidity}

\subsubsection{Fountain Pressure} \label{Sec:Fountain}

\begin{figure}[t]
\centerline{ \resizebox{8cm}{!}{ \includegraphics*{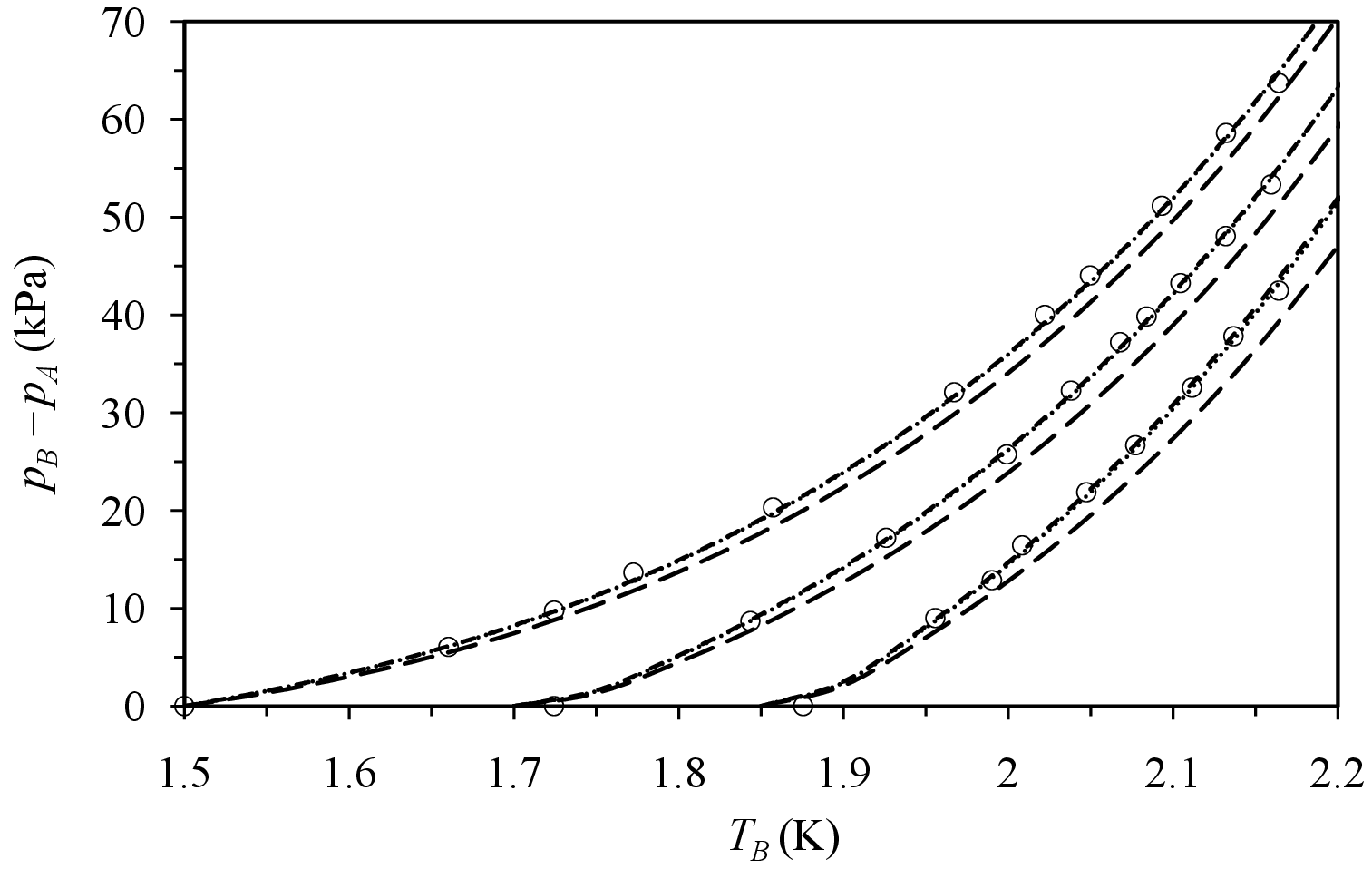} } }
\caption{\label{Fig:fountain} Measured and calculated fountain
pressure for $T_A =$ 1.502\,K (left), 1.724\,K  (middle), and
1.875\,K (right). The symbols are measured data (Hammel and Keller
1961), the short dashed curve is
the H. London (1939) expression, Eq.~(\ref{Eq:HL-ftn}),
the coincident dotted curve is for fixed chemical potential,
Eq.~(\ref{Eq:PA-mu}),
and the long dashed curve is for fixed fugacity, Eq.~(\ref{Eq:PA-mu/T}).
The calculated curves use measured thermodynamic data
(Donnelly and Barenghi 1998),
as corrected by Attard (2025a \S4.4.2).
From Attard (2022b, 2025a Fig.~4.4).}
\end{figure}

The fountain pressure refers to two chambers of liquid $^4$He
held at different temperatures below the $\lambda$-transition
and connected by a thin capillary or frit through
which superfluid flows.
The high temperature chamber attains a higher pressure
than the low temperature chamber,
which is often,  but not always, at  saturation.
If the high temperature chamber has a small opening in it,
then, due to the high pressure,
liquid eponymously spurts out.
If the high temperature chamber is closed, then a steady state develops,
with superfluid flowing 
from the low to the high temperature chamber through the capillary,
and normal viscous $^4$He flowing in the opposite direction
in the same capillary, driven by the pressure gradient.
It is in this latter arrangement
that the pressure difference is measured
as a function of the temperature difference (Fig.~\ref{Fig:fountain}).

H. London (1939) gave an expression
for the rate of change of the pressure with temperature
of the high temperature chamber B
at fixed temperature of the low temperature chamber A,
\begin{equation} \label{Eq:HL-ftn}
\frac{{\rm d}p_B}{{\rm d}T_B}
= \sigma_B   ,
\end{equation}
where $\sigma$ is the entropy density.
This expression, or its integrated version,
fits measured data with extraordinary accuracy (Fig.~\ref{Fig:fountain}).

Despite this,
the derivation of this expression given by H. London (1939) is faulty
(Attard 2025a \S4.3.3).
In his derivation H. London (1939) assumed the same properties
of superfluid $^4$He
as F. London (1938) in his ideal boson model for the $\lambda$-transition,
and as Tisza (1938) in his two-fluid model of superfluid hydrodynamics,
namely that the condensed bosons were in the ground energy state.
As mentioned in the introduction,
this erroneous picture of Bose-Einstein condensation
is ultimately due to Einstein (1924, 1925).
As a result of this assumption,
H. London (1939) asserted that the superfluid bosons
had zero energy, zero entropy, zero enthalpy, and zero chemical potential.
Tisza (1947 p.~845), perhaps unwisely,
claims `The assumption that the superfluid component
has the entropy zero [preprint, 17th~p.]\ has been
first advanced by the author (Tisza 1938)'.
Combining these with an artificial heat engine
and a dubious thermodynamic argument
that neglects the work done on the chambers in changing their entropy,
H. London (1939) purported to derive Eq.~(\ref{Eq:HL-ftn}).
In the opinion of the present author,
what likely happened was that H. London (1939) noticed
that  Eq.~(\ref{Eq:HL-ftn}) fitted the existing experimental data
(the left hand side has the units of Boltzmann's constant per unit volume,
and the right hand side is the simplest thermodynamic quantity
with these units),
and he cooked up a derivation that appeared to give this result.
The contradiction between the derivation
and the fountain pressure equation itself
will now be demonstrated by showing
that the chemical potential of the superfluid bosons
cannot be zero.

A thermodynamically equivalent expression
for the fountain pressure system
is that in the steady state
the chemical potentials of the two chambers are equal
\begin{equation} \label{Eq:PA-mu}
\mu_A = \mu_B .
\end{equation}
Writing this as the Gibbs free energy per particle for each chamber
and differentiating with respect to $T_B$ at constant $T_A$
gives
Eq.~(\ref{Eq:HL-ftn})
(Attard 2025a Eq.~(4.10)).
This is confirmed by the results in Fig.~\ref{Fig:fountain},
where the same pressure is given by these two equations.

This result proves the fallacy of the purported
derivation of H. London (1939),
specifically that the condensed bosons are in the ground state
and that they therefore have zero chemical potential.
For the most common case that the low temperature chamber
is at saturation, the measured chemical potential is strictly less than zero,
$\mu_A = \mu^{\rm sat}(T_A) < 0$
(Attard 2025a \S4.4.2, Donnelly and Barenghi 1998).
Since the condensed and uncondensed bosons are in equilibrium
(ie.\ they can swap identities),
the chemical potential of the superfluid $^4$He
must be the same as that of ordinary  $^4$He,
which is that of the system as a whole.
In other words, the fountain pressure expression
of H. London (1939) contradicts the properties
that he assumed in his published derivation of it.


It is impossible to overstate just how extraordinary
Eq.~(\ref{Eq:PA-mu}) is.
It represents a unique  characteristic of superfluid flow.
On first glance many might think it rather obvious
and only to be expected;
after all, equality of chemical potential is the standard condition
of equilibrium thermodynamics for particle exchange.
The point is, of course,
that the fountain pressure arrangement is a steady state system,
not an equilibrium system,
and the temperatures of the two chambers are not equal.
If one were to attempt to apply the equilibrium concept of maximizing
the total entropy with respect to particle exchange,
since $-\mu/T$ is the number derivative of the entropy
(Attard 2002 Table~3.1),
in the present case one would obtain
(Attard 2025a \S4.3.2)
\begin{equation} \label{Eq:PA-mu/T}
\frac{\mu_A}{T_A} = \frac{\mu_B}{T_B} .
\end{equation}
Since the fugacity is $z = e^{\mu/k_{\rm B}T}$,
this is the same as equating the fugacity of the two chambers, $z_A=z_B$.
The results in Fig.~\ref{Fig:fountain}
clearly show that this equilibrium result does not hold
for the fountain pressure arrangement.
Therefore, it cannot determine superfluid flow.

\subsubsection{Thermodynamic Principle of Superfluid Flow} \label{Sec:TDSF}

The general laws of physics
are usually formulated as variational principles.
The Second Law of Equilibrium Thermodynamics
---that entropy is maximized---
is the most well-known example.
The variational principle
for superfluid flow is now deduced
from the empirical equation for the fountain pressure.

The chemical potential is the number derivative of the energy
at constant entropy
(Attard 2002 Table~3.1),
\begin{equation}
\frac{\partial E(S,V,N)}{\partial N} = \mu .
\end{equation}
From this one can reasonably extrapolate
that the fountain pressure equation (\ref{Eq:PA-mu})
---that the chemical potentials of the two chambers are equal---
is equivalent to the principle that
\emph{superfluid flow minimizes the energy at constant entropy}.

The physical interpretation is this.
In general in a thermodynamic system
the average energy involves a statistical component,
the heat energy.
This expression says that the chemical potential
is the mechanical part of the energy per particle;
it is the change in energy with number at constant entropy.
This says that superfluid flow does not change the entropy of the subsystem.

Note for future reference that this is not the same as saying
that condensed bosons have zero entropy,
as Tisza (1938) asserted.
In fact, as will be discussed in detail below,
the origin of this principle is the exact opposite:
because condensed bosons have so much entropy
any change afforded by their flow would be to a lower value.
Fundamentally this is why they move at constant entropy.

Also note that this law that superfluid flow minimizes the energy
at constant entropy implies that condensed bosons carry energy.
In other words, condensed bosons do not have zero energy
and therefore they are not confined to the energy ground state.

One can in fact reconcile this principle
for superfluid steady state flow
with the Second Law of Equilibrium Thermodynamics
as follows.
In the fountain pressure arrangement,
we can regard the two chambers as being connected to
individual heat reservoirs at their respective temperatures.
The superfluid flow of $^4$He does not dissipate momentum
since it is inviscid,
which leaves the occupancies of the momentum states unchanged.
Hence the occupancy entropy is conserved
in the superfluid transfer
of a condensed boson from $A \Rightarrow B$.
The only change in entropy is due to the transfer of energy
between the chambers and their respective heat reservoirs.
Hence minimizing the total subsystem energy,
$E(S_A,V_A,N_A)+E(S_B,V_B,N_B)$,
maximizes the energy of the heat reservoirs, and hence their entropy.
(The total energy of the subsystems $A$ and $B$
and their reservoirs is fixed. The entropy of the reservoirs
is a monotonic increasing function of their energy.)
Because the subsystems' entropy is unchanged,
this increases the total entropy of the universe.
We conclude that the principle of subsystem energy minimization
at constant subsystem entropy for steady superfluid flow
is just a form of the Second Law of Equilibrium Thermodynamics.

Finally,
the chemical potential is the \emph{actual} change in energy
with the change in the number of the condensed bosons.
Its gradient will now be used to give the actual rate of change
of their momentum density, Eq.~(\ref{Eq:dotp0}).
In contrast,
due to entanglement with the environment,
only part of the gradient of the intermolecular potential energy,
(ie.\ a fraction of the mechanical force)
gives the rate of change of momentum of a condensed boson,
Eq.~(\ref{Eq:pnap}).

\subsubsection{Two-Fluid Model of Superfluid Flow} \label{Sec:TwoFluid}

The above thermodynamic principle for superfluid flow
applies to the steady state where the flux is constant.
The generalization to transient behavior
follows from the physical interpretation
that superfluid flow responds to the mechanical part of the energy,
namely the chemical potential.
Newton's second law of motion says that the rate of change of momentum
equals the force, which is the negative gradient of the energy.
The obvious generalization of this for superfluid flow
is to use the gradient of the chemical potential,
which leads to
\begin{equation} \label{Eq:dotp0}
\frac{\partial {\bf p}_0}{\partial t}
=
- \rho_0 \nabla \mu - \nabla \cdot ({\bf p}_0 {\bf v}_0 ).
\end{equation}
The second term on the right hand side is the convective rate of change,
which is just divergence of the momentum flux
(Attard 2012, de Groot and Mazur 1984).
Here  the momentum density of condensed (ie.\ superfluid) $^4$He
is ${\bf p}_0 = m \rho_0 {\bf v}_0$,
where $m$ is the atomic mass, $\rho_0$ is the number density,
and ${\bf v}_0$ the average velocity.
These and similar quantities are functions of position ${\bf r}$
and time $t$, and, like all hydrodynamics variables,
they are averaged over macroscopic volumes that are small on the scale
of the variations of the flows.

It should be mentioned that this expression neglects
the rate of change of momentum due to the `chemical' reaction
by which condensed and uncondensed bosons are interchanged,
$\dot \rho^{\rm react}_0 =  -\dot \rho^{\rm react}_*$
(Attard 2025e Eq.~(2.18)).
The time scales for this are on the order of tens of minutes
(Walmsley and Lane 1958).
This is important for the distinction
between steady and transient rotational motion (\S \ref{Sec:rot}).

Using the Gibbs-Duhem equation, the conservation laws, and linearizing,
it may be shown (Attard 2025e \S II)
that this yields the two-fluid model for superfluidity
\begin{equation} \label{Eq:TwoFluid0}
m \rho_0 \frac {\partial {\bf v}_0}{\partial t}
=
\frac {-\rho_0}{\rho} \nabla p + \frac{\rho_0}{\rho} \sigma \nabla T ,
\end{equation}
and
\begin{equation} \label{Eq:TwoFluid*}
m \rho_* \frac {\partial {\bf v}_*}{\partial t}
=
\frac {-\rho_*}{\rho} \nabla p - \frac{\rho_0}{\rho}  \sigma \nabla T
+ \eta \nabla^2 {\bf v}_* .
\end{equation}
(These equations appear in the literature
with myriad typographic errors.)
These were originally given by Tisza (1938),
who argued that $^4$He below the $\lambda$-transition
could be considered a mixture of two fluids:
the superfluid, subscript 0, which has no viscosity,
and the normal fluid, subscript $*$,
also known as helium~I, which has shear viscosity $\eta$.
In these $\rho=\rho_0+\rho_*$ is the total number density,
and $p$ is the pressure.

An early success of the two-fluid model
was the prediction of second sound
(Landau 1941, Tisza 1938).
This is essentially an entropy-temperature wave
that is unique to superfluidity (Donnelly 2009).

The two-fluid  equations give
(Attard 2025e Eq.~(2.23))
\begin{equation} \label{Eq:dot-sigma}
\frac{\partial \sigma}{\partial t}
=
- \nabla \cdot(\sigma {\bf v}_*).
\end{equation}
This gives the rate of change of the entropy  density
as the divergence of the  $^4$He entropy flux
due to the flow of normal fluid solely.
This is consistent with the fountain pressure principle
that the superfluid flow is at constant entropy.
To again  be clear, this result does not say
that superfluid $^4$He has no entropy.

\subsubsection{Non-Local Momentum Correlations and Plug Flow}
\label{Sec:super=nonlocal}

The occupation entropy for condensed bosons
does not depend upon their positions in space.
The permutations are composed of loops
of bosons all in the same momentum state,
and the corresponding Fourier factors in the symmetrization function
are unity independent of the positions of the bosons that comprise the loop.
Hence a pure momentum loop is not localized in space. Since the
$\lambda$-transition marks the change in dominance from position to
momentum permutation loops, and since the
former are localized in space, it is clear that non-localization
plays a fundamental r\^ole in superfluidity.

The consequences of non-locality become clearer
when viewed in the light of the physical origin of shear viscosity.
In classical shear flow,
the  momentum flux is inhomogeneous,
which non-uniformity is dissipated by molecular collisions.
The stratified fluid flow has longitudinal momentum transfer
between adjacent layers,
slowing the quick and accelerating the tardy.
The ultimate driver of this momentum dissipation
is the increase in the subsystem entropy,
since the order represented
by smooth spatial variations in momentum flux is a state of low
configurational entropy
(Attard 2012a \S9.6). 

In shear flow in a classical fluid,
the momentum correlations must be spatially localized.
(This is also true for $^4$He above the $\lambda$-transition
where position permutation loops dominate.)
A simple example is Poiseuille flow,
which is laminar flow in a pipe due to a pressure gradient.
In this the spatial correlations in momentum are manifest
as zero flow at the walls,
and a continuous increase in flow velocity toward the center.

For superfluid flow,
the non-local permutation entropy of bosons in the same momentum state
induces momentum correlations  without regard to  spatial position.
Such non-local momentum correlations
are inconsistent with shear flow;
the large momentum state in the center of the channel
induces the same state in the condensed bosons near the walls.
If the momentum correlations  are non-local,
then the momentum field must be spatially homogeneous.
In this case the only non-zero flow can be plug flow
in which the momentum state of the bosons is uniform across the channel.
Plug flow is the classical solution for inviscid hydrodynamic flow
down a channel.

\subsubsection{Critical Velocity} \label{Sec:Vcrit}

An upper limit is observed  for the velocity
of superfluid flow in a pore, capillary, or thin film,
which critical velocity increases
with decreasing pore diameter or film thickness.
One school of thought,
which the present author deprecates,
says that the critical velocity
enables the production and growth of excitations
 that  destroy the superfluid.
These excitations are said to be the rotons postulated by Landau (1941),
and they are pictured as vortex rings
(Feynman 1954,  Kawatra and Pathria 1966).
In contrast, the present analysis concludes that whilst superfluid flow
is destroyed at the critical velocity, Bose-Einstein condensation isn't.
It is shown that the phenomenon may be accounted
for by standard quasi-classical statistical mechanical analysis.

Consider a thin cylindrical capillary 
through which superfluid flows with velocity $v_z$,
and suppose that the most likely momentum state for the condensed bosons is
$\overline {\bf a} =  m v_z \hat{\bf z} $.
Suppose that the superfluid flow occurs
in a macroscopic number $M_A$ of momentum states
in the neighborhood $A$ about this value,
so that the total number of $^4$He involved in the superfluid flow is
$\overline N = \sum_{{\bf a} \in A} N_{\bf a}
\approx M_A \overline N_{\overline {\bf a}}$.
Because of the aspect ratio of the capillary,
the spacing between $z$-momentum states is much less
than that between the radial and angular momentum states,
and so we can take the ground states of the latter to be
the ones occupied.

The occupation entropy for these superfluid bosons is
\begin{equation}
S^{\rm occ}_{\overline{\bf a}}
= k_{\rm B} \sum_{{\bf a} \in A} \ln N_{\bf a}!
\approx
k_{\rm B} M_A  \ln \overline N_{\overline{\bf a}} !
\end{equation}
We suppose that beyond the critical velocity,
the condensed bosons are 
instead distributed
about the zeroth longitudinal momentum state,
with average $\langle a_z \rangle =0$.
The radial and angular momentum excited states may also be occupied.
The occupation entropy is largely unchanged
\begin{equation}
S^{\rm occ}_{0}
\approx
k_{\rm B} M_A  \ln \overline N_{1,0,0} !
\approx
S^{\rm occ}_{\overline{\bf a}} .
\end{equation}
Because the critical velocity is orders of magnitude smaller
than the thermal speed,
the occupancies of low-lying momentum states are more or less the same
before and after the critical velocity
(apart from which states are occupied).
This result reflects the idea
that condensation is determined by the occupancy
of individual momentum states
rather than by macroscopic flows.
To put it another way,
contrary to Einstein (1924, 1925),
condensation is not into a single quantum state.

For a particle in a cylinder of diameter $D=2R$,
we assume that the radial momentum states are
$p_r = 2\pi n \hbar/D$, $n=0,\pm 1,\pm2 ,\ldots$.
These are the same as for a rectangular channel or film of thickness $D$,
and in all cases the momentum eigenfunction for the
component orthogonal to the boundary is real on the boundary.
The relevant changes in energy eigenvalues are the square of these.


If all the superfluid bosons get knocked out of the flow
and into the first radial state
(and random low-lying longitudinal states, $\langle a_z \rangle =0$),
then the change in reservoir entropy is
\begin{equation}
\Delta S^{\rm r}
=
M_A  \overline N_{\overline{\bf a}}
\left\{\frac{-1}{2mT} \frac{(2\pi\hbar)^2}{D^2}
+
 \frac{ m v_z^2 }{2T} \right\} .
\end{equation}
Obviously this increases with increasing flow rate,
which makes it favorable to occupy the transverse momentum states
above the critical velocity.
Since the occupation entropy does not change,
the critical velocity is the one that makes this zero,
or
\begin{equation}
\frac{ m v_{\rm c}^2 }{2k_{\rm B}T}
=
\frac{1}{2m k_{\rm B}T} \frac{(2\pi\hbar)^2}{D^2} .
\end{equation}
Since the flow velocity is very much smaller than the thermal speed,
the left hand side of this is $ v_{\rm c}^2/2v_{\rm th}^2
\approx  10^{-6}$.
This is about six orders of magnitude smaller
than the occupation entropy per boson prior to the critical velocity,
$S^{\rm occ}_{\overline{\bf a}} /M_A \overline N_{\overline{\bf a}}k_{\rm B}
=  \ln \overline N_{\overline{\bf a}} - 1 $,
since 
$\overline N_{\overline{\bf a}} = {\cal O}(10^2)$.
This confirms that the occupation entropy remains relatively constant
through the critical velocity.
Condensed bosons remain condensed
whether or not they participate in macroscopic superfluid flow.

\begin{figure}[t!]
\centerline{ \resizebox{8cm}{!}{\includegraphics*{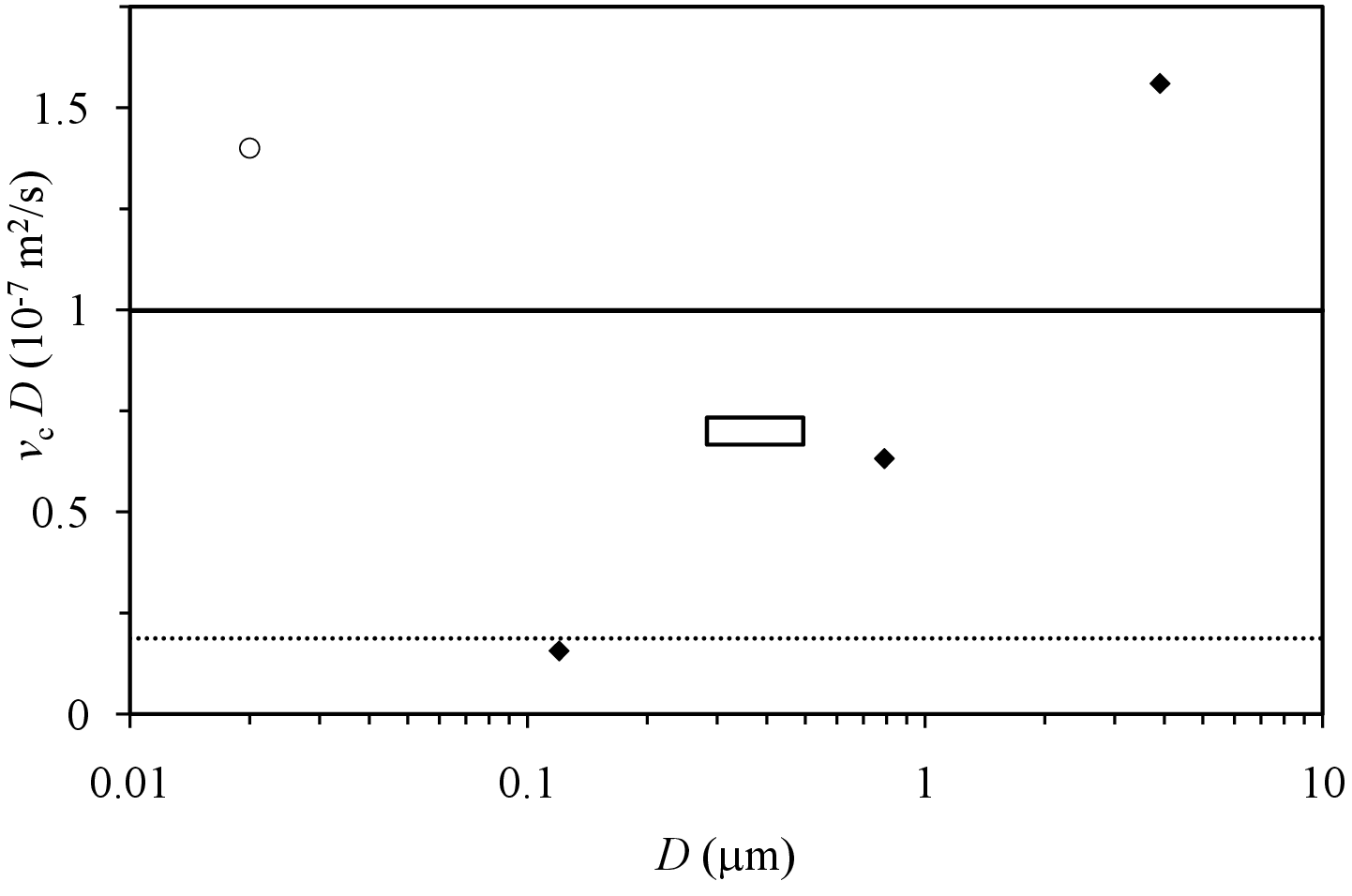} } }
\caption{
Superfluid critical velocity times the capillary diameter
or film thickness $D$.
The filled diamonds (Pathria 1972 \S 10.8)
and open circle (Allum \emph{et al.}\ 1977) are measured values
for a cylindrical capillary.
The rectangle covers measured values for planar films
with a range of thicknesses
(Ahlers 1969,  Clow and Reppy 1967).
The present prediction for cylinders and films,
$v_\mathrm{c} = 2 \pi \hbar/mD$, is shown by the solid line.
The vortex prediction for cylinders (Kawatra and Pathria 1966),
$v_\mathrm{c} = 1.18 \hbar/mD$, is shown by the dotted line.
}\label{Fig:vc}
\end{figure}


The critical velocity given by this,
$ v_{\rm c} = {2\pi\hbar}/{m  D}$,
is plotted in Fig.~\ref{Fig:vc},
along with measured data
and other predictions.
The most striking feature of the experimental data is that
the measured critical velocity varies inversely with the diameter
over several orders of magnitude.
Since generally the spacing between momentum states
can be expected to be inversely proportional to the pore diameter,
this provides strong qualitative confirmation of the present theory.
The results for planar films are consistent with this.
The present theory is obviously highly simplified
but it nevertheless could be described as quantitative:
over three orders of magnitude in diameter
the predicted result differs from the measured results
by less than a factor of four.

Landau gave a stability criterion for superfluid flow,
which originally predicted $v_{\rm c} \approx 60$\,m/s.
This is several orders of magnitude larger
than the measured values (Balibar 2017, Batrouni \emph{et al.}\ 2004).
Feynman (1954) suggested that Landau's (1941) rotons were in fact
quantized vortices.
Assuming that the excitation of such vortices
destroyed superfluid flow, Kawatra and Pathria (1966) calculated the
velocity of their onset, and their prediction is shown in the
figure.
One problem with the roton/vortex idea is that helium~II,
the mixture of normal and superfluid helium,
necessarily has rotons already in it,
which contradicts the axiomatic basis of the theory
that these rotons do not emerge in the
capillary until the critical velocity is achieved,
and when they are present superfluidity is destroyed.
Setting aside this problem with the vortex interpretation,
it can be seen that the present theory lies closer to
the experimental data in Fig.~\ref{Fig:vc}
than does the roton/vortex theory.
Detailed criticisms of the assumed r\^ole of vortices in superfluidity
are given in the following section.

The present theory for the critical velocity
is related to the principle of superfluid flow
derived from fountain pressure measurements
---energy is minimized at constant entropy---
via the conclusion that the occupation entropy does not change
through the critical velocity transition.
The reason for this is not that condensed bosons have no entropy,
but rather that the occupation entropy
is so much larger
than the ordinary thermodynamic entropy associated with the flow.

\subsection{Steady Rotational Motion}

\subsubsection{Background} \label{Sec:rot}

Rotational motion in helium~II,
specifically the damping of a torsional pendulum,
was the original method for measuring
the fraction of condensed bosons and the viscosity
(Andronikashvilli 1946).
Oscillatory and steady rotation are qualitatively different
for the superfluid.
As mentioned, the two-fluid model Eq.~(\ref{Eq:dotp0})
neglects the rate of change of momentum due to the chemical reaction
by which condensed and uncondensed bosons are interchanged
over tens of minutes (Walmsley and Lane 1958).
This is what creates the distinction
between steady and transient rotational motion.
The reason for the difference is that
the classical forces acting on condensed bosons are much reduced
(see \S\ref{Sec:SMSF})
and therefore they decouple from the transiently rotating system
in which the interchange rate is negligible.
This is consistent with
the path integral Monte Carlo simulations of $^4$He
by Ceperley (1995 \S IIIE),
who obtain a reduced moment of inertia below the $\lambda$-transition.
The moment of inertia plays no r\^ole in steady rotation.
In steady rotation chemical interchange between
condensed and uncondensed $^4$He is non-negligible.
Steady and transient superfluid flow are not comparable,
and it is steady rotation that is the sole focus here.

Tisza (1947 p.~854)  says
`According to Landau, the superfluid state is characterized
by the condition curl~${\bf v}_{\rm s} = {\bf 0}$.
The question has been further discussed by F. London (1945)
and Onsager (\emph{pers.\ com.})'.
Landau's assertion,
backed by F. London and Onsager, that superfluid flow
is irrotational 
(ie.\ zero vorticity),
has since been taken as a fundamental principle.
The notion is apparently motivated by two observations:
First,  inviscid classical fluids are irrotational,
and superfluids certainly lack viscosity.
Second,
a superconducting current is irrotational,
as shown by
the second London equation for superconductivity
(F and H London 1935),
and one expects superfluidity and superconductivity to be closely related.

We shall critically analyze the various theoretical justifications
that have been offered for irrotational flow below
(see \S\S \ref{Sec:nonVortex},
\ref{Sec:Landau}, \ref{Sec:QuantCirc}, and \ref{Sec:MacroWave}).
Here we point out the danger of relying upon analogy.
For example,
inviscid classical fluids are an idealization
for  gas-like densities,
whereas helium~II is a liquid.
Or the London equation for superconducting currents
is a direct consequence of Maxwell's equations,
and it is curly logic to assert that
these apply to superfluidity in helium~II,
which has no electromagnetic effects to speak of.

The issue of whether or not superfluid flow is irrotational
(ie.\ has zero vorticity)
is important for several reasons.
If true,
then the existence of a velocity potential restricts and simplifies
the allowed solutions to the two-fluid model hydrodynamic equations.
Another reason is that it gives insight into the behavior and understanding
of superfluidity at the molecular level.
In particular Landau's (1941) theory for the origin of superfluidity,
for which he was awarded the Nobel prize (Physics 1962),
introduces rotons
as a type of rotational first excited state
whose occupancy is taken to be inconsistent with superfluidity.
But if rotational superfluid flow is allowed,
then it is pointless to say that rotons destroy superfluidity.
Onsager (1949) (Nobel Laureate in Chemistry 1968)
suggested that
`Vortices in a suprafluid are presumably quantized',
and discussed Landau's condition $\nabla \times {\bf v}_{\rm s} = {\bf 0}$
(Tisza 1947 p.~854).
Feynman (1955) (Nobel Laureate in Physics 1965)
gave a quantized circulation theorem and
`was the first to suggest that the formation of vortices in liquid helium II
might provide the mechanism responsible
for the breakdown of superfluidity in the liquid'
(Pathria 1972 \S10.8).
This idea has been further developed,
successively improving agreement with measured data
(Fetter 1963, Kawatra and Pathria 1966, Pathria 1972 \S10.8).
(Landau's (1941) original roton formulation
overestimated  the critical velocity
by several orders of magnitude (see \S\ref{Sec:Vcrit}).)
Again, the idea that vortices destroy superfluidity
and give the critical superfluid velocity
is predicated on the assumption that superfluid flow must be irrotational.
Finally, Landau's (1941) macroscopic wavefunction,
which underpins much of the current theory
of superfluidity and superconductivity,
predicts irrotational flow.
Hence if the measured superfluid flow is rotational,
then the macroscopic wavefunction and the dependent theory must be incorrect.

The real puzzle is why the idea of irrotational superfluid flow
has persisted to this day
when in fact the experimental evidence is of rotational superfluid flow.
This is the simplest and most direct interpretation of the data,
as we shall now show.
It requires that the evidence be twisted
and that a fictional model be spun
in order to maintain that irrotational flow
is consistent with the measured data.
This is a second example
in the field of Bose-Einstein condensation and  superfluidity
where an authority figure has been exempt from the criticism
that would be expected in normal scientific debate.

\subsubsection{Steady Rotation}

Osborne (1950) measured the steady rotation of a bucket of helium~II
and established that the free surface of the liquid
had the classic  parabolic shape,
\begin{equation} \label{Eq:Zsurf}
z_{\rm surf}({\bf r}) = z_0 + \frac{\omega^2 r^2}{2g},
\end{equation}
where $\omega$ is the angular velocity,
$r$ is the radius from the $z$ axis,
and $g$ is the acceleration due to gravity.
This implies that the whole liquid is rotating rigidly.
If Landau's (1941) irrotational idea held,
only the normal liquid can rotate
and the quadratic term should be scaled by
the fraction of normal liquid, $\rho_*/\rho$.

It is straightforward to show (Attard 2025e \S IV)
that in the steady state
the two-fluid equations (\ref{Eq:TwoFluid0}) and (\ref{Eq:TwoFluid*})
give exactly this result with the whole fluid rotating
with equal velocities for the two components,
${\bf v}_0({\bf r}) = {\bf v}_*({\bf r}) = \omega r \hat{\bm \theta}$.
Evidently the superfluid is rotational,
\begin{equation}
\nabla \times {\bf v}_0({\bf r})
=
2 \omega \hat{\bf z} ,
\end{equation}
which  contradicts Landau's principle.
(Incidently, this is the same velocity field
given by Pathria (1972 prior to Eq.~(9)),
who nevertheless maintains that the superfluid is irrotational.)

To confirm this result
we can appeal to its physical plausibility
and to its consistency with the thermodynamic principle of superfluid flow.
The explicit form for the condensed boson velocity field means that
the chemical potential has gradient
\begin{equation}
\nabla \mu
=
- m  {\bf v}_0 \cdot  \nabla  {\bf v}_0
= -m \omega^2 [ x \hat{\bf x} + y\hat{\bf y}],
\end{equation}
or $\mu({\bf r}) = \mu({\bf 0}) - m\omega^2 r^2/2$.
(The vertical pressure gradient cancels with that of
the gravitational potential.)
This lateral gradient represents a centripetal force
that cancels the centrifugal force by changing the momenta
of the condensed bosons toward the central axis.
This centripetal force has the same magnitude and direction
as the force exerted on the normal fluid by the lateral pressure gradient.
The difference is that it is the chemical potential
that provides the driving force for the superfluid.
The thermodynamic principle of superfluid flow
is that it minimizes the energy at constant entropy, \S \ref{Sec:TDSF}.
This means that it is the gradient of the chemical potential
that is the statistical force experienced by condensed bosons.
Hence the present result for the centripetal force
is entirely consistent with this principle.

\subsubsection{Irrotational Vortices}
\label{Sec:nonVortex}

\begin{figure}[t]
\centerline{ \resizebox{8cm}{!}{ \includegraphics*{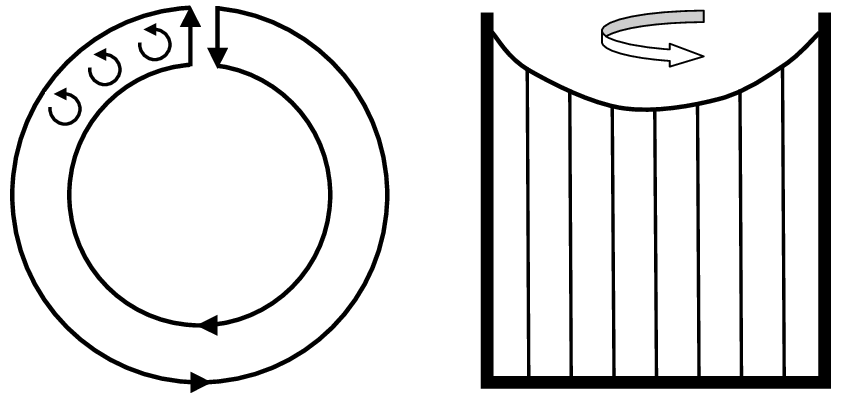} } }
\caption{\label{Fig:rot}
Model for rotating superfluid
as a macroscopic number of microscopic irrotational vortices.
After Pathria (1972 Fig.~10.11) and Annett (2004 Fig.~2.10).
}
\end{figure}

One attempt to reconcile
Landau's principle of irrotational flow
with the measured rotational flow just discussed
has become broadly accepted.
This model consists of a uniform distribution
of microscopic irrotational vortices
(Landau and Lifshitz 1955, Lifshitz and Kagenov 1955, Lane 1962),
and is sketched in Fig.~\ref{Fig:rot}.
The vortices, which have zero vorticity,
are said to reconcile the predicted irrotational superfluid flow
with the measured rigid rotation just discussed
(Annett 2004 \S 2.5, Pathria 1972  \S 10.7).

Occam's razor,
which says that a simple explanation
is to be preferred over a complicated one,
should cast doubt on this model.
The experiments are just as they appear and should be taken at face value:
the steady rotation of helium~II is rigid,
and it is unnecessarily complicated to invoke
a macroscopic number of microscopic, invisible vorteces
to explain the measured behavior.

But for the sake of the argument let us follow the published analysis anyway.
The so-called vorticity-free vortex,
in cylindrical coordinates ${\bf r} = \{r,\theta,z\}$,
is
\begin{equation}
{\bf v}({\bf r}) =
\left\{
\begin{array}{ll}
\displaystyle
\frac{K}{2\pi r}   \hat{\bm \theta}, & r > r_0 ,\\
\displaystyle \rule{0cm}{0.6cm}
\frac{K}{2\pi r_0} \hat{\bm \theta}, & r \le r_0 .
\end{array} \right.
\end{equation}
In general fluid dynamics,
such irrotational vorteces have the property
that the angular momentum is constant at each point
beyond the core region.
Such a vortex spontaneously arises when fluid flows toward a sink
conserving its angular momentum (eg.\ hurricanes, tornadoes).
The origin of microscopic sinks in superfluid flow is unclear.

In this expression the small radius cut-off $r_0$ has been invoked
to prevent the velocity diverging to infinity on the axis.
Continuity determines the initial value
of the hydrodynamic velocity field in this core region.
The curl of this velocity field is
\begin{equation}
\nabla \times {\bf v}({\bf r})
=
\frac{1}{r} \frac{\partial (r v_\theta)}{\partial r} \hat{\bf z}
=
\left\{
\begin{array}{ll}
\displaystyle
{\bf 0}, & r > r_0 ,\\
\displaystyle \rule{0cm}{0.6cm}
\frac{1}{r} \frac{K}{2\pi r_0} \hat{\bf z}, & r \le r_0 .
\end{array} \right.
\end{equation}
In the region where this vanishes, which is most of the range,
the flow is indeed irrotational.
But in the core region the curl is non-zero,
indeed divergent,
and the flow is definitely rotational.
To the present author it seems
a misnomer to call this a vortex with zero vorticity.

The circulation around a disc of radius $S \ge r_0$ enclosing the axis
of the so-called irrotational vortex is
\begin{eqnarray}
\oint {\rm d}{\bf l} \cdot {\bf v}({\bf r})
& = &
\int_{\rm S} {\rm d} {\bf S}  \cdot [\nabla \times {\bf v}({\bf r})]
\nonumber \\ & = &
2\pi \int_0^{r_0} {\rm d} r\, r
\frac{1}{r} \frac{K}{2\pi r_0}
\nonumber \\ & = &
K.
\end{eqnarray}
Thus the circulation 
is independent of the radius of any disc beyond the core
around which it is measured.

For the present case of a uniformly rotating bucket of helium~II,
a uniform distribution of irrotational vorteces,
as in Fig.~\ref{Fig:rot},
has been invoked
(Annett 2004 \S2.5, Lane 1962, Pathria 1972 \S 10.7).
Using the property of fixed circulation,
the conclusion is
that the overall superfluid velocity field
is that of rigid rotation,
${\bf v}_0({\bf r}) = \omega r \hat{\bm \theta}$,
where $r$ is measured from the axis
of the rotating system
(Annett 2004 Eq.~(2.46), Pathria 1972 Eqs~(10.7.7) and (10.7.9)).
But there are several issues with the analysis.
First, it avoids any discussion of the core region of the vortices,
where the flow is certainly rotational,
which contradicts the assertion that superfluid flow must be irrotational.
Second, it does not offer an explanation for the origin
of the macroscopic number of microscopic vorteces.
And third,
the model ultimately yields a rigidly rotating superfluid velocity field.
Surely this is in fact a simple and direct proof
that superfluid flow can be rotational.

Two further points about vorteces, irrotational or otherwise, can be made.
Abo-Shaeer \emph{et al.}\ (2001)
observed vortex lattices induced in spinning Bose-Einstein condensates.
These were transient with lifetimes from seconds to tens of seconds,
and so they are not directly relevant
to the present discussion of the steady state.
In any case, the fact that the condensate itself
(ie.\ almost all of the condensed bosons) is rotating
tends to support the present contention
that superfluid flow can be rotational.

In type II superconductors, the magnetic field
partially penetrates the sample
in the form of discrete flux quanta
(Annett 2004, Tinkham 2004).
Each flux quantum is believed to be surrounded by a vortex
of supercurrent of the above irrotational type.
Of course the existence of irrotational vorteces
of itself does not prove that a fluid \emph{must} be irrotational.
In any case, as pointed out above,
argument for superfluid flow by analogy with supercurrent behavior
is  unreliable.


\subsubsection{Landau's Irritational Principle}
\label{Sec:Landau}

Landau (1941) developed
his principle of irrotational superfluid flow
based on the fundamental belief
that the superfluid state is the ground energy state.
Landau followed
F. London (1938), Tisza (1938), and H. London (1939)
in this assumption
(without ever accepting Bose-Einstein condensation).
In any case,
based on a quantum formulation of hydrodynamics (see \S\ref{Sec:QuantCirc}),
Landau (1941)  purports to prove that in general
a uniform irrotational flow has lowest energy,
with an energy gap to rotational flow.
He concludes:
`The supposition that the normal level of potential motions
[ie.\ $\nabla \times {\bf v}_0 = {\bf 0}$]
lies lower [in energy]
than the beginning of the spectrum of vortex motions
[ie.\ $\nabla \times {\bf v}_0 \ne {\bf 0}$]
leads to the phenomenon of superfluidity' (Landau 1941 p.~356).

However, even if uniform irrotational flow
has the lowest energy with a gap to rotational flow,
the more numerous rotational states
are likely occupied for entropic reasons,
particularly given that the measurements are so far from absolute zero.
Further, since a macroscopic amount of $^4$He is involved in superfluid flow,
at these temperatures on the order of 50\% of the total number,
these cannot all be in the ground energy state.
Landau's (1941) assertion
that superfluid flow must be irrotational everywhere
is based on the unsound supposition that superfluid helium
must be in the ground energy state.

In addition, for the non-equilibrium case of forced flow,
such as the driven rotation of a bucket of helium~II,
analysis based on the equilibrium occupation of energy states
is largely irrelevant.
Landau's (1941 p.~357) conclusion
`when the walls of the vessel are in motion,
only a part of the mass of liquid helium is carried along by them,
and the other part remains stationary' has not aged well
(see the experimentally measured result, Eq.~(\ref{Eq:Zsurf})).

More generally, the present author deprecates Landau's (1941)
formualtion of quantum hydrodynamics
based on macroscopic quantum wave functions
(see \S\ref{Sec:MacroWave}).
The present author judges this to be fundamentally incompatible
with the principles of non-equilibrium thermodynamics (Attard 2012)
and with the principles of quantum statistical mechanics (Attard 2021).

Notably, Landau (1941) rejected F. London's (1938) theory
that the $\lambda$-transition and superfluidity
was a consequence of Bose-Einstein condensation:
`Tisza's' [ie.\ F. London's (1938)] well-known attempt to consider
helium II as a degenerate Bose gas cannot
be accepted as satisfactory' (Landau 1941 p.~356).
Amusingly, even today researchers continue to embrace both theories
despite their  mutual contradiction.
Bogulbov (1947) attempted to reconcile the two theories;
he, along with Landau, Tisza, F London and H London assumed
that superfluidity was due to $^4$He being in the energy ground state.

It was mentioned above that even if irrotational flow
was the flow with lowest energy,
this would not imply that superfluid flow had to be irrotational.
This logical inconsistency underlying Landau's (1941) principle
appears to have been recognized by Feynman (1954 p.~276),
 who says
`The third [question] is to describe states for which the superfluid velocity
is not vortex free. 
\ldots
A new element must presumably be added to our picture
[of vortex free motion]'.
Thus Feynman implies that
irrotational motion describes only some superfluid flow,
and that rotational flows are allowed.

\subsubsection{Quantized Circulation} \label{Sec:QuantCirc}

Hydrodynamics is defined over macroscopic volumes
that are large on the molecular scale
but small on the scale of variations in the fluxes and thermodynamic fields.
The superfluid velocity
${\bf v}_{{\rm s}}({\bf r}) \equiv {\bf v}_0({\bf r})$
is a hydrodynamic flux,
which is to say that it is the result of averaging the velocities
of a macroscopic number of condensed  $^4$He atoms
in a volume about ${\bf r}$.

Onsager (1949) and Feynman (1955)
argued that, like angular momentum,
superfluid  vortex motion is quantized.
Onsager's (1949 p.~281) derivation,
in its entirety, and leaving nothing out, is
`Vortices in a suprafluid are presumably quantized;
the quantum of circulation
is $h/m$, where $m$ is the mass of a single molecule.'
Onsager provided nothing further to support this supposition.

Feynman's (1955) circulation theorem
has been derived
starting from a perturbation on the ground state wavefunction
due to a uniform velocity field
(Pathria 1972 Eq.~(10.6.1)),
\begin{equation}
\Psi({\bf r}^N) =
\Psi_0^+({\bf r}^N) e^{i m \sum_i {\bf v}_{\rm s} \cdot {\bf r}_i/\hbar} .
\end{equation}
A related version is given by Feynman (1954 Eq.~(5)).
Pathria (1972 \S10.6) says that if the velocity field is non-uniform
then this ansatz `would still be good locally'.
However, in this case the symmetrization requirements for bosons
would be violated by the ansatz:
the exponent changes if the momenta of two bosons
in different parts of the system are swapped
if the velocity field is non-uniform.
The perturbation is symmetric if, and only if, the bosons
are all in the same momentum state,
which can only occur if  velocity field is uniform.

Pathria (1972 Eq.~(10.6.3)) considers a possibly macroscopic ring of bosons,
and the change when each is shifted by $\Delta {\bf r}_{k_i}$.
When this shift is onto its neighbor,
$\Delta {\bf r}_{k_i} = {\bf r}_{k_{i+1}} - {\bf r}_{k_{i}}$,
the ring is unchanged and
the change in phase must be an integer multiple of $2\pi$
for non-uniform velocity (Pathria 1972 Eq.~(10.6.3)),
but it is exactly zero  for uniform velocity,
\begin{equation}
\Delta \phi
=
\left\{ \begin{array}{ll}
\displaystyle
\frac{m}{\hbar}
\sum_i {\bf v}_{\rm s}({\bf r}_{k_i}) \cdot
[{\bf r}_{k_{i+1}} - {\bf r}_{k_{i}}]
= 2 \pi n , \\ 
\displaystyle
\frac{m}{\hbar} {\bf v}_{{\rm s}}\cdot
\sum_i [{\bf r}_{k_{i+1}} - {\bf r}_{k_{i}}]
= 0 .
\end{array} \right.
\end{equation}
Thus for a non-uniform velocity the ansatz says that
the circulation is quantized,
but the perturbing wave function is invalid.
Conversely, for uniform velocity
the perturbing wave function is valid, but the circulation is zero.
Unfortunately, a uniform velocity field
may well have zero circulation and zero curl,
but it also has zero interest.

There are two things wrong with
Feynman's (1955) circulation theorem
as derived by Pathria (1972 \S10.6).
First, it assumes that the superfluid is in the ground energy state,
with a small perturbing wavefunction for the flow.
Second, for non-uniform flow
the ansatz for the perturbing wave function is not fully symmetric.

If, for the sake of the argument,
we accept quantized circulation,
then we have
(Pathria 1972 Eq.~(10.6.7))
\begin{equation}
\oint {\rm d}{\bf l} \cdot {\bf v}_{\rm s}({\bf r})
=
\int_{\rm S} {\rm d} {\bf S}  \cdot (\nabla \times {\bf v}_{\rm s}({\bf r}))
= \frac{2\pi \hbar n}{m} .
\end{equation}
This says that the curl of the velocity field
through the area of integration has quantum number $n$,
where $n$ is meant to be an integer.
Pathria (1972 \S10.6) argues that
if the area of integration is shrunk continuously,
then the right hand side would change discontinuously unless $n=0$ always.
This implies that
$\nabla \times {\bf v}_{\rm s}({\bf r}) = {\bf 0}$
for all radii.
Thus Pathria says that quantization proves that superfluid flow
must be irrotational.

It is not clear to the present author
why the circulation around a macroscopic region
cannot change discontinuously,
particularly if $n$ is macroscopic.
Perhaps one might argue that as a hydrodynamic velocity field
$ {\bf v}_{\rm s}({\bf r})$ must be continuous.
But since this
is the macroscopically averaged hydrodynamic velocity field,
the quantum number $n$ (assuming that it exists)
must also be an average,
which means that it belongs to the continuum.
(For example, the average number of atoms
in any mathematical subvolume of a liquid is not an integer.)
This means that the right hand side can change continuously
as the area of integration is changed.
Hence there is no reason to insist that $n=0$,
and there is no proof on the basis of quantization that
$\nabla \times {\bf v}_{\rm s}({\bf r}) = {\bf 0}$.

Incidently, if one accepts Pathria's argument,
then it proves that the quantum number for superfluid circulation
must always be zero.
The same conclusion is reached from
the analysis of Annett (2004 \S2.5):
applying Stokes' theorem to the expression for the circulation
(Annett 2004 Eq.~(2.37))
with the  irrotational flow he assumes from the macroscopic wavefunction
(see next),
proves that the quantum number must be zero (Annett 2004 Eq.~(2.41)).
This is difficult to reconcile with Annett's claims
that flow quantization has been measured in helium~II
with both a non-zero quantum number
and the condensate rotating (Annett 2004 \S2.5).

\subsubsection{Macroscopic Wavefunction} \label{Sec:MacroWave}

Landau's (1941)
formulation of quantum hydrodynamics
is based on the  macroscopic wavefunction,
$\psi_0({\bf r})$,
in three-dimensional position space.
This is said to obey the operator relationships
of a normal quantum wavefunction (eg.\ Annett 2004 \S2.4).
Identifying the modulus squared with the condensed boson density
(Annett 2004 \S2.3, Pathria 1972, \S10.5),
gives superfluid flow
proportional to the gradient of a velocity potential,
which means that it is irrotational.
It is important to address the reliability of the macroscopic wavefunction
not only because it underpins current theories
for the irrotational nature of superfluidity,
but it is also applied more generally
to superfluidity and to superconductivity.

There are at least five objections to the macroscopic wavefunction approach.

First,
it is a sort of ideal gas approximation
in which the wave function of the whole system
is factorized as the product of identical single-particle wave functions,
$\psi({\bf r}_1,{\bf r}_2,\ldots,{\bf r}_N)
= \prod_{j=1}^N \psi_0({\bf r}_j)$,
(Annett 2004 Eq.~(2.61), Pathria 1972 \S 10.5).
This is presented as a type of mean field approximation.
However, it is a dubious approach to condensed matter
as it  ignores molecular structure and correlations between the particles
due to their interactions
(see point three).
It is particularly doubtful in the case that the wavefunction
is taken to be the energy eigenfunction (Pathria 1972 \S 10.5),
because for interacting particles this does not
factorize into single-particle energy eigenfunctions.

Second,
it applies only to the ground state.
This is because the factorized product
has to consist of identical energy eigenfunctions,
namely the macroscopic wavefunction,
and the ground state is the only state in which the particles
are in the same single-particle state
(Pathria 1972 \S 10.5).
The present author has presented extensive evidence
that Bose-Einstein condensation is not solely into the ground energy state
(see \S\S \ref{Sec:BEC} and \ref{Sec:Ideal},
and also Eq.~(\ref{Eq:PA-mu}) above,
as well as Attard (2025a \S\S1.1.4 and 2.5)).
In short,
it strains credulity to assert that the superfluid
and superconductor transitions,
which occur at temperatures far above absolute zero,
are dominated solely by particles in the ground energy  state.

Third,
identifying the square of the amplitude of the macroscopic wave function,
$|\psi_0({\bf r})|^2$,
with the condensed boson number density, $\rho_0({\bf r})$, is unrealistic.
It is not immediately obvious what
the single-particle mean-field ground-state energy eigenfunction
has to do with density.
Perhaps one might argue that,
apart from normalization,
both the Born probability,  $|\psi_0({\bf r})|^2$,
and the single-particle density, $\rho_0({\bf r})$,
give the probability of finding a particle
at ${\bf r}$ irrespective of the other particles.
But the macroscopic wave function formulation
implies that the $n$-particle density is
the product of single-particle densities,
$\rho^{(n)}_0({\bf r}_1,{\bf r}_2,\ldots,{\bf r}_n)
= \prod_{j=1}^n \rho_0({\bf r}_j)$,
which is a very poor approximation for a condensed liquid
because it neglects correlations and attractions between the particles,
and it allows particles to overlap.

Fourth,
if it exists, then the evolution of the macroscopic wave function
would be governed by the Schr\"odinger equation,
or its non-linear mean-field version
(Ginzburg and Pitaevskii 1958, Gross 1958, 1960),
as has been discussed (Annett 2004, Pathria 1972).
This implies that quantum mechanics also governs its flux,
as in
(Annett 2004 Eq.~(2.21), Tinkham 2004 Eq.~(4.14))
\begin{equation} \label{Eq:J0}
{\bf J}_0({\bf r})
 =  \frac{-{\rm i}\hbar}{2m}
[ \psi_0({\bf r})^*\, \nabla \psi_0({\bf r})
- \psi_0({\bf r}) \nabla \psi_0({\bf r})^*\,].
\end{equation}
This approach,
with the condensed boson density replacing the macroscopic wave function,
is said to form the basis for quantum hydrodynamics.
But quantum mechanics is fundamentally incompatible with hydrodynamics:
the former applies to a few particles isolated
from their surroundings,
whereas the latter deals with a macroscopic numbers of particles
contained in local volumes that are molecularly large
but thermodynamically small
(cf.\ the two-fluid model for flow in helium~II, \S\ref{Sec:TwoFluid}).
The hydrodynamic flux on the left hand side of this equation
has nothing to do with the single-particle quantum dynamics
on the right hand side.
Rather it should be the average 
over a macroscopic number of non-identical, interacting wavefunctions.
Indeed, quantum statistical mechanics
shows that 
the Schr\"odinger equation has to be modified
to account for condensation and environmental-induced decoherence
(see \S \ref{Sec:SMSF} and Attard (2025b)).

And fifth,
the macroscopic wave function
has been given different microscopic interpretations:
is it the gap parameter in BCS theory
(Gor'kov 1959, Tinkham 2004 \S1.5)?
Or is it the condensed boson density
(Annett 2004 \S2.3, Pathria 1972 \S10.5, Tinkham 2004 \S1.5)?
Or perhaps it is the single-particle mean-field ground-state energy
eigenfunction (Pathria 1972 \S10.5)?
Or even the expectation value of the field annihilation operator
in an unspecified wave-state
(or perhaps a single macroscopically-occupied momentum state)
(Annett 2004 Eqs~(5.67) and (5.72))?
The plethora of differing explanations
suggests that in fact it has no convincing basis in reality.
Perhaps these interpretations are not mutually exclusive,
but it is important to have a precise definition and understanding
in order to deduce the properties and dynamics
of the  macroscopic wave function,
and to assess its feasibility.
It is not at all clear
how the resultant  one-particle density operator
or one-particle density
can realistically describe the properties of $N$ interacting bosons
in a dense liquid.

To be clear, the present author
does not object to using the condensed boson density
$\rho_0({\bf r})$ as the order parameter
in Landau's (1937) phenomenological theory of second order phase transitions.
The objection is to the macroscopic wave function $\psi_0({\bf r})$
and to its use in the quantum flux equation
to predict superfluid or superconductor flow.
Of course if $\rho_0({\bf r})$ is used directly as the order parameter,
then there is no recourse to the  Schr\"odinger equation
or the quantum flux equation.
In this case the order parameter approach \emph{per se} provides no basis
for the dynamics of superfluid flow or for supercurrents.

There are several predictions of the macroscopic wave function
that directly contradict measured data.

First,
using the macroscopic wavefunction,
$\psi_0({\bf r}) = \sqrt{ \rho_0({\bf r})} \, e^{{\rm i}\theta({\bf r})}$,
in the quantum mechanical   expression for the flux,
Eq.~(\ref{Eq:J0}),
gives
(Annett 2004 Eq.~(2.21))
\begin{eqnarray}
{\bf J}_0({\bf r})
& = &
 \frac{\hbar}{m} \rho_0({\bf r}) \nabla \theta({\bf r}).
\end{eqnarray}
This says that the local superfluid velocity is  proportional
to the gradient of the phase of the macroscopic wave function,
${\bf v}_0({\bf r}) = (\hbar/m) \nabla \theta({\bf r})$.
Since the curl of the gradient of a scalar vanishes,
this would make superfluid flow irrotational,
$\nabla \times {\bf v}_0({\bf r}) = {\bf 0}$.
But, as discussed in detail above,
measurement shows
that superfluid flow has non-zero rotation
(Osborne 1950, Walmsley and Lane 1958).

Second,
in the case of superconductivity
it predicts a temperature scaling
$\rho_{20}(T) \sim  [1-T/T_{\rm c}]$ (Tinkham 2004 Eq.~(4.6)).
But the experimentally measured temperature dependence
of the London penetration length is
$ \lambda(T) \sim \lambda(0) [1-(T/T_{\rm c})^4]^{-1/2}$
(Tinkham 2004 Eq.~(1.7)),
which implies that $\rho_{20}(T) \sim [1-(T/T_{\rm c})^4]$.
There is a clear contradiction between the predicted and the measured
temperature dependence
that is only resolved in the limit $T \to T_{\rm c}^-$.
This is a greatly restrictive limit
that casts doubt on the physical reality of equating $|\psi_0({\bf r})|^2$
with $\rho_{20}({\bf r})$ in the condensed bosonic electron pair regime.
It means that the macroscopic wave function
and the consequent velocity field
do not apply anywhere in the condensed regime
except possibly in the immediate vicinity of the transition,
$T \to T_{\rm c}^-$.

This second point reflects
the limits of Landau's (1937) phenomenological theory,
since the second London equation itself implies
that supercurrents are irrotational in magnetic field-free regions
(\S\ref{Sec:LondonEqns}).
The macroscopic wavefunction and the implied fluxoid quantization
appear to have some utility for superconductivity (Tinkham 2004).
The long-range Coulomb repulsion between electrons
is suited to a mean-field treatment,
which may explain the utility of one-electron hydrogen-like orbitals
for the electronic structure of atoms.
This and the fact that at room temperature
electrons lie deep in the quantum regime
possibly justifies the macroscopic wavefunction
as a more suitable approximation for superconductivity
than for superfluidity.

\subsection{Molecular Dynamics of Superfluidity} \label{Sec:SMSF}

In this section the molecular equations of motion
for condensed bosons are derived.
How superfluidity arises from them is explained.
We follow and extend Attard (2025b),
which work supersedes  Attard (2025a Ch.~5).

The two key points to understand
are that we are dealing with an open macroscopic quantum system,
which means that the motion must be compatible with decoherence.
And we are at the interface between the quantum and classical worlds,
which means that the equations of motion are intermediate between
Schr\"odinger's equation and Hamilton's equations.

\subsubsection{Hamilton's Equations}

In the discussion of Bose-Einstein condensation in \S\ref{Sec:BEC},
the symmetrization factor for the momentum state occupancies,
$\chi^+({\bf p}) = \prod_{\bf a} N_{\bf a}!$,
was introduced as ensuring the normalization of the symmetrized wavefunction,
$\Phi_{\bf p}^+({\bf q}) =
(N!\chi^+({\bf p}))^{-1/2}
\sum_{\hat{\rm P}} \Phi_{\hat{\rm P}{\bf p}}({\bf q})$.
With it,
the Born probability associated with a point in classical phase space
for the subsystem in a symmetrized decoherent momentum state is
(Attard 2025b Eq.~(2.3))
\begin{eqnarray}
\lefteqn{
\Phi^+_{\bf p}({\bf q})^*\,\Phi^+_{\bf p}({\bf q})
} \nonumber \\
& = &
\frac{V^{-N}}{N!\chi^+_{\bf p}}
\sum_{\hat{\mathrm P}',\hat{\mathrm P}''}
e^{-(\hat{\mathrm P}'{\bf p}-\hat{\mathrm P}''{\bf p})\cdot{\bf q}
/\mathrm{i}\hbar}
\nonumber \\ & \approx &
\frac{V^{-N}}{N!\chi^+_{\bf p}}
\sum_{\hat{\mathrm P}',\hat{\mathrm P}''}
\!\!\!^{(\hat{\mathrm P}'{\bf p} \approx \hat{\mathrm P}''{\bf p})}\;
e^{-(\hat{\mathrm P}'{\bf p}-\hat{\mathrm P}''{\bf p})\cdot{\bf q}
/\mathrm{i}\hbar} .
\end{eqnarray}
This retains only permutations  between bosons
in nearly the same momentum state,
in which case the exponent is close to zero.
Since these similar state permutations dominate,
particularly on the low temperature side of the $\lambda$-transition,
and since
$\sum_{\hat{\mathrm P}',\hat{\mathrm P}''}
\!^{(\hat{\mathrm P}'{\bf p} = \hat{\mathrm P}''{\bf p})}
= N! \prod_{\bf a} N_{\bf a}({\bf p})!
= N!\chi^+_{\bf p} $,
explicit symmetrization is redundant,
$\Phi^+_{\bf p}({\bf q})\approx\Phi_{\bf p}({\bf q})$.

This also follows from the fact
that an open quantum system is decoherent
(Attard 2018, 2021, Joos and Zeh 1985, Schlosshauer 2005, Zurek 1991).
Decoherence means that the only allowed permutations
must satisfy $\hat{\mathrm P}{\bf p} ={\bf p}$,
else the symmetrized momentum eigenfunction,
$\Phi_{\bf p}^+({\bf q})$,
would be a superposition of states.
There is a decoherence time
(Caldeira  and Leggett  1983,
Schlosshauer 2005, Zurek \emph{et al.}\ 2003),
which  likely decreases with increasing distance
between permuted momentum states.

Schr\"odinger's equation
for the time evolution of the momentum eigenfunction
in a decoherent system for a small time step $\tau$ gives
(Attard 2023d, 2025a, 2025b),
\begin{equation}
\left[\hat{\mathrm I}
+ \frac{\tau}{\mathrm{i}\hbar}\hat{\cal H}({\bf q})\right]
\Phi_{\bf p}({\bf q})
= \Phi_{{\bf p}'}({\bf q}').
\end{equation}
Notice that this links two specific points in classical phase space,
${\bf \Gamma} = \{{\bf q},{\bf p}\}$ and
${\bf \Gamma}' = \{{\bf q}',{\bf p}'\}$.
This is the difference from Schr\"odinger's equation
for a closed quantum system,
which would instead give a superposition of momentum eigenfunctions
on the right hand side,
$\sum_{{\bf p}''} C_{{\bf p},{\bf p}''} \Phi_{{\bf p}''}({\bf q})$.
Demanding time reversible and continuous evolution
the present expression gives
\begin{eqnarray}
{\bf q}'
& = &
{\bf q} + \tau \nabla_p {\cal H}({\bf q},{\bf p}),
\nonumber \\
\mbox{ and }
{\bf p}'
& = &  {\bf p} - \tau \nabla_q {\cal H}({\bf q},{\bf p}).
\end{eqnarray}
These are Hamilton's classical equations of motion.
The second is just Newton's second law of motion:
for particle $j$ the rate of change of momentum
is the classical force,
$\dot {\bf p}_j
= -  \nabla_{q,j} {\cal H}({\bf q},{\bf p})
\equiv {\bf f}_j$.

\subsubsection{The Condensed Law of Motion}
\label{Sec:EoM-Cond}

\begin{figure}[t]
\centerline{ \resizebox{8cm}{!}{ \includegraphics*{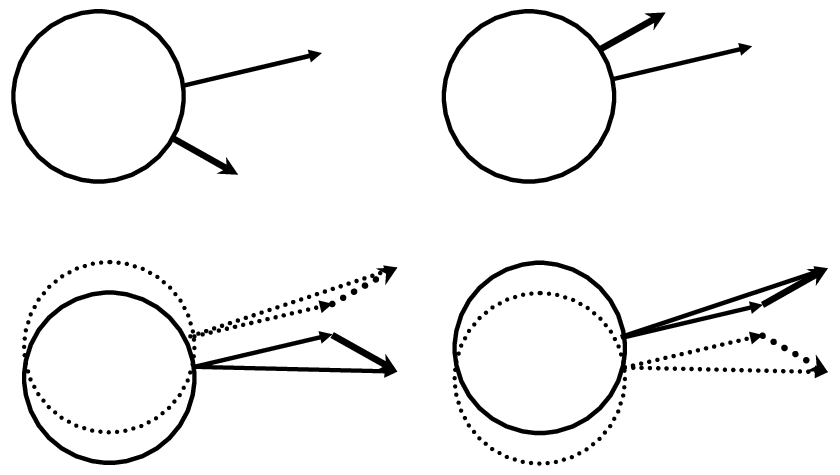} } }
\caption{\label{Fig:sup}
Two bosons initially in the same momentum state
(upper, thin arrow), acted upon by different forces
(times the time step, bold arrow)
evolve to two superposed momentum configurations (lower).
}
\end{figure}

Figure~\ref{Fig:sup} is a sketch
of the evolution over a time step
of two bosons initially in the same momentum state.
The two superposed resultant configurations
collapse into a single configuration due to decoherence.
The one of these that survives
is most likely the unpermuted one (solid lines and curves),
partly because this has unit weight in the symmetrization function.
The full argument for this conclusion is both detailed and subtle.
First we give the result,
and then we give the argument.

Suppose that in a configuration ${\bm \Gamma} = \{{\bf q},{\bf p}\}$
the momentum state ${\bf a}$ is occupied by $N_{\bf a}$ bosons.
In view of the fact that we have a decoherent macroscopic system,
for each boson in it
all but one of the $N_{\bf a}$ possible superposed evolved configurations
for each boson are suppressed.
Since the identity permutation has greatest real symmetrization weight,
it survives but with force reduced by the occupancy.
That is, for boson $j$ with momentum ${\bf p}_j$
acted upon by the force ${\bf f}_j $,
the change in momentum over a time step $\tau$ is given by
\begin{equation} \label{Eq:pnap}
{\bf p}_j'
=
{\bf p}_j
+
 \frac{\tau}{N_{{\bf p}_j}}  {\bf f}_j .
\end{equation}
The justification for this expression is given next
and in \S\ref{Sec:AdTrans}.
This says that compared to Newton's second law of motion,
a force causes the momentum of a boson in the condensed regime
to change at a rate
in inverse proportion to the occupancy of its momentum state.
This means that highly occupied momentum states persist
because the bosons in them are less likely to change their momentum.
This is the key to understanding the reduction in viscosity in superfluidity.
Note the difference between this result and Eq.~(\ref{Eq:dotp0}),
where it is the gradient of the chemical potential that gives
the actual rate of change of momentum for condensed bosons.

There are $N_{\bf a}$! permutations of the bosons
in the momentum state ${\bf a}$,
and each evolves to a separate configuration
that is superposed with all the rest.
Due to entanglement decoherence, only one of these configurations survives.
In each permutation $\hat{\rm P}$, the boson $j$
with initial momentum ${\bf p}_{j} = {\bf a}$,
evolves classically according to the force ${\bf f}_{j'}$,
where the original change in momentum of $j'$
is assigned to $j$ by the particular permutation,
$\hat{\rm P}j' = j$.
The force decorrelation time
is much longer than the decoherence time.
A small but finite time step $\tau$ can be defined
in which the classical force on each boson hardly changes
due to its and other bosons' motion.
But since the spacing between momentum states is infinitesimal,
this time step can be subdivided into infinitesimal increments
over which the boson of interest, $j$, visits many different
momentum states.
Since the specific bosons in these states vary,
and since the random permutation
that survives decoherence at each infinitesimal time step varies,
the force at each infinitesimal instant, ${\bf f}_{j'}$, also varies.
These forces average to zero
because the occupancies of the momentum states are non-local;
the forces on two different bosons in the state,
${\bf f}_k$ and ${\bf f}_l$, are uncorrelated.
Only the self-correlation ${\bf f}_{j}$
(or a macroscopic driving force, which will be discussed shortly)
does not average to zero over the permutations.
In a given permutation, $j' = j$ if, and only if,
the boson of interest belongs to a monomer loop.
There are exactly  $(N_{\bf a}-1)!$ permutations out of $N_{\bf a}!$
in which this is the case.
Hence the average force  is
\begin{eqnarray}
\langle {\bf f}_{j'}\rangle
& = &
\frac{1}{N_{\bf a}!} \sum_{\hat{\rm P}} {\bf f}_{j'} ,
\quad \hat{\rm P} j' = j
\nonumber \\ & = &
\frac{1}{N_{\bf a}!}
\sum_{\hat{\rm P}}\!^{(j'\ne j)}\, {\bf f}_{j'}
+
\frac{1}{N_{\bf a}!}
\sum_{\hat{\rm P}}\!^{(j' = j)}\, {\bf f}_{j'}
\nonumber \\ & = &
\frac{(N_{\bf a}-1)!}{N_{\bf a}!} {\bf f}_{j}
= \frac{1}{N_{\bf a}} {\bf f}_{j} .
\end{eqnarray}
This explains the equation of motion
for condensed bosons, Eq.~(\ref{Eq:pnap}).

In the case of a macroscopic driving force $-\nabla \mu$ (\S\ref{Sec:TDSF}),
the average force on boson $j$ after the random permutations
and decoherence culling of superposition states
is ${\bf f}_{j}/N_{\bf a} - \nabla \mu$.
The driving force $-\nabla \mu$ is felt by all the bosons;
it is that averaged over the subsystem volume,
since the occupancy of the momentum state is non-local.
This latter term, if present,
will usually dominate the intermolecular force term.

In the condensed regime,
the momentum evolution Eq.~(\ref{Eq:pnap}),
in combination with the position evolution
${\bf q}_j' = {\bf q}_j + ({\tau}/{m}) {\bf p}_j $,
gives a change in total energy of
$[N_{{\bf p}_j}^{-1} - 1 ] {\bf p}_j \cdot {\bf f}_j$.
This is only zero in the classical regime, $N_{{\bf p}_j}=1$;
unlike Newton's equations of motion,
energy is not individually conserved in the condensed regime.
However, since force and momentum have opposite time parity
they are uncorrelated,
and so on average this is zero.
Similarly the total change in momentum
averages to zero because the force is uncorrelated with the occupancy.
For each individual transition,
the excess energy or momentum
is presumably dissipated
to or from the neighborhood or environment
by the entanglement that leads to the decoherence.


\subsubsection{Adiabatic Stochastic Transition} \label{Sec:AdTrans}

The change in position of the bosons over a time step $\tau$ is
deterministic,
\begin{equation}
{\bf q}(t+\tau) = {\bf q}(t) + \frac{\tau}{m} {\bf p}(t).
\end{equation}
The momenta are quantized,
with ${\bf p}$ being a $3N$-dimensional vector
integer multiple of $ \Delta_p$.

The present configuration transition
must account for changes in occupation entropy.
The configuration probability density in the condensed regime is
(Attard 2025a)
\begin{equation}
\wp({\bf \Gamma}) =
\frac{1}{Z} e^{-\beta {\cal K}({\bf p})}  e^{-\beta U({\bf q})}
\prod_{\bf a} N_{\bf a}! ,
\end{equation}
where $\beta=1/k_\mathrm{B}T$ is the inverse temperature,
${\cal K}({\bf p})$ is the kinetic energy,
and $U({\bf q})$ is the potential energy.
The Wigner-Kirkwood (ie.\ commutation) function has been neglected,
as has position permutation loops and chains.
The pure momentum permutations are retained,
with the occupancy of the momentum state ${\bf a}$ being
$N_{\bf a} = \sum_{j=1}^N \delta_{{\bf p}_j,{\bf a}}$.
Also, a point in quantized phase space
is ${\bf \Gamma} = \{{\bf q},{\bf p}\}$,
and the conjugate point with momenta reversed is
${\bf \Gamma}^\dag = \{{\bf q},-{\bf p}\}$.

We seek the conditional transition probability
of boson $j$ in the momentum state  ${\bf p}_j = {\bf a}$
to the neighboring momentum state
in the direction of the $\alpha$ component of the force  ${\bf f}_j$,
namely from ${\bf a}$ to
${\bf a}'_\alpha = {\bf a}
+ \mbox{sign}(\tau f_{j\alpha})\Delta_p \widehat{\bf x}_\alpha$.
Microscopic reversibility
(ie.\ detailed balance),
which guarantees that the probability distribution is stationary,
gives the ratio of conditional transition probabilities,
\begin{eqnarray}
\frac{ \wp({\bf \Gamma}'|{\bf \Gamma};\tau)
}{ \wp({\bf \Gamma}^\dag|{\bf \Gamma}'^\dag;\tau) }
& = &
\frac{\wp({\bf \Gamma}')}{\wp({\bf \Gamma})}
 \\ & = &\nonumber
\frac{N_{{\bf a}'}+1}{ N_{{\bf a}}}
e^{(-\beta /2m) [ a'^2-a^2] }
e^{({ \beta \tau }/{m}) {\bf f}_{j} \cdot {\bf a} }.
\end{eqnarray}
By inspection,
this is satisfied by the conditional transition probability
\begin{eqnarray} \label{Eq:wploop}
\lefteqn{
\wp_{j\alpha}({\bf a}'_\alpha|{\bf a})
}  \\
& = &
\frac{\lambda_{j\alpha} }{ N_{{\bf a}}}
\left\{ 1
- \frac{\beta \Delta_p}{2m} \mbox{sign}(\tau {F}_{j\alpha}) a_\alpha
+ \frac{ \beta \tau a_\alpha }{2m} {f}_{j\alpha}  \right\} .\nonumber
\end{eqnarray}
The exponential of the change in energy has been linearized here.

With this conditional transition probability,
the average rate of change of momentum
in the direction $\alpha$
for boson $ j \in {\bf a}$ to leading order is
\begin{equation}
\left\langle \dot p_{j\alpha}^0 \right\rangle
=
\frac{\Delta_p\lambda_{j\alpha}}{\tau N_{\bf a} }
\mbox{sign}(\tau f_{j\alpha}) .
\end{equation}
The classical regime is $N_{{\bf a}}  = 1$,
and in order to satisfy  Newton's second law of motion in this case
we must have
\begin{equation}  \label{Eq:lambda}
\lambda_{j\alpha}
\equiv
\frac{|\tau f_{j\alpha}|}{\Delta_p} .
\end{equation}

With this result for $\lambda_{j\alpha}$
and the conditional transition probability,
Newton's second law does not hold
in the quantum condensed  regime, $N_{\bf a} > 1$.
As a consequence neither energy nor momentum are conserved,
which is not unexpected as these results apply
to an \emph{open} quantum subsystem.
For an individual boson with transition probability as here,
its rate of change of momentum is reduced
by a factor of $N_{\bf a}^{-1}$ from the value
given by Newton's second law of motion.
Typically for low lying momentum states in the condensed regime,
$N_{\bf a} = {\cal O}(10^2)$.

The suppression of superposed states
effectively reduces the force on individual bosons
by a factor of their occupation number.
This explains at the molecular level
the reduction in viscosity in the condensed superfluid.

Microscopic reversibility conserves the probability distribution,
and hence the entropy, on a trajectory.
It follows from the present result
that the conservation of occupation entropy
reduces  the rate of change of momentum in the condensed regime.
This leads to the loss of shear viscosity in superfluidity.
The conservation of entropy in superfluid flow
is consistent with the principle of superfluid flow
---energy is minimized at constant entropy---
empirically deduced from fountain pressure measurements
in \S\ref{Sec:TDSF}.

In the results obtained with the computer algorithm below,
sequential transitions
for each component of momentum of each boson in the momentum state
are attempted at each time step.
There appears to be little difference
in whether the occupancy is updated after each successful transition,
or only at the end of the time step.
There also appears little difference if the quadratic term in the change in
kinetic energy is added.
And it is also possible to use the exponential form of the term in braces.

\subsubsection{Dissipative Transition}

The dissipative transitions complement the adiabatic stochastic transitions,
acting like a thermostat
and providing another  mechanism
for the change in occupancy of the momentum states
and for the equilibration of the occupancy distribution.
The computer algorithm for boson $j$ with ${\bf p}_j = {\bf a}$
uses the following conditional transition probability
to the 27 near neighbor states ${\bf a}'$
(including the original state ${\bf a}$).

The dissipative transition is irreversible,
which means that the forward and backward unconditional transitions
are equally likely.
Hence  for the transition to a neighboring momentum state
${\bf a} \stackrel{j}{\to} {\bf a}'$,
the ratio of conditional transition probabilities is
\begin{eqnarray}
\frac{\wp_j({\bf a}'|{\bf a})}{\wp_j({\bf a}|{\bf a}')}
& = &
\frac{\wp_j({\bf a}')}{\wp_j({\bf a})}
\nonumber \\ & = &
\frac{N_{{\bf a}'}+1}{N_{\bf a}}
\left[ 1  - \frac{\beta (a'^2-a^2)}{2m} \right] .
\end{eqnarray}
Here the change in kinetic energy has been expanded to quadratic order.
This is satisfied by
\begin{equation} \label{Eq:DissQuad}
\wp_j({\bf a}'|{\bf a})
=
\left\{ \begin{array}{ll}
\displaystyle
\frac{\varepsilon}{N_{\bf a}}
\left[ 1  - \frac{\beta (a'^2-a^2)}{4m} \right],
& {\bf a}' \ne {\bf a}  \\
\displaystyle
1 - \frac{26 \varepsilon}{N_{\bf a}}
+  \frac{54\beta \Delta_p^2}{4m} \frac{\varepsilon}{N_{\bf a}} ,
& {\bf a}' = {\bf a}  .
\end{array} \right.
\end{equation}
For the following results, $\varepsilon=1/27$.
The dissipative transitions were attempted one boson at a time,
for all $N$ bosons in a cycle,
typically once every 10 time steps.
Less frequent attempts would probably suffice.

The present algorithm
has proven adequate to ensure the equilibrium distribution,
although it is not actually clear that a dissipative thermostat is required
because unlike the classical adiabatic equations of motion,
temperature already appears
in the present adiabatic conditional transition probability
(cf.\ Attard 2012 Ch.~11).

\subsubsection{Quantum Molecular Dynamics Results} \label{Sec:QMD}

Results are now presented for classical and quantum Lennard-Jones $^4$He.
For the classical liquid,
results are obtained with the same algorithm as the quantum liquid
but as if the momentum states were solely occupied.
Hence the momenta are quantized,
and a transition probability without the factor of $N_{\bf a}^{-1}$ is used.
This factor is also dropped
in the rate of change of the first momentum moment in the shear viscosity
(see below).
However, the momenta are still quantized
and the adiabatic transitions are still stochastic,
exactly as in the quantum case.
A Lennard-Jones saturated homogeneous liquid is simulated,
as in \S\ref{Sec:QMCCPS}.
Neither the Wigner-Kirkwood (ie.\ commutation) function
nor position permutation loops are used.
The number of $^4$He atoms is $N=1,000$;
for further details see Attard (2025b).

\begin{figure}[t]
\centerline{ \resizebox{8cm}{!}{ \includegraphics*{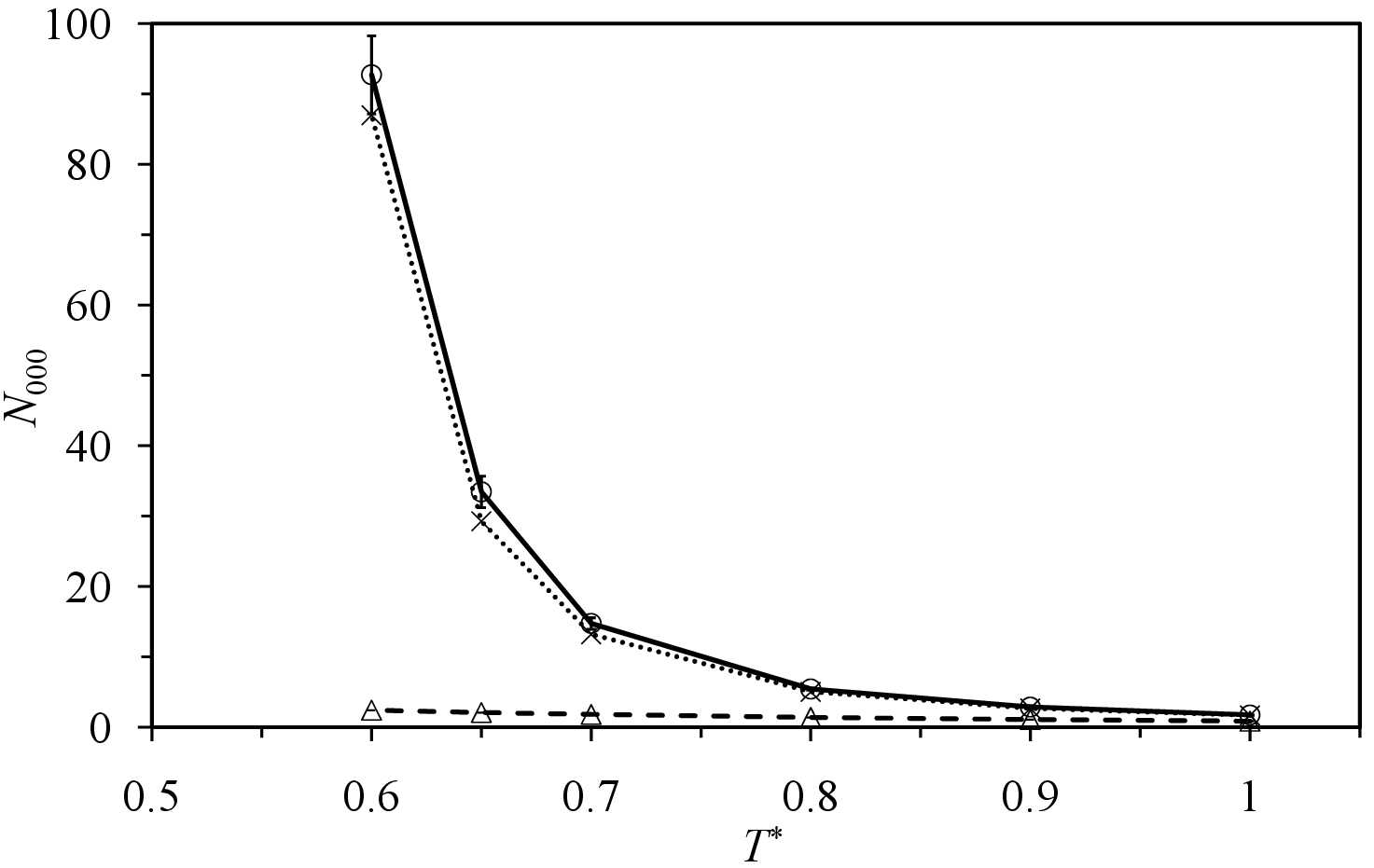} } }
\caption{\label{Fig:N000}
Ground state occupancy in the saturated Lennard-Jones liquid
($N=1,000$).
The circles are the quantum liquid,
the triangles are the classical liquid,
and the crosses are the exact result for ideal bosons.
The error bars are less than the symbol size
and the lines are an eye guide.
}
\end{figure}

Figure~\ref{Fig:N000} shows the ground state occupancy
on the saturation curve for the Lennard-Jones liquid.
It can be seen that the simulated occupancy in the quantum liquid
is in good agreement with the analytic result calculated
for non-interacting bosons.
Comparable if not better agreement holds for the first several
excited momentum states.
The ideal boson result should apply to interacting bosons
on the far side of the $\lambda$-transition
(Attard 2025 \S5.3).
The slightly larger than ideal value
for the quantum liquid ground momentum state occupancy
is probably a finite size effect.

At lower temperatures
the occupancy  in the quantum liquid is much larger
than for the classical liquid.
This is of course due to the r\^ole of the occupation entropy
on the transition probability.
However, the occupancy of the ground state in the quantum liquid
is a small fraction of the total number of bosons in the subsystem.
This fraction decreases with increasing subsystem size.
Obviously this means that ground state condensation
cannot account for the $\lambda$-transition
or for superfluidity.

This fraction of bosons in the quantum liquid
that are in the ground momentum state
is much less less than  the measured fraction of condensed bosons in He~II
(Donnelly and  Barenghi 1998).
By coincidence,
for this system size it is comparable to the results of
the  path integral quantum Monte Carlo simulations of Ceperely (1995),
who find that the condensate  fractions is less than 10\%,
based on the definition that
`the condensate fraction [is] the probability of finding
an atom with precisely zero momentum' (Ceperley 1995 p.~297).
This is consistent with estimates by others:
Penrose and Onsager (1956) using Feynman's approximation,
McMillan (1965) using Monte Carlo calculations,
and Kalos \emph{et al.}\ (1981) and Whitlock and Panoff (1987)
using Green's-function Monte Carlo calculations
all estimate the ground-state condensate as close to 8--9\%
(Ceperley 1995 p.~18).
The main reason for the discrepancy between these estimates
of condensation and the measured values
is that the computed condensation is solely into the ground state.

\begin{figure}[t]
\centerline{ \resizebox{8cm}{!}{ \includegraphics*{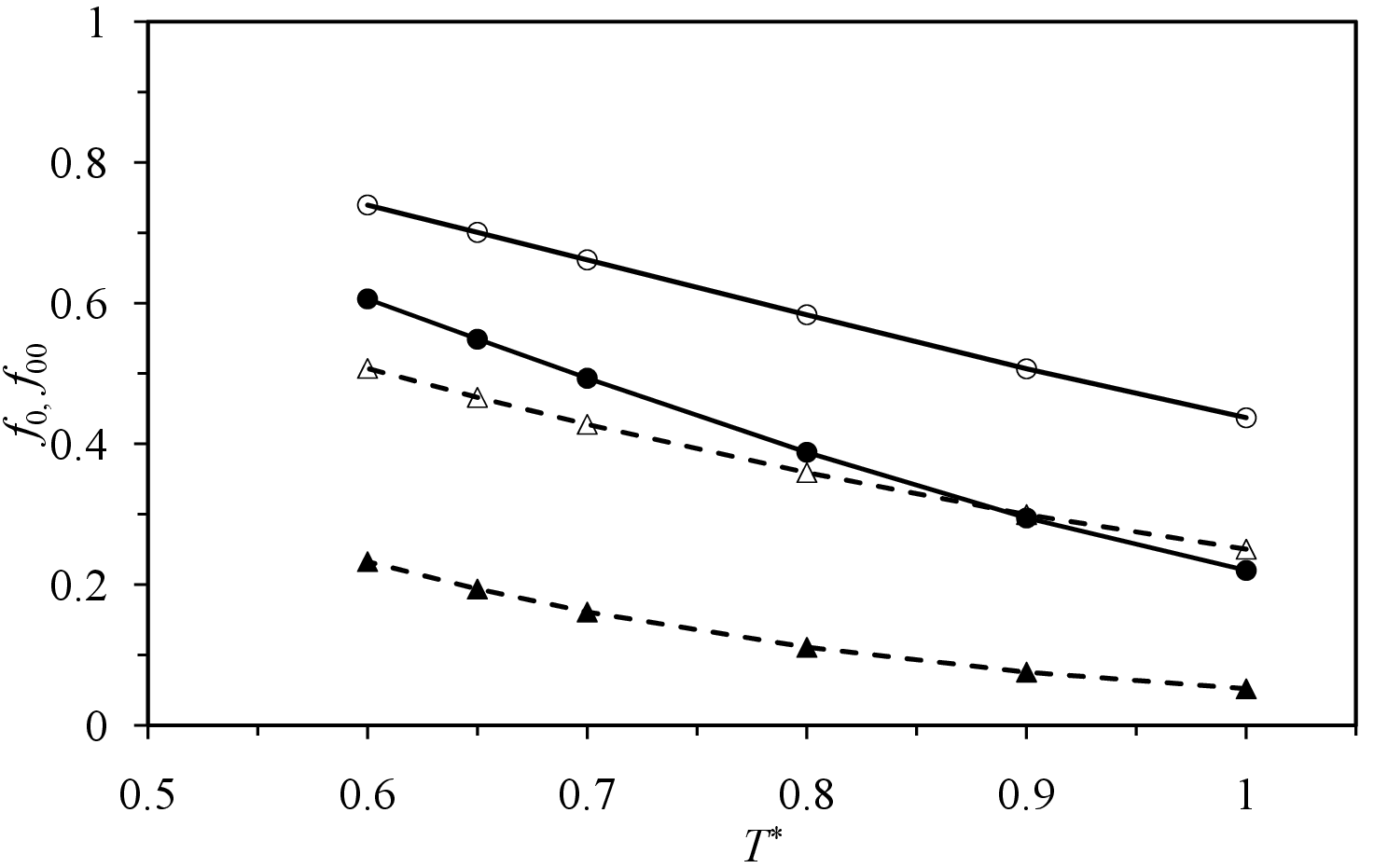} } }
\caption{\label{Fig:f0}
Fraction of bosons condensed in the saturated Lennard-Jones liquid.
The circles are the quantum liquid,
the triangles are the classical liquid.
The open symbols are $f_0$,
the fraction in states occupied by two or more bosons,
and the filled symbols are $f_{00}$,
the fraction in states occupied by three or more bosons.
The error bars are less than the symbol size.
The lines are an eye guide.
}
\end{figure}

Figure~\ref{Fig:f0} gives the fraction of condensed bosons,
defined as being in multiply occupied momentum states.
This fraction of bosons that are condensed in the quantum liquid
is comparable to the measured fraction in helium~II
(Donnelly and  Barenghi 1998).
The threshold for condensation was set at an occupancy of 2 for $f_0$,
and at 3 for $f_{00}$.
In the quantum liquid at the lowest temperature studied
about 74\% of the bosons are in states with two or more,
and about 61\% are in states with three or more.
In contrast the fraction for the classical liquid
is 51\% and 23\%, respectively.
These specific results are for $N=1,000$,
but other simulations show that these fractions are quite insensitive
to the system size.
The conclusion is that at these temperatures
Bose-Einstein condensation is substantial,
and that multiple momentum states are multiply occupied.
That the majority of the bosons in the system can be considered
to be condensed explains the macroscopic nature
of the $\lambda$-transition and superfluidity.

It can be seen that at higher temperatures in Fig.~\ref{Fig:f0}
the condensation in the quantum liquid
is approaching that in the classical liquid,.
However even at the highest temperature studied, $T^*=1.00$,
there is still excess condensation in the quantum liquid,
$f_0^\mathrm{qu}=44\%$ compared to $f_0^\mathrm{cl}=25\%$.
That there remains condensation in the quantum liquid
well-above the superfluid transition temperature
is likely due to the neglect in the present calculations
of position permutation loops,
which suppress condensation
(cf.\ Fig.~\ref{Fig:N0barA} and also Attard (2025a \S3.2)).

The kinetic energy per boson in the classical liquid is
$\beta {\cal K}/N=1.4513(4)$ at $T^*=1.00$
and 1.4048(2) at $T^*=0.60$.
The equipartition theorem gives the exact classical value as 3/2.
Clearly the present stochastic equations of motion
that use the transition probability for quantized momentum
are close to the continuum classical equations of motion.
The discrepancy is probably an effect of finite size.
The kinetic energy per boson in the quantum liquid is
$\beta {\cal K}/N=1.2248(5)$ at $T^*=1.00$
and 0.778(2)  at $T^*=0.60$.
The decrease in kinetic energy with decreasing temperature
is a manifestation of the increasing condensation
in the quantum liquid
that preferentially occurs in the low lying momentum states.

The shear viscosity can be expressed as an integral of
the momentum-moment time-correlation function
(Attard 2012 Eq.~(9.117)),
\begin{equation} \label{Eq:eta(t)}
\eta_{\alpha\gamma}(t)
=
\frac{1}{2 V  k_\mathrm{B} T}
\int_{-t}^{t} \mathrm{d}t'\,
\left< \dot P_{\alpha\gamma}^0({\bm \Gamma})
\dot P_{\alpha\gamma}^0({\bm \Gamma}(t'|{\bm \Gamma},0))
\right> .
\end{equation}
This is called a Green-Kubo expression (Green 1954, Kubo 1966),
although it was Onsager (1931) who originally gave
the relationship between the transport coefficients
and the time correlation functions.

The first $\alpha$-moment of the $\gamma$-component of momentum is
$P_{\alpha\gamma} = \sum_{j=1}^N q_{j\alpha} p_{j\gamma}$,
and its classical adiabatic rate of change is
\begin{equation}
\dot P^0_{\alpha\gamma}
=
\frac{1}{m} \sum_{j=1}^N p_{j\alpha} p_{j\gamma}
+
\sum_{j=1}^N q_{j\alpha} f_{j\gamma} .
\end{equation}
This can be symmetrized using the gradient of the pair potential,
which enables the minimum image convention for periodic boundary conditions
to be applied.

\begin{figure}[t]
\centerline{ \resizebox{8cm}{!}{ \includegraphics*{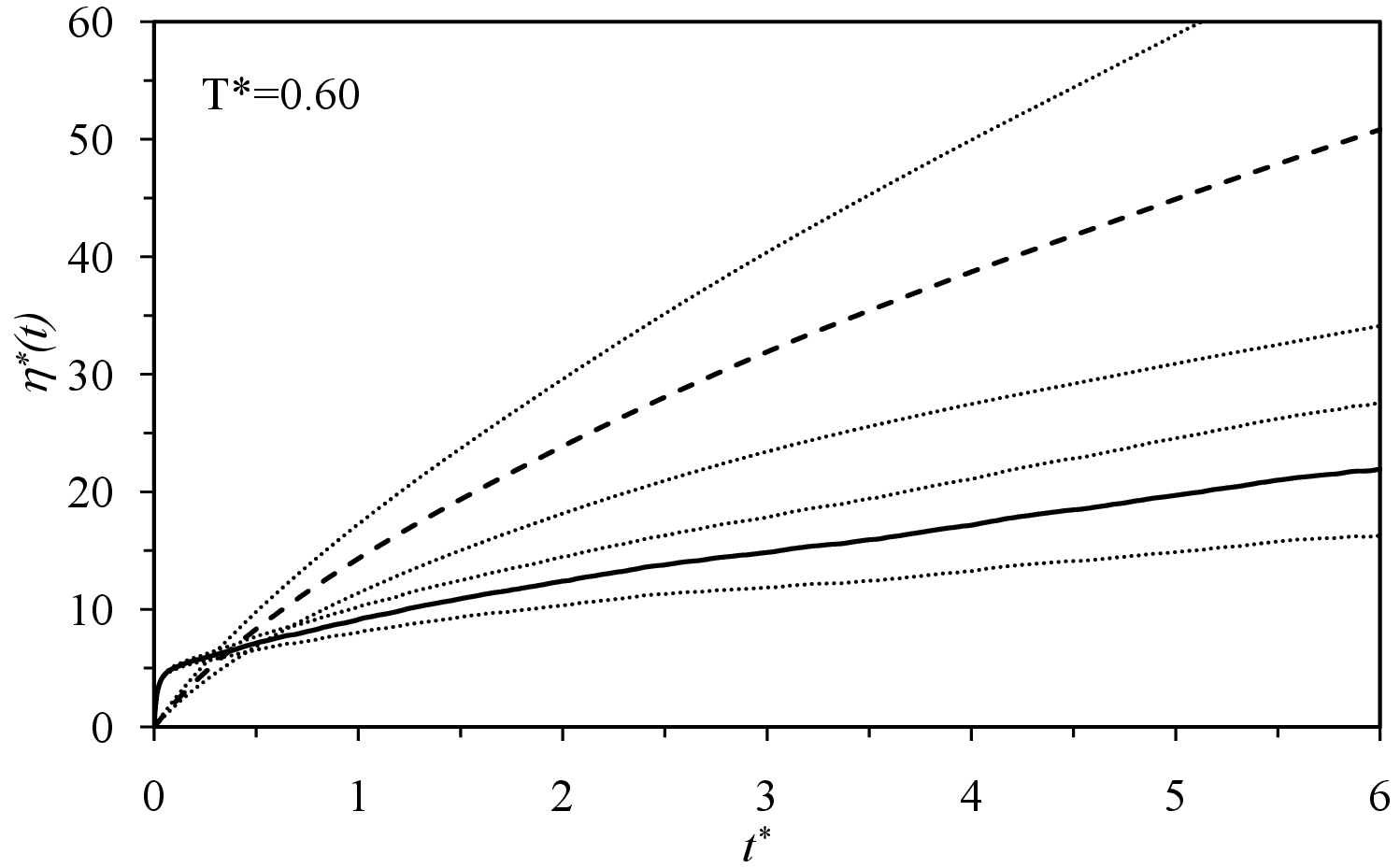} } }
\caption{\label{Fig:eta60}
Shear viscosity time function for the Lennard-Jones liquid
at $T^*=0.60$ and $\rho^*=0.8872$.
The solid curve is the quantum liquid,
the dashed curve is the classical liquid,
and the dotted curves give the 95\% confidence level.
The unit of time is
$t_\mathrm{He} =
\sqrt{m_\mathrm{He}\sigma_\mathrm{He}^2/\varepsilon_\mathrm{He}}$,
and the shear viscosity is
$\eta^* = \eta \sigma_\mathrm{He}^3/\varepsilon_\mathrm{He} t_\mathrm{He}$.
}
\end{figure}

In the condensed regime,
the average rate of change of momentum
for boson $j$ to leading order is
$\left\langle \dot {\bf p}_{j}^0 \right\rangle
=
{\bf f}_j /{N_{{\bf p}_j}}  $.
With this
the adiabatic rate of change of the first momentum moment is
\begin{eqnarray} \label{Eq:.mtm-mmt}
\dot{\underline {\underline P}}^0
& = &
\frac{1}{m} \sum_{j=1}^N {\bf p}_{j} {\bf p}_{j}
+
\sum_{j=1}^N
{\bf q}_{j} \frac{1}{N_{{\bf p}_j}} {\bf f}_j
\nonumber \\ & = &
\frac{1}{m} \sum_{j=1}^N {\bf p}_{j} {\bf p}_{j}
+
\frac{1}{2} \sum_{j,k} \tilde {\bf q}_{jk} {\bf f}_{j,k} .
\end{eqnarray}
Here ${\bf f}_{j,k}$ is the force on boson $j$ due to boson $k$,
so that the total force on boson $j$ is
${\bf f}_{j} = \sum_k {\bf f}_{j,k}$,
and
\begin{equation}
\tilde {\bf q}_{jk}
\equiv
\frac{1 }{N_{{\bf p}_j}} {\bf q}_j
-
\frac{1 }{N_{{\bf p}_k}} {\bf q}_k .
\end{equation}
For the  periodic boundary conditions
usually invoked in computer simulations,
the minimum image convention may be applied to this modified separation
to guarantee that  $|\tilde q_{jk,\alpha}| \le L/2$.

Figure~\ref{Fig:eta60} shows the viscosity time function
at the lowest temperature studied.
In general this reaches a plateau,
which maximum value is called `the' shear viscosity.
The extrapolated maximum viscosity of the classical liquid is
$\eta^{\rm cl}(20) = 91(53)$,
which is over four times larger than the quantum value
$\eta^{\rm qu}(6) = 21.9(57)$,
which is close to its maximum.
It is emphasized that the only difference between the classical
and quantum programs was whether or not the factor of $N_{{\bf p}_j}^{-1}$
was applied to the force on atom $j$.

\begin{figure}[t]
\centerline{ \resizebox{8cm}{!}{ \includegraphics*{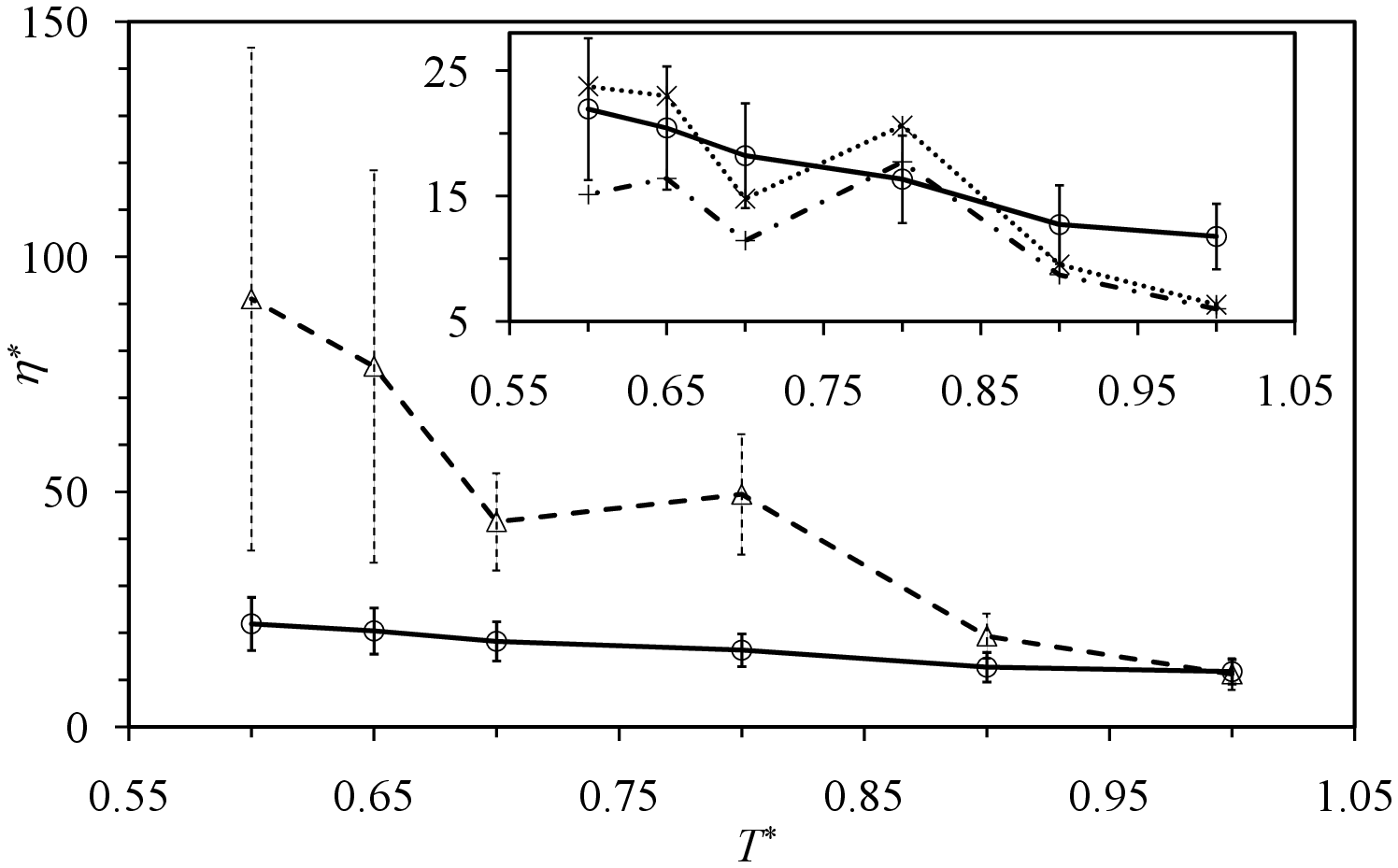} } }
\caption{\label{Fig:eta}
Shear viscosity on the Lennard-Jones liquid saturation curve.
The circles are for the quantum liquid
$\eta^\mathrm{qu}(6)$,
the triangles are for the classical liquid
$\eta^\mathrm{cl}_\mathrm{max}$.
The error bars give the 95\% confidence level,
and the lines are an eye guide.
\textbf{Inset.}
Viscosity of the quantum liquid (circles) on a magnified scale
compared to two predictions based on the classical viscosity.
The asterisks connected by dotted lines are  Eq.~(\ref{Eq:etaT}),
and the plus symbols connected by dash-dot lines
are Eq.~(\ref{Eq:etaA}).
The statistical error of these is larger
than that shown for the quantum liquid.
}
\end{figure}

Figure~\ref{Fig:eta} shows the shear viscosity
as a function of temperature for the saturated liquid.
It can be seen that at higher temperatures
the classical and quantum viscosities converge.
At the lowest temperature studied the classical viscosity is about
four times larger than the quantum viscosity.
Whereas the classical viscosity increases by a factor of eight
over the temperature range,
the quantum viscosity only increases by a factor of two.
The not quite doubling in condensation in the quantum liquid,
Fig.~\ref{Fig:f0},
is sufficient to cancel almost completely
the classical viscosity increase.
Evidently condensation reduces the rate of change of momentum
via the factor of $1/N_{\bf a}$,
and this directly reduces the shear viscosity.

The inset to the figure includes results for
\begin{equation} \label{Eq:etaT}
\eta_\mathrm{T}
=
(1-f_0^\mathrm{qu})\eta^\mathrm{cl}_\mathrm{max} .
\end{equation}
This viscosity is the analogue
of Tisza's two-fluid model of superfluidity
(Attard 2025e, Landau 1941, Tisza 1938),
which was outlined in  \S\ref{Sec:TwoFluid}.
In this case we use the fraction of uncondensed bosons
(ie.\ those in singly-occupied momentum states)
in the quantum liquid
times the viscosity in the classical liquid
(ie.\ the viscosity calculated as if all the bosons
were in singly-occupied momentum states).
Essentially this assumes that the viscosity of condensed bosons is zero,
and it assumes that the actual viscosity is a linear combination
of that of the individual components of a two-component mixture.

It can be seen that the two-fluid approximation is surprisingly good.
For $T^*=0.60$, $\rho^*=0.8872$,
the quantum viscosity is $\eta^\mathrm{qu}(6)=21.9(57)$,
and the linear binary mixture result is
$\eta_\mathrm{T}=24(14)$.
For $T^*=1.00$, $\rho^*=0.7009$,
the quantum viscosity is $\eta^\mathrm{qu}(6)=11.8 (26)$,
and the linear binary mixture result is
$\eta_\mathrm{T}=6.3(19)$.

Of course in the present equations of motion
the rate of changed of momentum of condensed bosons is not zero,
but is rather reduced by the occupancy of their respective momentum states.
For the lowest temperature studied,
$T^*=0.60$ $\rho^*=0.8872$,
the average occupancy of occupied momentum states
in the quantum case was $\overline N_\mathrm{occ}  = 2.455(4)$.
For such a small average occupancy it is perhaps surprising
that the reduction of the rate of change of momentum
is sufficient to reduce the superfluid viscosity by so much.
In fact however a plausible model for the viscosity
in the condensed regime is
\begin{eqnarray} \label{Eq:etaA}
\eta_\mathrm{A}
& = &
\frac{1}{ \overline N_\mathrm{occ}^{2} }
\eta^\mathrm{cl}_\mathrm{max}.
\end{eqnarray}
The average occupancy of occupied states is
$\overline N_\mathrm{occ} \equiv  N/\overline M_\mathrm{occ}$,
where $\overline M_\mathrm{occ}$ is the number of occupied states.
This factor gives the reduction in the rate of change
of the first momentum moment in the force term in Eq.~(\ref{Eq:.mtm-mmt}).
This neglects the diffusive (ie.\ ideal) contribution.
This is squared because the viscosity time function
is the \emph{pair} time correlation
of the rate of change of the first momentum moment.
In the case $T^*=0.60$ $\rho^*=0.8872$ this formula gives
$\eta_\mathrm{A}  =15.1(89)$,
to be compared with the two-fluid model $\eta_\mathrm{T}  =23.7(139)$
and the actual simulated value $\eta^\mathrm{qu}(6)=21.9(57)$.
At $T^*=1.00$ $\rho^*=0.7009$,
with $\overline N_\mathrm{occ} = 1.3712(3)$,
the respective values are
$\eta_\mathrm{A} = 6.0(18)$, $\eta_\mathrm{T} = 6.3(19)$,
and $\eta^\mathrm{qu}(6) = 11.8(26)$.
The inset to Fig.~\ref{Fig:eta} shows
that the accuracy of this second model is comparable
to that of the two-fluid model.
This second model has a quantitative justification
that does not insist that the viscosity of condensed bosons is zero.
This formula explains how the seemingly small average occupancy
of condensed bosons is sufficient to cause the reduction
in superfluid viscosity comparable to the measured values.
In addition,
it is likely that bosons in momentum states with above average occupancies,
due to fluctuations,
dominate the reduction in the viscosity at each instant.

\begin{figure}[t]
\centerline{ \resizebox{8cm}{!}{ \includegraphics*{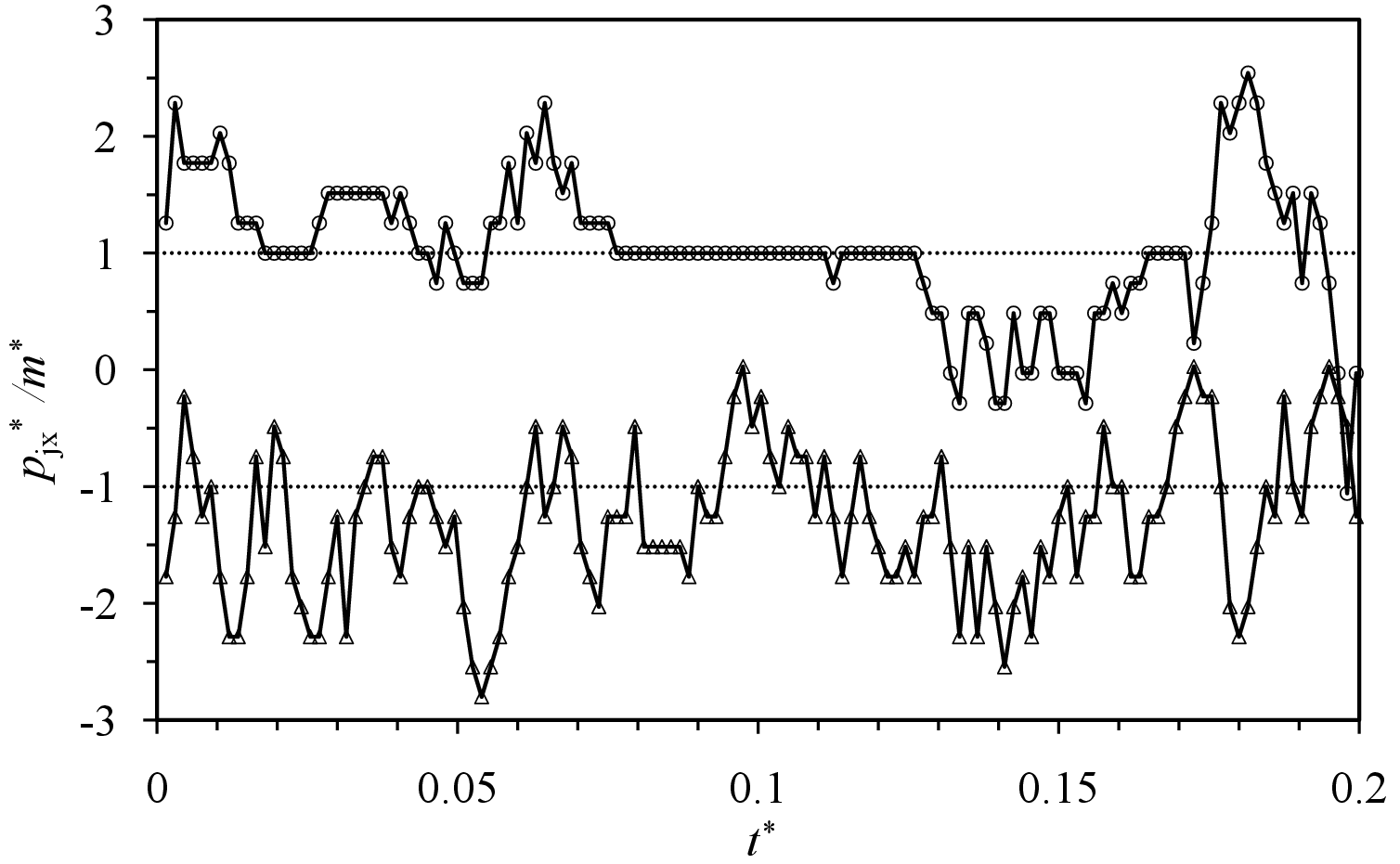} } }
\caption{\label{Fig:traj}
Component of velocity of a typical boson on a trajectory
at $T^*=0.60$ and $\rho^*=0.8872$, plotted once every 75 time steps.
The circles are for the quantum liquid
(offset by $+1$)
and the triangles are for the classical liquid
(offset by $-1$).
The dotted lines give the zero momentum state for the component.
The dimensionless spacing between momentum states is 0.26.
}
\end{figure}

Figure~\ref{Fig:traj}
shows a quantum and  classical trajectory
for a component of momentum of a typical boson.
It can be seen that there is a qualitative difference between
the trajectory in the quantum liquid and in the classical liquid.
The quantum equations of motion yield
a smoother curve, with smaller fluctuations,
and noticeable stretches of constant momentum.
These are correlated with the boson being
in a highly occupied momentum state,
which are noticeably the low-lying momentum states.
On this portion of the quantum trajectory,
the occupancy of the momentum state that this boson is in
ranges up to 91, and averages 19.4.
The conclusion is that the occupancy factor, $1/N_{\bf a}$,
damps the accelerations experienced by condensed bosons.
It is not hard to imagine that the more frequent changes in momentum
evident in the classical liquid dissipate momentum more efficiently
and give rise to the non-zero viscosity of everyday experience.

%
\section{Superconductivity}
\setcounter{equation}{0} \setcounter{subsubsection}{0}
\renewcommand{\theequation}{\arabic{section}.\arabic{equation}}
%

The modern theory of Bose-Einstein condensation 
is most relevant for high temperature superconductivity.
The BCS theory (Bardeen Cooper and Schrieffer 1957)
gives a successful quantitative account
of low-temperature superconductivity
(Annett 2004, Kittel 1976, Tinkham 2004).
What the modern theory adds to this is of a more conceptual nature.
One advantage of the modern theory  of superconductivity is its coherence
with the theory of superfluidity.
In particular the thermodynamical and statistical mechanical techniques
used in the earlier sections of this review
are applied with minor adjustments to superconductivity.
Two other advantages that the modern theory
has over BCS theory
are that it provides molecular explanations
for the Meissner-Ochsenfeld effect (Meissner and Ochsenfeld 1933)
and for the London equations (F and H London 1935) (Attard 2025f),
and it gives a physical mechanism for the formation of bosonic electron pairs
in the high temperature case
(Attard 2022b, 2025a Ch.~6).

The connection between superconductivity
and Bose-Einstein condensation
lies in the formation of electron pairs (Cooper 1956).
These are effective bosons formed from electrons
with opposite spin.
Cooper pairs are defined as having zero nett momentum,
motivated by the belief that Bose-Einstein condensation
was confined to the ground state.
A more general approach is to define bosonic pairs
as consisting of electrons with opposite spin and non-zero nett momentum
(Attard 2022b, 2025a \S6.2.2).
(We distinguish between pairs
that are bosonic (ie.\ opposite spin)
or fermionic (ie.\ the same spin).
The former are essential for superconductivity;
the latter are not always negligible.)
In the statistical mechanical theory of superconductivity
we shall elucidate the nature and importance of bosonic pairs
(\S\ref{Sec:2e}).

What is novel about the Cooper pairs in the BCS theory
of low-temperature superconductivity is the binding mechanism and size.
Unlike $^4$He,
which is an effective boson composed of three atomic-sized,
strongly bound fermion pairs,
the separation of the electrons in Cooper pairs
can be hundreds of nanometers,
on the order of the wavelengths of lattice vibrations,
and the potential that binds them  is very weak.
The statistical mechanical theory of high-temperature superconductivity
invokes electron pairs that are akin to the pairs of fermions in $^4$He,
namely they are tightly bound
at much shorter separations than in BCS theory,
possibly even sub-nanometer (\S\S\ref{Sec:2e} and \ref{Sec:HighT-SC}).
In explaining the phenomena of superconductivity,
it is important to address any questions
that arise from the difference in sizes
of the two types of bosonic electron pairs.

\subsection{Thermodynamics of Superconductivity}
\subsubsection{Meissner-Ochsenfeld Effect}

The Meissner-Ochsenfeld (1933) effect
refers to the expulsion of a magnetic field from the interior
of a superconductor (Annett 2004, Kittel 1976, Tinkham 2004).
When a critical field is reached
the magnetic field penetrates the sample
and superconductivity is destroyed,
either entirely (Type~I superconductors),
or partially  (Type~II superconductors).
The expulsion of a magnetic field
is the test that is often used to identify superconductors
and the superconducting transition.
The degradation of the supercurrent by the penetration of the field
is what limits the power of superconducting electromagnets,
which is one of the main applications of superconductivity.
For these reasons, as well as a general curiosity,
understanding the physical and molecular basis of the effect
is of some value.

The conventional theory of superconductivity
does not give the cause of the Meissner-Ochsenfeld effect.
It begs the question
to assert that a magnetic field is spontaneously expelled
from a superconductor because doing so lowers its free energy.
This is, in essence, the conventional approach
that obtains the free energy of the superconducting state
by equating it to the energy of the critical magnetic field
that destroys it  (Annett 2004, Kittel 1976, Tinkham 2004).
Likewise it puts the cart before the horse
to offer the second London equation  (F and H London 1935)
as proof that a magnetic field is expelled:
\emph{if} the second London equation is true
\emph{then} certainly supercurrents and  magnetic fields are incompatible.
But \emph{why} is the second London equation true?

The answer to this question lies
in the magnetic nature of bosonic electron pairs.
Because the electrons have equal and opposite spin,
a bosonic electron pair with separation
${\bf q}_2={\bf q}_+-{\bf q}_-$
is a magnetic quadrupole.
In a local magnetic field ${\bf B}({\bf r})$
the magnetic contribution to the pair energy is (Attard 2025f)
\begin{eqnarray}
\varepsilon_2({\bf r})
& = &
\mu_{\rm B} {\bf q}_2 \cdot \nabla {B}({\bf r})
\nonumber \\ & = &
\frac{-\beta \mu_{\rm B}^2 \overline{q}_2^2}{3}(\nabla B({\bf r}))^2 .
\end{eqnarray}
Here in SI units
$\mu_{\rm B} = e \hbar /2m$ is the Bohr magneton,
The second equality follows after a classical average
and linearization for weak fields,
with the inverse temperature being $\beta = 1/k_{\rm B}T$.

This expression for the magnetic quadrupole energy
assumes that any variation in the gradient of the magnetic field
is negligible over the size of the bosonic electron pair.
It is difficult to foresee an experimental situation
that would violate this condition.
But even if it were violated
in the case of the Cooper pairs of BCS theory,
the following argument would still hold
with the gradient replaced by the magnetic field difference
experienced by the pair,
$ {\bf q}_2 \cdot \nabla {B}({\bf r})
\Rightarrow  {B}({\bf q}_+) - {B}({\bf q}_-)   $.

Since in general the energy of a region with constant external potential
can be written
$ E(S,V,N;\varepsilon) = E(S,V,N) + N \varepsilon $,
and since the number derivative of this is the chemical potential
(cf.\ \S\ref{Sec:TDSF}),
a slowly varying one-body potential
can be incorporated into a local chemical potential
(cf.\ Attard 2025e Eq.~(2.5), de Groot and Mazur 1984).
In the present case for bosonic electron pairs this is
\begin{equation}
\mu_2({\bf r})
=
\mu_2^{(0)}
- \frac{\beta \mu_{\rm B}^2 \overline{q}_2^2}{3}(\nabla B({\bf r}))^2   ,
\end{equation}
where $\mu_2^{(0)}$ is the chemical potential
for the same density of condensed bosonic electron pairs
in the absence of any magnetic field.

The thermodynamic principle that determines superfluid flow
is that energy is minimized at constant entropy
(\S\ref{Sec:TDSF}).
At equilibrium this
is equivalent to the local chemical potential being the same
in all connected superfluid regions,
\begin{equation}
\mu({\bf r}) = \mu.
\end{equation}
The experimental and theoretical evidence
for this 
was discussed in detail in \S\S\ref{Sec:TDSF} and \ref{Sec:TwoFluid}.

As a general thermodynamic principle
it must also apply to superconductor currents.
In this case connected regions of superconductor
must have 
$\mu_2({\bf r}) = \mbox{const}.$,
or
\begin{equation}
 \nabla {B}({\bf r}) = \mbox{const}.
\end{equation}
Since in macroscopic volumes the magnetic field would diverge
if it had constant gradient everywhere,
the constant gradient must be zero.
Hence the magnetic field itself must  be constant,
\begin{equation}
{\bf B}({\bf r})
=
(1+\chi) {\bf B}_{\rm ap}({\bf r})
= \mbox{const} .
\end{equation}
Since this must hold for applied magnetic fields ${\bf B}_{\rm ap}$
with arbitrary spatial variation,
and since the  magnetic susceptibility per unit volume $\chi$
has to be a property of the superconductor
that is independent of the applied field,
this gives
\begin{equation}
\chi = -1,
\mbox{ and }
{\bf B}({\bf r}) = {\bf 0} .
\end{equation}
This is the Meissner-Ochsenfeld (1933) effect
that was to be obtained.
The conclusion is that magnetic fields must be canceled
in the interior of a superconductor
due to the requirement that regions connected by supercurrents
must have the same chemical potential.
Thus the fountain pressure observed in superfluidity
and the Meissner-Ochsenfeld  effect observed in superconductivity
are two sides of the same coin:
both minimize the energy at constant entropy,
which requires that the chemical potential of the condensed bosons
be everywhere equal.

It seems that the mechanism by which the applied magnetic field is canceled
in the interior of the superconductor
is the creation of a solenoidal supercurrent in the surface
regions of the superconductor.
One would guess that there is little thermodynamic cost to this.
Presumably other mechanisms for equalizing the chemical potential
(eg.\ increasing the condensed bosonic pair density
in regions of non-zero field gradient)
have higher cost.

For a type~II superconductor
with partial penetration of the magnetic field,
bosonic electron pairs
are attracted to high gradients.
In this case the cost of density inhomogeneities must be less
than the cost of the surface supercurrent required
to completely cancel the applied field.
High magnetic field gradients occur in the vicinity
of the flux lines,
and therefore bosonic pairs of electrons move toward them.
One could well imagine that irrotational vorteces
develop about the flux lines
in order to conserve the angular momentum during the inflow.
If angular momentum is not conserved,
or if it dissipates over time once the inflow has stopped,
then there is no requirement that the vorteces be irrotational,
or that there be vortex flow about the flux lines.

\subsubsection{London Equations} \label{Sec:LondonEqns}

The second London equation has been taken as the axiomatic basis
for the behavior of supercurrents
(Annett 2004, Kittel 1976, Tinkham 2004).
It was derived by the London brothers (F and H London 1935)
starting from the Drude model of an electric current
with the resistivity set to zero (ie.\ a perfect conductor).
For reasons discussed below, the London brothers argued that
the first equation that resulted was too general
and that it contradicted the  Meissner-Ochsenfeld effect
in certain respects.
For this reason they focussed upon a particular solution
that is now called the second London equation,
or more simply the London equation,
that was more consistent with the  Meissner-Ochsenfeld effect
and that predicted quantitatively other known phenomena
in superconductors.
This equation has come to dominate the analysis of superconductors
ever since (Annett 2004, Kittel 1976, Tinkham 2004).
The London equations are now derived
from the thermodynamic principle of superfluid flow.

In \S\ref{Sec:TwoFluid}
we derived the two-fluid equations of Tisza (1939)
for superfluid flow
from the principle that the energy is minimized at constant entropy.
This is equivalent  to the notion that the force
on condensed bosons is the gradient of the mechanical part of the energy,
which is the chemical potential.
This was given as Eq.~(\ref{Eq:dotp0}),
${\partial {\bf p}_0}/{\partial t}
=
- n_0 \nabla \mu - \nabla \cdot ({\bf p}_0 {\bf v}_0 )$.
Here and below all quantities
are functions of position ${\bf r}$ and time $t$.

In the present case of superconductivity,
the momentum density for the condensed bosonic pairs
involves the canonical momentum, 
which includes the contribution of the magnetic field,
$ {\bf p}_{20} = 2m n_{20} {\bf v}_{20}
- 2 e n_{20} {\bf A}$,
where $n_{20}$ is the number density of the pairs
and ${\bf v}_{20}$ is their local velocity.
(The subscript 2 designates pairs;
the subscript 0 designates condensed bosonic.)
The magnetic field is given by the magnetic vector potential,
${\bf B} = \nabla \times {\bf A}$.
The superconducting current is ${\bf j}_{20} = -2 e n_{20}{\bf v}_{20}$.
The chemical potential for the condensed bosonic pairs is
$ \mu_2 = \mu_2^{0}
- {\beta \mu_{\rm B}^2 \overline{q}_2^2} (\nabla B)^2/{3}
-2e \phi$,
where $\phi$ is the electrostatic potential.
This is the electrochemical potential (de Groot and Mazur 1984 Eq.~(XIII.42))
with the  magnetic quadrupole contribution added,
and no velocity-dependent terms.

With these
Eq.~(\ref{Eq:dotp0}) becomes (Attard 2025f)
\begin{eqnarray}
\lefteqn{
\frac{\partial {\bf j}_{20}}{\partial t}
+ \frac{2e^2}{m}
\frac{\partial (n_{20} {\bf A}) }{\partial t}
} \nonumber \\
& = &
\frac{ -e\beta \mu_{\rm B}^2 \overline{q}_2^2}{3m}
n_{20} \nabla (\nabla B)^2
- \frac{2e^2}{m} n_{20} \nabla \phi .
\end{eqnarray}
Here we have neglected the convective term
$(e/m)\nabla \cdot ({\bf p}_{20} {\bf v}_{20} )$,
which is second order in the velocity.

Taking the curl of this equation,
and using the facts that, by a Maxwell equation,
$\nabla \times {\bf B} = \mu_0 {\bf j}_{20}$,
as well as
$\nabla \times \nabla \times {\bf B} = - \nabla^2 {\bf B}$,
and that the curl of the gradient of a scalar is zero,
upon rearrangement we obtain
\begin{eqnarray}
\lefteqn{
\frac{\partial }{\partial t}
\left[ - \mu_0^{-1} \nabla^2 {\bf B}
+
\frac{ 2e^2 n_{20}}{m} {\bf B}
\right]
} \nonumber \\
& =&
\left[ \frac{e\beta \mu_{\rm B}^2 \overline{q}_2^2}{3m}\nabla (\nabla B)^2
+ \frac{2e^2}{m} \nabla \phi
\right] \times \nabla n_{20}
\nonumber \\ & & \mbox{ }
+ \frac{2e^2}{m} \frac{\partial ({\bf A}\times \nabla n_{20})}{\partial t} .
\end{eqnarray}

The first London equation,
which is derived from the Drude equation with zero resistivity,
has the same left hand side as this,
but it is zero on the right hand side (F and H London 1935).
(Of course the London brothers used the charge and mass of an electron
and the electron density,
whereas we use the charge and mass of an electron pair, and the density
of condensed bosonic pairs.)
The first London equation
predicts that a magnetic field in a superconductor cannot change with time
(Annett 2004, Kittel 1976, F and H London 1935, Tinkham 2004).
This would mean that a pre-existing magnetic field would remain trapped
inside an initially normal sample that transitioned
to the superconducting state,
which is contrary to the Meissner-Ochsenfeld (1933) effect.
This is the reason why the first London equation
has been rejected as too general.
Instead the  London brothers (1935)  focussed upon the particular solution
that is the second London equation (see below).

The starting point of the present equation
---that the time rate of change of the momentum flux
is given by the gradient of the chemical potential---
is better justified than the Drude equation assumed by the London brothers
(F and H London 1935).
The non-zero right hand side in the present result says that it is possible
for a magnetic field to change with time,
and thereby avoid being trapped,
if the density of condensed bosonic electron pairs is inhomogeneous,
$\nabla n_{20}({\bf r},t) \ne {\bf 0}$.
It is reasonable to assume that this is the case
during the superconducting transition.
If a sample in a magnetic field
is cooled below  the superconducting transition temperature,
then obviously the density of superconducting electrons must go from zero
to some finite value, which is to say that it is time dependent.
Also, in the process of the transition
the superconducting electrons are nucleated at different points in space
due to temperature and magnetic field inhomogeneities,
which means that $\nabla n_{20}({\bf r},t) \ne {\bf 0}$.
Indeed the magnetic quadrupole energy for bosonic electron pairs
is one source of nucleation inhomogeneity.

Of course after equilibration the magnetic field is independent of time
and both sides of this equation must be zero.
In this case the gradient of the magnetic field
and the gradient of the potential vanish
in the macroscopic interior of the sample.
In the surface region
the gradient of the condensed bosonic pair density either vanishes
or else lies parallel to the other gradients.

Setting the right hand side to zero, and integrating over time gives
for the steady state
\begin{equation} 
 -\mu_0^{-1}  \nabla^2 {\bf B}({\bf r})
+
\frac{ 2e^2 n_{20}}{m} {\bf B}({\bf r})
=
{\bf C}({\bf r}).
\end{equation}
But from the results in the previous section,
${\bf C}({\bf r}) \to 0$ in the interior of the superconductor.
Choosing the initial condition ${\bf C}({\bf r}) = {\bf 0}$ everywhere
gives a particular solution that is consistent with the
Meissner-Ochsenfeld (1933) effect.
This is the second London equation,
and it is believed to give
the decay of the magnetic field in the surface region
(Annett 2004, Kittel 1976, F and H London 1935, Tinkham 2004).

The second London equation gives the supercurrent
as proportional to the magnetic vector potential
(F and H London 1935, Tinkham 2004).
Therefore, the curl of the supercurrent
must vanish wherever the magnetic field vanishes,
such as in the interior of the superconductor.
In such regions the supercurrent is irrotational.

\subsubsection{Critical Magnetic Field} \label{Sec:Bcrit}

This section identifies the physical mechanism
by which superconductivity is destroyed
when the magnetic field exceeds a critical value.
We use an ideal electron model,
which, in the linear regime, can be shown to give
the known result for the Pauli paramagnetic susceptibility
(Pathria 1972 \S8.2).
We combine this with a treatment of Bose-Einstein condensation
using ideal bosons
(Attard 2025a Ch.~2,  Pathria 1972 \S7.1, F. London 1935),
as summarized in \S\ref{Sec:Ideal}.

The unpaired electrons, labeled $1\pm$,
are neglected  for the present purposes.
The paired electrons consist of fermionic pairs
in which both electrons have the same spin, $2\pm$,
and bosonic pairs in which the electrons have opposite spin,
and which are either condensed, $20$, or uncondensed, $2*$.
There is no prohibition on electrons in a fermionic pair
having the same spin as they are in different momentum states.
Also the Fermi repulsion
(cf.\ Attard 2025a \S2.2.5, Pathria 1976 \S5.5)
does not apply because the momentum is not integrated over,
and it is in any case weaker than the Coulomb repulsion
that is overcome by the binding potential.
The propensity to form electron pairs is determined
by the characteristics of the material and the thermodynamic state,
and is different for low- and for high-temperature superconductors
(Annett 2004, Attard 2025a \S6.5,
Bardeen \emph{et al.}\ 1957,  Tinkham 2004).

We treat the case that the applied magnetic field
partially or entirely penetrates the sample,
${\bf B} = [1+\chi] {\bf B}_{\rm ap}$,
with $\chi > -1 $.
Since the goal is restricted to discovering the electronic mechanism
by which superconductivity is destroyed,
for simplicity we take the magnetic field
to be uniform over the region being considered.
This means that we can neglect the magnetic quadrupole contribution
of the paired electrons.
We do not consider the effects of magnetic field inhomogeneity,
quantized flux tubes,
etc.
The effective fugacity for the unpaired electrons is
$z_\pm = z e^{\pm \beta \mu_0 \mu_{\rm B} B}$,
and that for the paired electrons is
$z_{2\pm} =  z_2 e^{\pm 2\beta \mu_0 \mu_{\rm B} B}$.
Below the superconducting transition,
$z_2 \equiv e^{\beta \mu_2^{(0)}} = 1^-$.
It is emphasized that this is an artefact of the ideal boson model.

By standard methods
(cf.\ Pathria 1972 \S8.2),
the average number of fermionic electron pairs is
\begin{equation}
\overline N_{2\pm}(z_{2\pm},V,T)
=
V 2^{-3/2} \Lambda^{-3} f_{3/2}(z_{2\pm}) ,
\end{equation}
where the Fermi-Dirac integral appears (Pathria 1976 Appendix~E).
The thermal wavelength for single electrons is
$\Lambda = \sqrt{2\pi \beta \hbar^2 /m}$.
Similarly
(cf.\ Attard 2025a Ch.~2, Pathria 1972 \S7.1),
the average number of uncondensed bosonic electron pairs is
\begin{eqnarray}
\overline N_{2*}(z_2,V,T)
& = &
V 2^{-3/2} \Lambda^{-3}   g_{3/2}(z_{2}),
\end{eqnarray}
where the Boise-Einstein integral appears (Pathria 1976 Appendix~D).
It is an artefact of the ideal boson model
that at a given temperature this has a maximum value,
$g_{3/2}(1) = \zeta(3/2) =2.612$.
This average number is independent of the magnetic field
and below the transition
it is insensitive to the actual value of the pair fugacity.

The number of condensed bosonic electron pairs is
\begin{equation}
\overline N_{20} 
= N_2 - \overline N_{2+} - \overline N_{2-} -  \overline N_{2*} .
\end{equation}
This is used below the superconducting transition.
Given the fixed number of electrons pairs $N_2$,
from measurement or other,
this determines the pair fugacity,
$z_2 = \overline N_{20}/[1+\overline N_{20}] \to 1^-$.
This is an artefact of the ideal electron model
and the treatment is analogous to that of F. London (1938)
for superfluidity (Attard 2025a Ch.~2, Pathria 1972 \S7.1).
We could include the unpaired electrons in this
without changing the conclusion.

At high fields the grand potential is dominated by $\Omega_{1+}$
and by $\Omega_{2+}$,
which favor the penetration of the magnetic field.
One can show that the total number of fermionic electron pairs
increases with increasing magnetic field,
\begin{equation}
\frac{\partial [\overline N_{2+} + \overline N_{2-}] }{\partial B}
> 0,
\end{equation}
and similarly for the unpaired electrons.
This is a non-linear effect
in which the increase in spin-up electrons is greater than
the decrease in spin-down electrons.
Since the  relatively small number of uncondensed bosonic pairs,
$\overline N_{2*} $,
is independent of the magnetic field,
and since the total number of pairs is determined
by the material and the thermodynamic state,
this shows that the number of condensed bosonic electron pairs
must decrease with increasing magnetic field,
\begin{equation}
\frac{\partial \overline N_{20}  }{\partial B} < 0.
\end{equation}
Hence there exists a critical field
at which the number of condensed bosonic electron pairs goes to zero
and superconductivity is annihilated.


The conclusion is that a magnetic field destroys superconductivity
because it non-linearly favors spin-up electrons,
which reduces the number of spin-down electrons available
for bosonic electron pairs.
When the bosonic electron pair density  falls below
their transition density in the absence of a field,
then superconductivity is destroyed.
Results of computer simulations in Fig.~\ref{Fig:fvsB} below
confirm this picture.

\subsection{Statistical Mechanics of Superconductivity}

\subsubsection{Fermion Pairs} \label{Sec:2e}

\begin{figure}[t]
\centerline{ \resizebox{8cm}{!}{ \includegraphics*{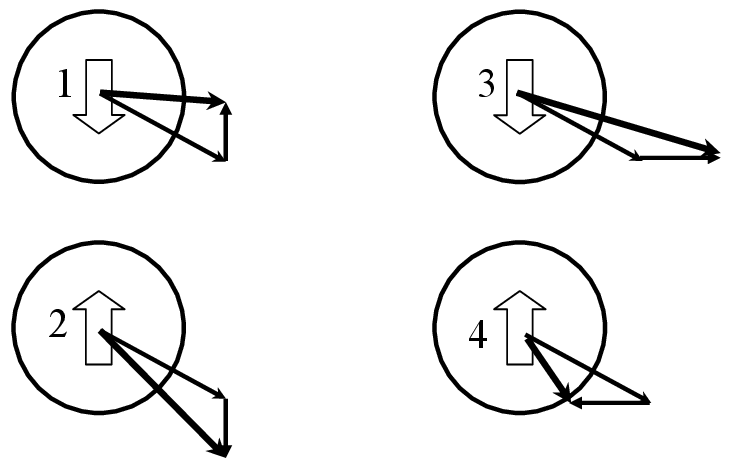} } }
\caption{\label{Fig:pair}
Paired electrons, $\{1,2\}$  (left), and $\{3,4\}$ (right),
in the total momentum state ${\bf P}$.
The block arrows indicate the spin state, $s_j$,
the thick arrows show each electron's momentum
${\bf p}_j = {\bf P}/2 + {\bm \pi}_j$,
the long thin arrow is the common value ${\bf P}/2$,
and the short thin arrows are the excess,
${\bm \pi}_1 = -{\bm \pi}_2$ and ${\bm \pi}_3 = -{\bm \pi}_4$,
with ${\bm \pi}_1 \ne \pm{\bm \pi}_3$.
}
\end{figure}

Cooper (1956) defined a Cooper pair of electrons
as having equal and opposite spin, $s_1 = -s_2$,
and equal and opposite momenta, ${\bf p}_1 = -{\bf p}_2$.
In the more general case,
the bosonic electron pairs are grouped into sets
with the same non-zero total momentum ${\bf P}$
(Fig.~\ref{Fig:pair}).
In general, permitted permutations
are those between individual fermions with the same spin.
Bosonic permutations are further restricted
to those between pairs in the same total momentum state.

For the four electrons in Fig.~\ref{Fig:pair},
there are four permitted permutations:
the identity,
the transpositions $\hat{\mathrm P}_{13}$ and $\hat{\mathrm P}_{24}$,
and their composition $\hat{\mathrm P}_{13}\hat{\mathrm P}_{24}$.
The symmetrization function
for these four fermions is therefore
\begin{eqnarray}
\lefteqn{
\sum_{\hat{\mathrm P}}
(-1)^p
e^{-{\bf q} \cdot [{\bf p}-{\bf p}']/\mathrm{i}\hbar}
\delta_{{\bf s}',{\bf s}}
} \nonumber \\
\! & = & \!
1
- e^{- {\bf q}_{13} \cdot {\bf p}_{13} /\mathrm{i}\hbar}
- e^{- {\bf q}_{24} \cdot {\bf p}_{24} /\mathrm{i}\hbar}
\nonumber \\ && \mbox{ }
+ e^{- {\bf q}_{13} \cdot {\bf p}_{13} /\mathrm{i}\hbar}
e^{- {\bf q}_{24} \cdot {\bf p}_{24} /\mathrm{i}\hbar}
\nonumber \\ \! & \approx & \!
1
+
e^{ - {\bf q}_{12} \cdot {\bf p}_{13}/\mathrm{i}\hbar}
e^{ - {\bf q}_{34} \cdot {\bf p}_{31}/\mathrm{i}\hbar}.
\end{eqnarray}
The two terms with a negative prefactor,
each of which corresponds to a single transposition,
have been neglected in the final equality.
This is justified because they oscillate much more rapidly
than the two terms that are retained.
To see this we simply note that
the neglected fermionic terms
have an exponent that depends upon the separation
between the  pairs,
which is with overwhelming probability  macroscopic.
The exponent of the final retained bosonic term
depends only upon the internal separations of the electrons
in each pair,
and these are of molecular size, as we shall see.
Simple algebra confirms the equality of the two ways of writing
the exponent for the double transposition in the above equation,
\begin{eqnarray}
\lefteqn{
{\bf q}_{13} \cdot {\bf p}_{13}
+
{\bf q}_{24} \cdot {\bf p}_{24}
} \nonumber \\
& = &
{\bf Q}_{13} \cdot {\bf P}_{13}
+
\frac{1}{2} ({\bf q}_{12}-{\bf q}_{34})
\cdot ({\bf p}_{13}-{\bf p}_{24})
\nonumber \\ & = &
\frac{1}{2} ({\bf q}_{12}-{\bf q}_{34})
\cdot ({\bm \pi}_{13} + {\bm \pi}_{13})
\nonumber \\ & = &
{\bf q}_{12} \cdot {\bf p}_{13}
+ {\bf q}_{34} \cdot  {\bf p}_{31}.
\end{eqnarray}
The center of mass separation for the pairs is
${\bf Q}_{13} = {\bf Q}_{1} - {\bf Q}_{3}
= ({\bf q}_{1}+{\bf q}_{2})/2 - ({\bf q}_{3}+{\bf q}_{4})/2 $,
and their total momentum difference is
${\bf P}_{13}
= ({\bf p}_{1}+{\bf p}_{2}) - ({\bf p}_{3}+{\bf p}_{4})$.
In the final equality,
the size of each pair,
which is the separation of the two electrons,
${\bf q}_{12} = {\bf q}_{1} - {\bf q}_{2}  $
and ${\bf q}_{34} = {\bf q}_{3} -{\bf q}_{4} $,
plays the r\^ole of its location as an effective boson.
That is, two bosons,
one located at ${\bf r}_1$ with momentum ${\bf p}_1$
and the other located ${\bf r}_3$ with momentum ${\bf p}_3$
would have symmetrization dimer
$ 1 +
e^{ - {\bf r}_{1} \cdot {\bf p}_{13}/\mathrm{i}\hbar}
e^{ - {\bf r}_{3} \cdot {\bf p}_{31}/\mathrm{i}\hbar}$,
which is the same as the final equality above
if one identifies ${\bf r}_{1} \equiv {\bf q}_{12}$
and ${\bf r}_{3} \equiv {\bf q}_{34}$.
As mentioned, we shall show that the size of the pair is molecular
and relatively constant, ${q}_{12} \approx {q}_{34} \approx \overline q$,
which means that the Fourier factors for these particular permutations
oscillate relatively slowly.

The Boltzmann-weighted momentum average
of the bosonic symmetrization function
shows that a bound fermion pair behaves
as a boson molecule with average internal weight
(Attard 2025a Eq.~(6.15))
\begin{equation} \label{Eq:numf}
\nu_\mathrm{mf}
\approx
\frac{\Lambda e^{-\pi \overline q^2/\Lambda^2} }{\overline q \sqrt{2\pi} }
\sqrt{ \sinh ( 2\pi \overline q^2/\Lambda^2 ) }.
\end{equation}
We expect this to hold when the binding potential
has a relatively narrow minimum at $\overline q$.
In the regime where the thermal wavelength
exceeds the mean size of the pairs,
this weight approaches unity.
This result has been derived for Cooper pairs with zero momentum
and uncorrelated orientations;
for non-zero momentum states it is reduced by a factor
of $e^{-\beta { P}^2/4m}$, as well as by an orientation factor.

In addition to this weight
each pair picks up bound volume factor (Attard 2025a Eq.~(6.19)),
\begin{equation} \label{Eq:vbnd}
v_\mathrm{bnd}
\approx
4\pi \overline q^2  e^{-\beta \overline w} \sqrt{2\pi/\beta \overline w''},
\end{equation}
where $w(q)$ is the pair potential of mean force.
This reflects the loss of configurational volume
by an electron bound in a pair.

With these the so-called monomer grand potential,
which includes $N_0$ condensed bosonic electron pairs,
which for simplicity are taken to have zero momentum,
and $N_1$ unpaired electrons,
so that the total number of electrons is $N =N_1 + 2 N_0$,
is given by (Attard 2025a Eq.~(6.28)),
\begin{eqnarray}
\lefteqn{
e^{ -\beta \Omega^{(1)}(z,V,T) }
}  \\ \nonumber
& = &
\sum_{N_{0},N_{1}}
\frac{z^N}{V^N} Q(N,V,T)
\left( \frac{ \nu v_\mathrm{bnd}
}{ 2^{3/2} \Lambda^{3} } \right)^{N_{0}}
\frac{ V^{N_{1}} }{ \Lambda^{3N_{1}} N_{1}!}.
\end{eqnarray}
The unpaired loop grand potentials $l \ge 2$
can be written
\begin{equation} 
-\beta \Omega^{-,(l)}_{1,\pm}
=
 (-1)^{l-1} N_{1,\pm}\left(\frac{N_{1,\pm}}{N}\right)^{l-1} g^{(l)}  ,
\end{equation}
where the intensive loop gaussians $g^{(l)}$ are as for bosons,
Eq.~(\ref{Eq:gl}).
The anti-symmetrization factor for fermions,
$(-1)^{l-1}$,
alternates the sign of successive terms,
which creates problems for the convergence of the series
approaching the transition.

\begin{figure}[t]
\centerline{ \resizebox{8cm}{!}{ \includegraphics*{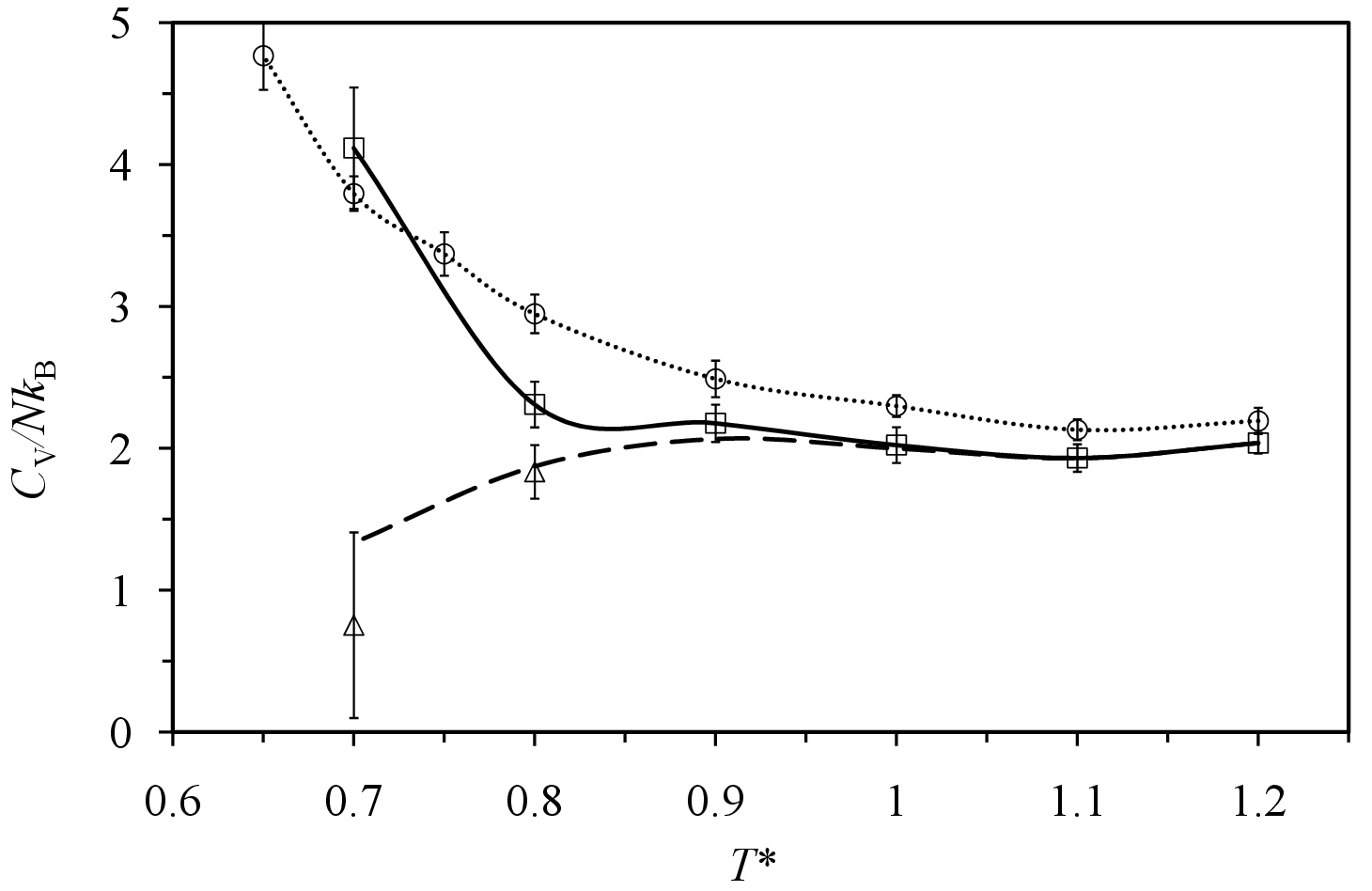} } }
\caption{\label{Fig:Cv10}
Specific heat capacity for Lennard-Jones $^3$He
along the saturation curve
with $l^\mathrm{max} =5$ (squares on solid curve),
$l^\mathrm{max} =4$ (dashed curve),
and $l^\mathrm{max} =6$ (triangles).
The circles on the dotted curve are for $^4$He with $l^\mathrm{max} =5$.
Data obtained with canonical Monte Carlo simulation
of an homogeneous system with $N_*=N=5,000$ (Attard 2022c).
The error bars give the 95\% confidence interval.
}
\end{figure}

\subsubsection{Computational Results for $^3$He}

Monte Carlo computer simulation results
have been obtained for Lennard-Jones $^3$He,
which is a fermion,
using the same algorithm as in \S\ref{Sec:QMCCPS}
(Attard 2022c, 2025a \S6.3).
Results for the heat capacity are shown in Fig.~\ref{Fig:Cv10}.
These include  the position permutation loops,
and it can be seen that these have an even/odd effect.
This makes it difficult to be sure of the convergence of the loop series,
particularly at lower temperatures.
The contrast with $^4$He is marked,
as the best estimate is that the specific heat capacity for  $^3$He
is actually decreasing approaching the lowest temperature  studied.
Certainly it is much lower than the diverging heat capacity
of  $^4$He as the $\lambda$-transition is approached,
which is no doubt due to the fact that only a fraction
of the $^3$He atoms are within the thermal energy of the Fermi surface.

In general terms the loop series approach is problematic for fermions.
Basically it is trying to satisfy the Fermi exclusion principle
by a numerical series
that can only be guaranteed exact if an infinite number of terms are retained.
In addition, the Lennard-Jones model
and the neglect of the commutation function
introduce approximations that challenge the quantitative applicability
of the results.
For example,
the lowest temperature studied here, $\approx 7$\,K,
is about three orders of magnitude higher
than the measured superfluid transition temperature in $^3$He, 2.5\,mK
(Osheroff {\emph{et al.}\ 1972a, 1972b).

\begin{figure}
\centerline{ \resizebox{8cm}{!}{ \includegraphics*{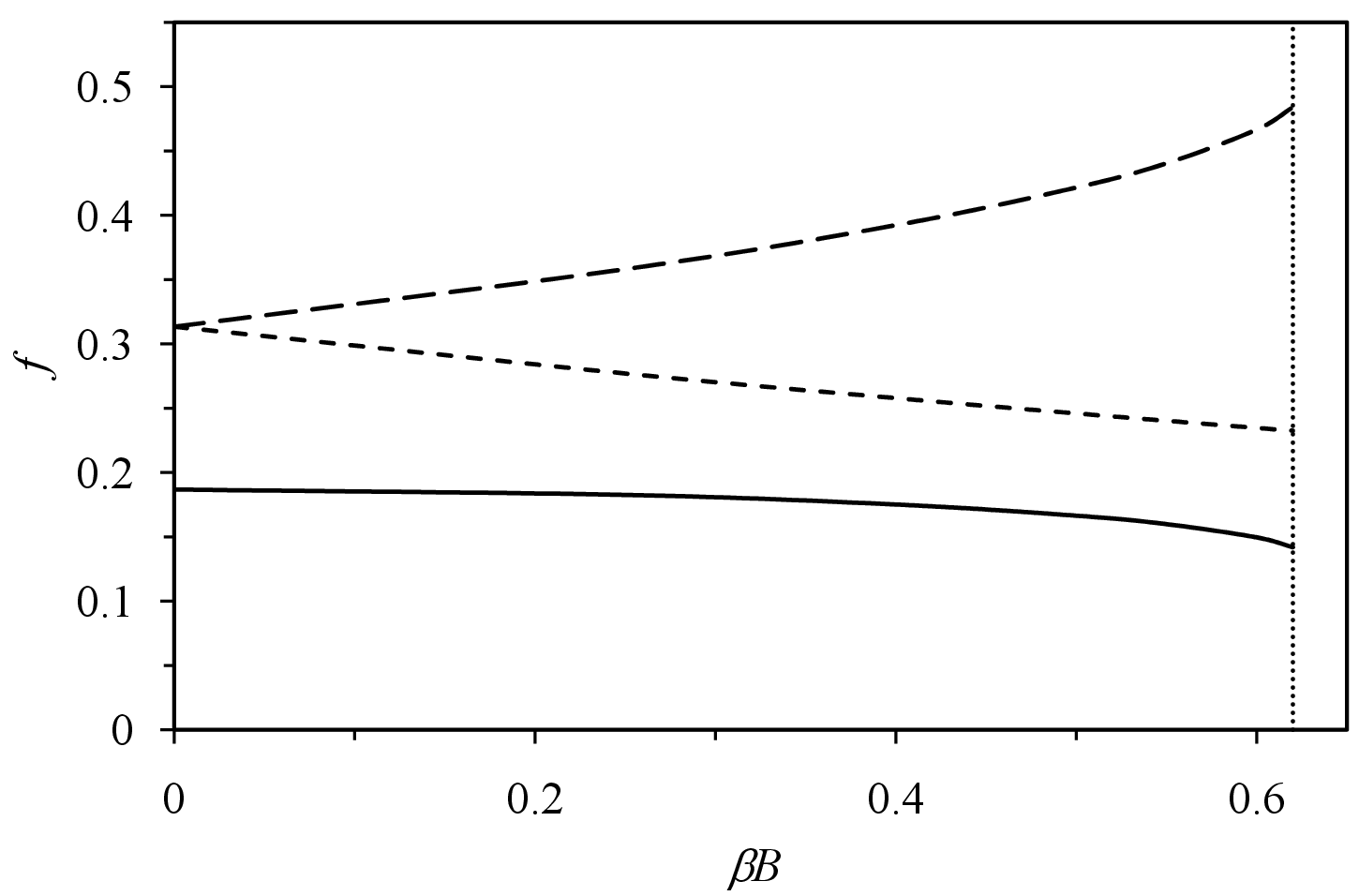} } }
\caption{\label{Fig:fvsB}
Optimum fraction of fermions
as a function of the magnetic energy per spin
($T^*=0.5$,
saturated liquid Lennard-Jones $^3$He, $S=1/2$ ).
The long dashed curve is for spin-up unpaired fermions,
the short dashed curve is for spin-down unpaired fermions,
and the solid curve is for paired fermions.
The dotted line indicates the critical magnetic field strength,
 $\beta B_\mathrm{crit} = 0.62$. 
Data from Attard (2022c).
}
\end{figure}

Adding a magnetic field to the analysis,
which favors unpaired electrons with spin-up,
shows that there is a critical magnetic field
that destroys condensation.
This can be seen in Fig.~\ref{Fig:fvsB},
where $B$ is the magnetic energy per spin
(see also Attard 2022c, 2025a \S6.2.4).
This finding is consonant with the ideal bosonic pair
analysis in \S\ref{Sec:Bcrit}.
The results in Fig.~\ref{Fig:fvsB} for interacting Lennard-Jones $^3$He
include the position permutation loops up to $l_{\rm max} = 5$.

\subsubsection{High-Temperature Superconductivity} \label{Sec:HighT-SC}

\begin{figure}[t!]
\centerline{ \resizebox{8cm}{!}{ \includegraphics*{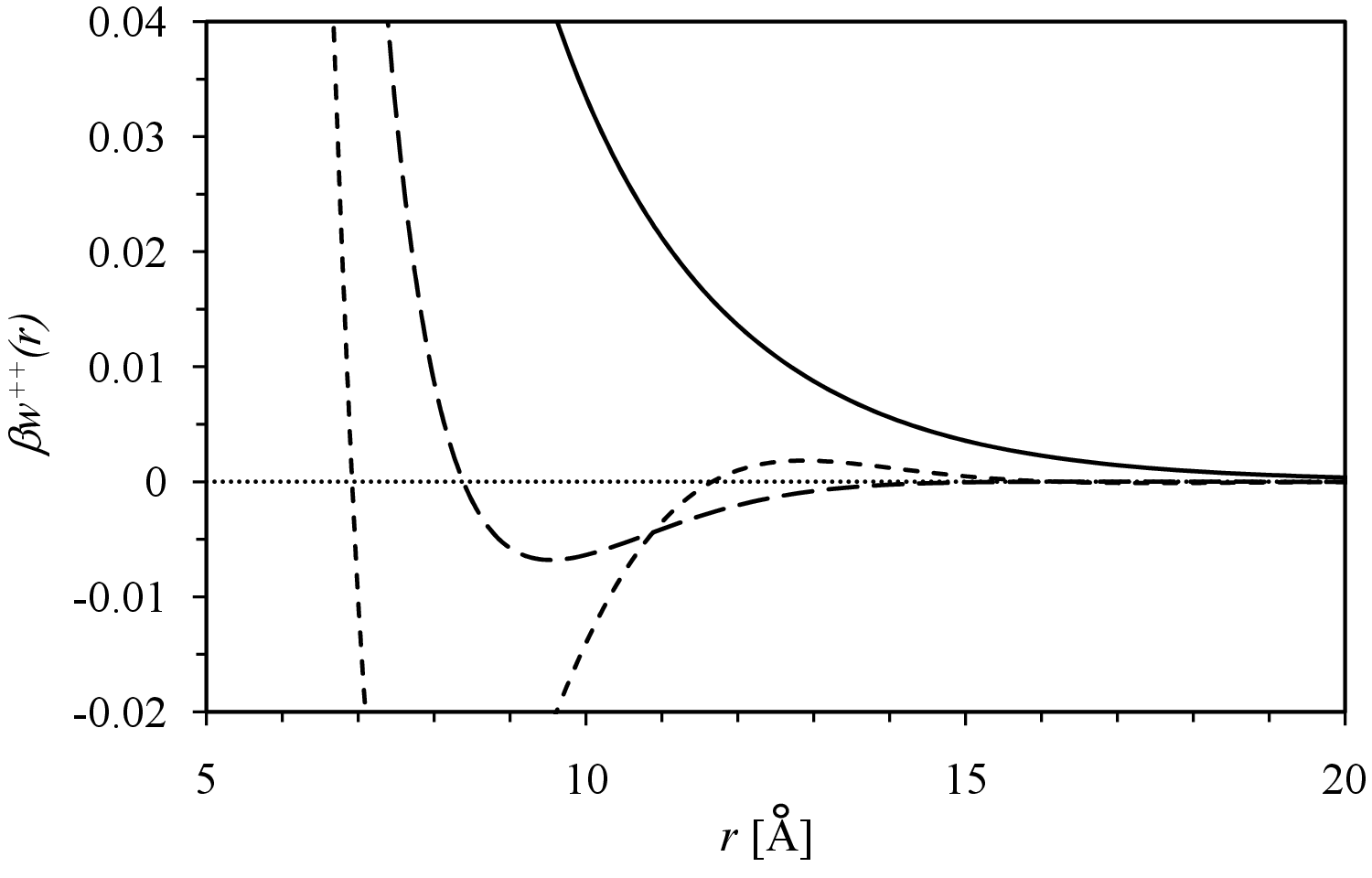} } }
\caption{\label{Fig:pmf2}
Pair potential of mean force between coions
in a symmetric binary monovalent electrolyte
($d=3.41$\,\AA, $\epsilon_\mathrm{r} = 100$,  $T=100$\,K,
hypernetted chain approximation).
The solid curve is $\Gamma = 1.8$
(0.5\,M, $\kappa_\mathrm{D}^2d^2 = 1.5$), 
the long-dash curve is $\Gamma = 2.3$
(1.0\,M, $\kappa_\mathrm{D}^2d^2 = 2.9 $), 
the short-dash curve is $\Gamma = 2.9$
(2.0\,M, $\kappa_\mathrm{D}^2d^2= 5.9 $). 
The dotted line is a guide to the eye.
From Attard (2025a Fig.~6.10).
}
\end{figure}

We now turn to the subject of high-temperature superconductivity
(Bednorz and  M\"oller 1986, Wu \emph{et al.}\ 1987),
and specifically to the nature of the potential
that binds the electron pairs.
Low-temperature superconductors,
for which BCS theory is appropriate,
have transition temperatures below 23\,K.
The first reports of high-temperature superconductors
showed a significant increase in transition temperature to 35\,K,
which has since been extended to 90--130\,K
in copper oxide materials (Tinkham 2004).
That these are above the temperature of saturated liquid nitrogen
at atmospheric pressure is obviously significant for practical applications.

The statistical mechanical theory of condensation in fermionic systems
requires the formation of bosonic pairs.
As discussed in connection with Eqs~(\ref{Eq:numf}) and (\ref{Eq:vbnd}),
in order for the pairs to have a meaningful statistical weight
the binding potential has to be localized at short range,
with a narrow, deep potential well.
Since quantum statistical mechanics
applies at high-temperatures,
this rules out the BCS binding potential,
which is long-ranged and diffuse.
Indeed, the fact that the measured transition temperatures
are independent of the isotopic masses of the solid
also demonstrates that BCS theory
does not apply to high-temperature superconductors.

The BCS theory is a quantum mechanical approach,
and on general grounds quantum mechanics works when entropy is immaterial,
namely at low temperatures.
Conversely, quantum statistical mechanics accounts for entropy,
and it applies at high temperatures.
It is obvious that the former,
irrespective of the actual binding potential,
has no relevance to high-temperature superconductivity.
Accordingly we pursue a quantum statistical mechanical approach,
which, as discussed in the preceding sections,
requires a binding potential with a deep, narrow minimum
at short-range.

The challenge with postulating a binding potential for electrons
is the Coulomb repulsion.
But in fact
attractive interactions between like-charged particles do exist,
as is well-known in charge fluids, such as the one-component plasma
and electrolytes.
It has long-been established that at high coupling
the static pair correlation function becomes oscillatory
(Attard 1993,
Brush \emph{et al.}\ 1966,
Ennis \emph{et al.}\ 1995,
Fisher and Widom 1969,
Outhwaite 1978,
Parrinello and Tosi 1979,
Stell  \emph{et al.}\ 1976,
Stillinger and Lovett 1968).
For such an oscillatory structure
the pair potential of mean force must have minima.
The question is whether the primary minimum
has sufficient depth, width, and location
to create bound fermion pairs according to the criteria given above.

The transition from monotonic behavior at low coupling
to oscillatory behavior at high coupling occurs at
(Attard 1993, Brush 1966)
\begin{equation}
\kappa_{\rm D} d = \surd 2 ,
\mbox{ or }
\Gamma = 2.
\end{equation}
Here $\kappa_{\rm D} \equiv \sqrt{8\pi\beta\rho z^2 e^2/\epsilon} $
is the Debye screening length,
which is analogous to the Thomas-Fermi screening length,
$d$ is the effective repulsive diameter of the charged species,
and $\Gamma \equiv \beta z^2 e^2/\epsilon(3/4\pi\rho^{1/3})$
is the plasma coupling parameter.
In general, coupling increases with
increasing ionic valence,
increasing number density,
increasing diameter,
decreasing dielectric constant,
and decreasing temperature.

\begin{figure}[t!]
\centerline{ \resizebox{8cm}{!}{ \includegraphics*{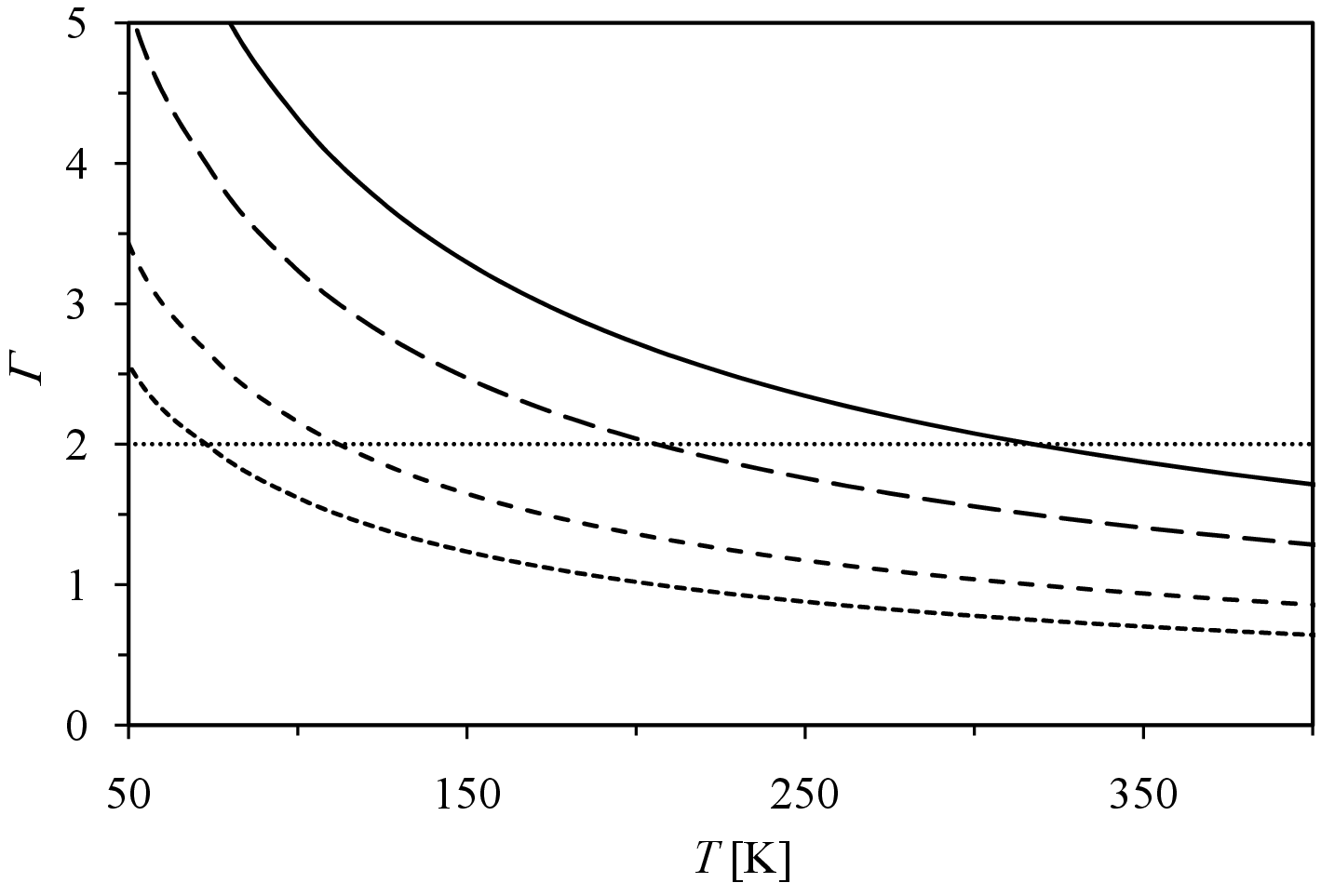} } }
\caption{\label{Fig:GvsT}
The plasma coupling parameter for different dielectric constants
for typical ceramics parameters.
From top to bottom, the relative permittivity is
$\epsilon_\mathrm{r} =$ 75, 100, 150 and 200.
The dotted line marks the oscillatory transition.
From Attard (2025a Fig.~6.11).
}
\end{figure}

The monotonic-oscillatory transition is graphically illustrated
in Fig.~\ref{Fig:pmf2}.
It can be seen that the coion pair potential of mean force
acquires a minimum at short-range
whose depth increases and separation decreases with increasing coupling.
The physical origin of the oscillatory behavior and the potential minimum
is easily understood:
for ions with finite size,
in order to maintain electroneutrality at high densities,
packing constraints demand above average placement
at nearest neighbor spacing.
The size can be due to a hard core,
or Coulomb, or other repulsion.

Qualitatively at least the evolution of the minimum
in the pair potential of mean force
beyond the monotonic-oscillatory transition
satisfies the requirements
for the bosonic binding potential and superconducting transition.
Quantitatively,
Fig.~\ref{Fig:GvsT} shows that for parameters typical of ceramic materials
the monotonic-oscillatory transition temperature is within the range measured
for the superconducting transition temperature
for high-temperature superconductors
(Annett 2004, Bednorz and  M\"oller 1986,
Tinkham 2004, Wu \emph{et al.}\ 1987).

The superconducting transition requires
the existence of a short-ranged potential minimum
and also that the density of paired electrons
relative to the thermal wavelength be large enough
for condensation to occur.
Identifying the superconducting transition
with the monotonic-oscillatory transition
assumes that the second condition is also satisfied,
which may not always be the case.
That is,
the monotonic-oscillatory transition is a necessary
but not sufficient condition
for the superconducting transition in high-temperature superconductors.
Also, the anisotropy
of the layered copper oxide high-temperature superconductors
may modify the present prediction for the transition
in particular cases.


%
\setcounter{equation}{0} \setcounter{subsubsection}{0}
\renewcommand{\theequation}{\arabic{section}.\arabic{equation}}
%

\section*{References}


\begin{list}{}{\itemindent=-0.5cm \parsep=.5mm \itemsep=.5mm}

\item 
Abo-Shaeer J R, Raman C, Vogels J M, and Ketterle W 2001
Observation of vortex lattices in Bose-Einstein condensates
\emph{Science}  {\bf 292} 476

\item 
Ahlers G 1969
Critical heat flow in a thick He II film near the superfluid transition
\emph{J.\ Low Temp.\ Phys.}\  {\bf 1} 159

\item 
Allen M P and Tildesley D J 1987
\emph{Computer Simulation of Liquids}
(Oxford: Clarendon Press)

\item 
Allum D R, McClintock P V E, and Phillips A 1977
The breakdown of superfluidity in liquid He:
an experimental test of Landau's theory
\emph{Phil.\ Trans.\ R.\ Soc.\ A} {\bf 284} 179

\item 
Andronikashvilli E L 1946
\emph{J.\ Phys.\ USSR} {\bf 10} 201

\item 
Annett J E 2004
\emph{Superconductivity, Superfluids and Condensates}
(Oxford: Oxford University Press)

\item 
Attard P 1993
Asymptotic analysis of primitive model electrolytes
and the electrical double layer
\emph{Phys.\ Rev.\ E} {\bf 48} 3604

\item 
Attard P 2002
\emph{Thermodynamics and Statistical Mechanics:
Equilibrium by Entropy Maximisation}
(London: Academic)

\item 
Attard  P 2012
\emph{Non-equilibrium Thermodynamics and Statistical Mechanics:
Foundations and Applications}
(Oxford: Oxford University Press)

\item
Attard P 2017
Quantum statistical mechanics results for argon, neon, and helium using
classical Monte Carlo
arXiv:1702.00096

\item 
Attard P 2018b
Quantum statistical mechanics in classical phase space. Expressions for
the multi-particle density, the average energy, and the virial pressure
arXiv:1811.00730

\item 
Attard P  2021
\emph{Quantum Statistical Mechanics in Classical Phase Space}
(Bristol: IOP Publishing)


\item  
Attard P 2022b
Attraction between electron pairs in high temperature superconductors
arXiv:2203.02598

\item  
Attard P 2022c
New theory for Cooper pair formation and superconductivity,
 arXiv:2203.12103v2

\item 
Attard P 2023d
Hamilton's equations of motion from Schr\"odinger's equation
arXiv:2309.03349

\item 
Attard P 2025a
\emph{Understanding Bose-Einstein Condensation,
Superfluidity, and High Temperature Superconductivity}
(London: CRC Press)

\item 
Attard P 2025b
The molecular nature of superfluidity: Viscosity of helium from quantum
stochastic molecular dynamics simulations over real trajectories
arXiv:2409.19036v5

\item 
Attard P 2025d
Bose-Einstein condensation and the lambda transition
for interacting Lennard-Jones helium-4
arXiv:2504.07147v3

\item 
Attard P 2025e
The two-fluid theory for superfluid hydrodynamics
and rotational motion
arXiv:2505.08826v6

\item 
Attard P 2025f
Thermodynamic explanation of the Meissner-Ochsenfeld effect
in superconductors
arXiv:2509.14247

\item 
Balibar S 2014
Superfluidity: how quantum mechanics became visible
pages 93--117
in
\emph{History of Artificial Cold, Scientific, Technological
and Cultural Issues}
(Gavroglu K editor)
(Dordrecht: Springer)

\item 
Balibar S 2017
Laszlo Tisza and the two-fluid model of superfluidity
\emph{C.\ R.\ Physique} {\bf 18}  586

\item 
Bardeen J, Cooper L N, and Schrieffer J R 1957
Theory of superconductivity
\emph{Phys.\ Rev.}\ {\bf 108} 1175

\item  
Batrouni G G, Ramstad T,  and Hansen A 2004
Free-energy landscape and the critical
velocity of superfluid films
\emph{Phil.\ Trans.\ R.\ Soc.\ Lond.}\ A {\bf 362} 1595

\item 
Bednorz J G and  M\"oller K A 1986
Possible high $T_\mathrm{c}$ superconductivity in the Ba-La-Cu-O system
\emph{Z.\ Phys.\ B} {\bf 64} 189

\item 
Bogolubov N 1947
On the theory of superfluidity
\emph{J. Phys.}\ {\bf 11} 23

\item 
Brush S G, Sahlin H L, and Teller E 1966
Monte Carlo study of a one-component plasma. I.
\emph{J.\ Chem.\ Phys.}\ {\bf 45} 2102

\item 
Caldeira A O and Leggett A J 1983
Quantum tunnelling in a dissipative system
\emph{Ann.\ Phys.}\ {\bf 149} 374

\item 
Ceperley  D M  1995
Path integrals in the theory of condensed helium
\emph{Rev.\ Mod.\ Phys.}\ {\bf 67} 279

\item 
Clow J R and Reppy J D 1967
\emph{Phys.\ Rev.\ Lett.}\ {\bf 19} 291

\item 
Cooper L  N 1956
Bound electron pairs in a degenerate Fermi gas
\emph{Phys.\ Rev.}\ {\bf 104} 1189

\item 
de Groot S R and Mazur P 1984
\emph{Non-equilibrium Thermodynamics}
(New York: Dover)

\item 
Donnelly R J 2009
The two-fluid theory and second sound in liquid helium
\emph{Physics Today} {\bf 62} 34

\item 
Donnelly R J and  Barenghi C F 1998
The observed properties of liquid helium at the saturated vapor pressure
\emph{J.\ Phys.\ Chem.\ Ref.\ Data} {\bf 27} 1217

\item 
Einstein A 1924
Quantentheorie des einatomigen idealen gases
Sitzungsberichte der Preussischen Akademie der Wissenschaften
{\bf XXII} 261 

\item 
Einstein A  1925
Quantentheorie des einatomigen idealen Gases. Zweite abhandlung.
Sitzungsberichte der Preussischen Akademie der Wissenschaften
{\bf I} 3

\item 
Ennis J,  Kjellander R, and Mitchell D J 1995
Dressed ion theory for bulk symmetric electrolytes
in the restricted primitive model
\emph{J.\ Chem.\ Phys.}\ {\bf 102} 975

\item 
Fetter A L 1963
Vortex rings and the critical velocity in helium II
\emph{Phys.\ Rev.\ Lett.}\ {\bf 10} 507

\item 
Feynman R P 1953
The $\lambda$-transition in liquid helium
\emph{Phys.\ Rev.}\ {\bf 90} 1116.
Atomic theory of the $\lambda$-transition in helium
\emph{Phys.\ Rev.}\ {\bf 91} 1291.
Atomic theory of liquid helium near absolute zero
\emph{Phys.\ Rev.}\ {\bf 91} 1301.

\item 
Feynman R P 1954
Atomic theory of the two-fluid model of liquid helium
\emph{Phys.\ Rev.}\ {\bf 94} 262

\item 
Feynman R P 1955
\emph{Progress in Low Temperature Physics}
ed.\ C J Gorter
(Amsterdam: North Holland) {\bf 1} 17

\item 
Fisher M E  and Widom B 1969
Decay of correlations in linear systems
\emph{J.\ Chem.\ Phys.}\ {\bf 50} 3756

\item 
Ginzburg V L and Pitaevskii L P 1958
\emph{Zh.\ Eksperim.\ i.\ Teor.\ Fiz.}\ {\bf 34} 1240.
\emph{Sov.\ Phys.\ JETP} {\bf 7} 858).

\item 
Gor'kov L P 1959
\emph{Zh.\ Eksperim.\ i.\ Teor.\ Fiz.}\ {\bf 36} 1918.
Microscopic derivation of the Ginzburg-Landau equations
in the theory of superconductivity
\emph{Sov.\ Phys.\ JETP} {\bf 9} 1364.

\item 
Green M S 1954
Markoff random processes and the statistical mechanics
of time-dependent phenomena.
II. Irreversible processes in fluids.
\emph{J.\ Chem.\ Phys.}\ {\bf 23} 298

\item 
Gross E P 1958
Classical theory of boson wave field
\emph{Annals of Phys.}\ {\bf 4} 57

\item 
Gross E P 1960
Quantum theory of interacting bosons
\emph{Annals of Phys.}\ {\bf 9} 292

\item 
Hammel E F  and Keller W E 1961
Fountain pressure measurements in liquid He II
\emph{Phys.\ Rev.}\ {\bf 124} 1641

\item 
Joos E and Zeh H D 1985
The emergence of classical properties through
interaction with the environment
\emph{Z.\ Phys.}\  B {\bf 59} 223

\item 
Kalos M H Lee M A  Whitlock P A and Chester G V 1981
Modern potentials and the properties of condensed He 4
\emph{Phys.\ Rev.}\ B {\bf 24} 115

\item 
Kawatra M P and Pathria R K 1966
Quantized vortices in an imperfect
Bose gas and the breakdown of superfluidity in liquid helium II
\emph{Phys.\ Rev.}\ {\bf 151}  132

\item 
Kirkwood J G 1933
Quantum statistics of almost classical particles
\emph{Phys.\ Rev.}\ {\bf 44}, 31

\item  
Kittel C 1976
\emph{Introduction to Solid State Physics}
(New York: Wiley 5th edn)

\item 
Kubo R 1966
The fluctuation-dissipation theorem
\emph{Rep.\ Prog.\ Phys.}\ {\bf 29} 255

\item 
Landau L D 1937
On the theory of phase transitions
\emph{Zh.\ Eksperim.\ i.\ Teor.\ Fiz.}\ {\bf 7} 19

\item 
Landau L D 1941
Theory of the superfluidity of helium II
\emph{Phys.\ Rev.}\ {\bf 60} 356.
Two-fluid model of liquid helium II
\emph{ J.\ Phys.\ USSR} {\bf 5} 71

\item 
Landau L D and  Lifshitz E 1955
\emph{Doklady Akad.\ Nauk USSR} {\bf 100} 669

\item 
Lane C T 1962
\emph{Superfluid Physics} (McGraw-Hill: New York)

\item 
Lifshitz E  and Kaganov M 1955
\emph{J.\ Exptl.\ Theoret.\ Phys.\ USSR} {\bf 29} 257.
Translated in \emph{Soviet Phys.\ JETP} {\bf 2} 172.

\item 
Lipa J A, Swanson D R, Nissen J A, Chui T C P, and Israelsson U E 1996
Heat capacity and thermal relaxation of bulk helium
very near the lambda point
\emph{Phys.\ Rev.\ Lett.}\ {\bf 76} 944

\item  
London F 1938
The $\lambda$-phenomenon of liquid helium and the Bose-Einstein degeneracy
\emph{Nature} {\bf 141} 643

\item 
London F 1945
Planck's constant and low temperature transfer
\emph{Rev.\ Mod.\ Phys.}\ {\bf 17} 310

\item 
London F and London H 1935
The electromagnetic equations of the supraconductor
\emph{Proc.\ Royal Soc.\ (London)} {\bf A149} 72.
Supraleitung und diamagnetismus
\emph{Physica} {\bf 2} 341. 

\item 
London H 1939
Thermodynamics of the thermomechanical effect of liquid He II
\emph{Proc.\ Roy.\ Soc.}\ {\bf  A171} 484

\item 
McMillan W L 1965
Ground state of liquid $^4$He
\emph{Phys.\ Rev.}\ A {\bf 138} 442

\item 
Meissner W and Ochsenfeld R 1933
Ein neuer effekt bei eintritt der supraleitf\"ahigkeit
\emph{Die Naturwissenschaften} {\bf 21} 787

\item 
Onsager L (1931)
Reciprocal relations in irreversible processes. I.
\emph{Phys.\ Rev.}\ {\bf 37} 405.
Reciprocal relations in irreversible processes. II.
\emph{Phys.\ Rev.}\ {\bf 38} 2265.

\item  
Onsager L 1949
Statistical hydrodynamics
\emph{Nuovo Cim.}\ {\bf 6} Suppl.~2 279

\item  
Osborne D V 1950
The rotation of liquid helium II
\emph{Proc.\ Phys.\ Soc.\ London} A {\bf 63} 909

\item
Osheroff D D, Richardson R C, and Lee D M 1972a
Evidence for a new phase of solid He3
\emph{Phys.\ Rev.\ Lett.}\ {\bf 28}  885

\item
Osheroff D D, Gully W J, Richardson R C, and Lee D M 1972b
New magnetic phenomena in liquid He3 below 3 mK
\emph{Phys.\ Rev.\ Lett.}\ {\bf 29} 920

\item 
Outhwaite C W 1978
Modified Poisson-Boltzmann equation in electric double layer theory
 based on the Bogoliubov-Born-Green-Yvon integral  equations
\emph{J.\ Chem.\ Soc.\ Faraday Trans.\ 2}  {\bf 74} 1214

\item 
Parrinello M and Tosi M P 1979
Structure and dynamics of simple ionic liquids
\emph{Rev.\ Nuovo Cimento} {\bf 2}  1

\item 
Pathria R K 1972
\emph{Statistical Mechanics} (Oxford: Pergamon Press)

\item  
Penrose O and Onsager L 1956
Bose-Einstein condensation and liquid helium
\emph{Phys.\ Rev.}\ {\bf 104} 576

\item 
Schlosshauer M 2005
Decoherence, the measurement problem,
and interpretations of quantum mechanics
arXiv:quant-ph/0312059v4

\item 
van Sciver  S W 2012
\emph{Helium Cryogenics}
(New York: Springer 2nd edition)

\item 
Stell G, Wu K C, and Larsen B 1976
Critical point in a fluid of charged hard spheres
\emph{Phys.\ Rev.\ Lett.}\ {\bf 37} 1369

\item 
Stillinger F H and Lovett R 1968
Ion-pair theory of concentrated electrolytes. I. Basic concepts
\emph{J.\ Chem.\ Phys.}\ {\bf 48} 3858

\item 
Tinkham M 2004
\emph{Introduction to Superconductivity}
(New York: Dover 2nd edn)

\item  
Tisza L 1938
Transport phenomena in helium II
\emph{Nature} {\bf 141} 913

\item  
Tisza L 1947
Theory of liquid helium
\emph{Phys.\  Rev.}\ {\bf 72} 838

\item  
Walmsley R H and Lane C T 1958
Angular momentum of liquid helium
\emph{Phys.\ Rev.}\ {\bf 112} 1041

\item  
Whitlock P A and Panoff R M 1987
Accurate momentum distributions from computations on $^3$He and $^4$He
\emph{Can.\ J.\ Phys.}\ {\bf 65} 1409

\item 
Wigner E 1932
On the quantum correction for thermodynamic equilibrium
\emph{Phys.\ Rev.}\ {\bf 40}, 749

\item 
Wu M K, Ashburn J R, Torng C J, Hor P H, Meng R L,
Gao L, Huang Z J, Wang Y Q, and Chu C W  1987
Superconductivity at 93\,K in a new mixed-phase Y-Ba-Cu-O compound system
at ambient pressure
\emph{Phys.\ Rev.\  Lett.}\ {\bf 58} 908


\item 
Zurek W H 1991
Decoherence and the transition from quantum to classical
\emph{Phys.\ Today} {\bf 44} 36

\item 
Zurek W H,  Cucchietti F M, and  Paz J P 2003
Gaussian decoherence from random spin environments
arXiv:quant-ph/0312207


\end{list}



%
%

\end{document}